\documentclass[prd,aps,floats,floatfix,eqsecnum,nofootinbib]{revtex4}
\usepackage{array,amsmath,amssymb,verbatim}
\usepackage{bm}
\usepackage{graphicx}
\usepackage{psfrag}
\newcommand{\avg}[1]{\langle #1 \rangle}

\newcommand{\bds}[1]{\boldsymbol{#1}}
\newcommand{\rotatefigure}[2]

\newcommand{\be}{\begin{equation}}
\newcommand{\ee}{\end{equation}}
\newcommand{\bea}{\begin{eqnarray}}
\newcommand{\eea}{\end{eqnarray}}

\begin{document}

\def\E{{\cal E}}
\def\H{{\hat{\cal H}}}
\def\hphi{{\hat\phi}}
\def\dm{\partial_\mu}
\def\udm{\partial^\mu}
\def\dk{\frac{d^3k}{(2\pi)^3}}
\def\bk{\overline{k}}
\def\Teff{T_{\rm eff}(t)}
\def\Tpeff{T_{\rm p,eff}(t)}

\def\ephik2{\avg{|{\tilde\phi}_\vk|^2}}
\def\epik2{\avg{|{\tilde\pi}_\vk|^2}}
\def\phivk2{\overline{|{\tilde\phi}_\vk|^2}(t)}
\def\pivk2{\overline{|{\tilde\pi}_\vk|^2}(t)}
\def\phik2{\overline{|{\tilde\phi}_k|^2}(t)}
\def\pik2{\overline{|{\tilde\pi}_k|^2}(t)}

\def\dk3{\frac{d^3k}{(2\pi)^3}}

\def\vx{{\bds x}}
\def\vn{{\bds n}}
\def\vk{{\bds k}}
\def\vs{{\bds\sigma}}
\def\vt{{\bds\tau}}

\date{}

\title{Ultraviolet cascade in the thermalization of the classical 
  $\phi^4$ theory in $3+1 $ dimensions.}
\author{C. Destri$^{(a)}$}
\email{Claudio.Destri@mib.infn.it}
\author{H. J. de Vega$^{(b)}$}
\email{devega@lpthe.jussieu.fr} \affiliation{ 
$^{(a)}$ Dipartimento di Fisica G. Occhialini, Universit\`a
Milano-Bicocca Piazza della Scienza 3, 20126 Milano and
INFN, sezione di Milano, via Celoria 16, Milano Italia\\
$^{(b)}$ LPTHE, Universit\'e Pierre~et~Marie Curie,~~Paris VI et
Denis Diderot, Paris VII, Laboratoire Associ\'e au CNRS UMR 7589,
Tour 24, 5\`eme.~~\'etage, 4, Place
Jussieu, 75252 Paris, Cedex 05, France}
\affiliation{Observatoire de Paris, LERMA.
Laboratoire Associ\'e au CNRS UMR 8112.
 \\61, Avenue de l'Observatoire, 75014 Paris, France.}
\begin{abstract}
  We investigate the dynamics of thermalization and the approach to
  equilibrium in the classical $\phi^4$ theory in $3+1$ spacetime
  dimensions.  The non-equilibrium dynamics is studied by numerically
  solving the equations of motion in a light--cone--like discretization of
  the model for a broad range of initial conditions and energy densities. A
  smooth cascade of energy towards the ultraviolet is found to be the basic
  mechanism of thermalization.  After an initial transient stage, at a time
  scale of several hundreds inverse masses, the squared magnitude of the
  field spatial gradient becomes larger than the nonlinear term and there
  emerges a stage of universal cascade, independent on the details of the
  initial conditions. As the cascade progresses, the modes with higher
  wavenumbers, but well behind the forefront of the cascade, exhibit weaker
  and weaker nonlinearities well described by the Hartree approximation,
  while the infrared modes retain strong selfinteractions. As a
  consequence, two timescales for equilibration appear as characteristic of
  two time-dependent wavenumber regions. For $ k^2 \gtrsim
  \overline{\phi^2}(t) $, we observe an effective equilibration to a
  time--dependent power--like spectrum with a time scale in the hundreds of
  inverse masses; cutoff effects are absent and the Hartree approximation
  holds for $ k^2 \gg \overline{\phi^2}(t) $.  On the other hand, infrared
  modes with $ k^2 \lesssim \overline{\phi^2}(t) $ equilibrate only by time
  scales in the millions of inverse masses when cutoff effect become
  dominant and complete thermalization is setting in. Accordingly, we
  observe in the field correlator a relatively large and long--lived
  wavefunction renormalization of nonperturbative character. There
  corresponds an effective mass governing the long distance behaviour of
  the correlator which turns to be significantly smaller than the Hartree
  mass which is exhibited by the modes with $ k^2 \gtrsim
  \overline{\phi^2}(t)$.  Virialization and the equation of state start to
  set much earlier than thermalization. The applicability of these results
  in quantum field theory for large occupation numbers and small coupling
  is analyzed.
\end{abstract}
\date{\today}
\maketitle \tableofcontents

\section{Introduction}

The understanding of the dynamics of thermalization and relaxation in a
field theory is a subject of critical importance both in early cosmology as
well as in ultrarelativistic heavy ion collisions.

Pioneering work in this topic was initiated by Fermi, Pasta and
Ulam\cite{FPU} for a chain of coupled oscillators.  Since then, this
problem has been studied within a variety of models\cite{varios} with the
goal of answering fundamental questions on ergodicity, equipartition and in
general the approach to equilibrium in non-linear theories with a large but
finite number of degrees of freedom.

By understanding the dynamics of thermalization we mean to uncover in
detail the mechanisms that leads to thermal states starting from arbitrary
initial conditions as well as to determine the time scales of these
phenomena.  In cosmology the inflationary paradigm \emph{assumes} that
after the inflationary stage a period of particle production and relaxation
leads to a state of local thermal equilibrium thus merging inflation with
the standard hot big bang cosmology\cite{kolbook,lidlyt}.

Inflationary scenarios lead to particle production either via parametric
amplification of fluctuations of the inflaton (in the case of an
oscillating inflaton) or by spinodal instabilities during phase
transitions\cite{cosmo,ours,ours2,Ngran}. In both cases the non-equilibrium
dynamics is non-perturbative and results in a large population of soft
quanta whose dynamics is nearly classical. The non-equilibrium dynamics of
particle production and eventual thermalization are non-perturbative and
the resulting fluctuations contribute to the evolution of the scale factor,
namely the back-reaction from the fluctuations becomes important in the
evolution of the cosmological space-time\cite{cosmo,ours,ours2,Ngran}.  Both
parametric amplification or spinodal decomposition lead to non-perturbative
particle production of a band of wave-vectors, typically for soft
momenta\cite{cosmo,ours,ours2,Ngran}. This non-perturbatively large population
allows a classical treatment of the non-equilibrium evolution.
Thermalization may be described in this classical framework or one can 
implement the quantum 2PI framework in cosmological spacetimes.

\medskip

In this article we study the non-equilibrium dynamics going towards
thermalization in the classical $\phi^4$ theory in $3+1$ dimensions. As
mentioned above the initial stages of non-equilibrium dynamics either in
cosmology or in ultrarelativistic heavy ion collisions is mainly
\emph{classical}.  Classical field theory must be understood with an
ultraviolet cutoff, or equivalently an underlying lattice spacing, to avoid
the Rayleigh-Jeans catastrophe.

The advantage of studying the classical field theory is that the equation
of motion can be solved exactly. Contrary to quantum theory, there is no
need of making approximations. Classical field theory is expected to be a
good approximation to the quantum theory for large occupation numbers and
small coupling as we argue in this paper.

We provide a detailed understanding of the \emph{main dynamical
  mechanisms} that lead to thermalization and to study the approach to
thermalization by several different observables. We address several
questions on the equilibrium and non-equilibrium aspects: i) what are the
criteria for thermalization in an interacting theory?, ii) what is the
mechanism that leads to thermalization?, iii) what is the dynamics for
different observables, how they reach thermal equilibrium and which are the
relevant time scales?.

We focus our study on these issues within the context of classical field
theory, which is an interesting and timely problem all by itself. It is
also \emph{likely} to describe the initial stages of evolution strongly out
of equilibrium in quantum theory for the relevant cases of very large
occupation numbers and small coupling.

\bigskip

The main results of this work are the following:

\begin{itemize}

\item We implement a discretization of the $\varphi^4$ model in any number
  of space-time dimensions which is very accurate and stable, maintains the
  relativistic symmetry between space and time and conserves energy exactly
  (that is to machine accuracy on a computer). The same discretization
  scheme, restricted to space alone, can be used to study the model at
  canonical equilibrium through Monte Carlo simulations.

\item We provide the main physical properties of the classical $\varphi^4$
  model in thermal equilibrium using low temperature perturbative
  expansions as well as Monte Carlo simulations. These simulations show that
  the leading order perturbative results have a large domain of validity,
  further extended by the Hartree approximation, except for infrared modes,
  which retain a strong nonperturbative character.
 
\item We extensively investigate the dynamics of the $ \varphi^4 $ model
  for a wide range of (infrared supported) initial conditions and energy
  densities. After a first stage with relatively important fluctuations
  whose precise structure depends on the details of the initial conditions,
  the model evolves towards a {\bf universal stage} where energy transfer
  from low to high wavenumbers becomes steady and very effective, resulting
  in a steady smooth {\bf ultraviolet cascade}. Namely, the power spectrum
  of the field $\phi$ and its canonical momentum $\pi$ acquire support in a
  monotonic and slow way over larger and larger values of the wavevectors.
  This ultraviolet cascade leads to a efficient transfer of energy from
  $\pi^2$, $\phi^2$ and the interaction terms $\phi^4$ to the spatial
  gradient $\left( {\nabla \phi}\right)^2 $ which grows monotonically.
  Therefore, there is a crossover during this stage from a strongly to a
  weakly interacting theory since the non-linear term $\phi^4$ becomes much
  smaller than the spatial gradients. The average wavenumber $\bk(t)$ of
  the modes grows monotonically with time approximately as $ t^{1/3} $.
  The {\bf universal stage} is well established by a time $t_0 \simeq 500$
  (in inverse mass units) which does not depend on the lattice spacing and
  depends weakly on the energy density $E/V$ as long as $E/V$ is not very
  small. The study of the $\phi$ and $\pi = {\dot \phi} $ correlators
  allows us to obtain the power spectrum of $\phi$ and $\pi$ which exhibits
  universal scaling properties during the smooth cascade. We find that
  virialization sets rather fast at times $t\gtrsim t_0$.  Parallel to
  virialization, the equation of state approaches the radiation equation.
  Namely, the ratio of the pressure divided by the energy density approaches
  $\frac13$ for energy densities not too small as the nonlinearities weaken.
  The ultraviolet cascade continues till the lattice cutoff is reached
  and therefore the universal properties w.r.t. the discretization method
  are lost. These cutoff effects become important long before full
  thermalization of the power spectra sets in, but late enough, for small
  enough lattice spacing, for the window of the universal cascade to be
  clearly visible. 

\item The infrared modes with $ k^2 \lesssim {\overline{\phi^2}}(t) $
  exhibit a much slower dynamics than the rest of the modes. The modes with
  $\bk(t)\gg k^2 \gg \phi^2 $ are well described by the Hartree
  approximation and equilibrate effectively by times $ \sim t_0 \simeq 500
  $ (in inverse mass units) while the infrared modes reach an equilibrium
  state only by times $t \sim 10^6$, when $\bk(t)$ is very close to the UV
  cutoff. In this equilibrium state a rather large wavefunction
  renormalization remains for low wavenumbers. In configuration space the
  correlation length turns to be $ 1/M_{\rm eff}(t) $ where the effective
  mass of the infrared modes $ M_{\rm eff}(t) $ is substantially smaller
  than the Hartree mass.  Therefore, there are {\bf two} scales for
  equilibration: the shorter one $ t_0 $ characterizes the UV cascade
  evolution, the local magnitudes as the time average $
  {\overline{\phi^2}}(t) $ and the modes with $ k^2 \gtrsim
  {\overline{\phi^2}}(t) $, while the longer scale governs the evolution of
  the infrared modes and the wavefunction renormalization for $ k^2
  \lesssim {\overline{\phi^2}}(t)$.

\item The thermalization process in $3+1$ dimensions is quite different
  from that in $1+1$, studied with the same light--cone--like
  discretization approach in ref.\cite{uno}, although a universal smooth
  ultraviolet cascade is present in both cases. In the $1+1$ case the
  cascade of the $\pi$ power spectrum is characterized by a single
  universal shape function with the time evolution reducing to a scale
  transformation on such function. On the contrary, in $D=3+1$, one can
  still define a shape function for the cascade but it turns out to be
  time--dependent; in particular, there exist a window at intermediate
  values of the scaled wavenumber $k/\bk(t)$ with a power--like behaviour,
  but the power depends markedly on time even when cutoff effects are fully
  negligible. Most remarkably, the shape function is almost flat near the
  origin in $1+1$ dimensions, allowing to consistently define an effective
  thermalization which, starting from the deep infrared where power is
  concentrated at zero time, progresses to higher wavenumbers with a well
  defined effective temperature monotonically decreasing in time. In
  $D=3+1$ instead, the deep infrared is the last to thermalize and the mode
  equilibration in the bulk of the cascade, where power--like spectra are
  observed, cannot be characterized only by a time--dependent effective
  temperature.

  Nonetheless, the notion of a time--dependent effective temperature
  keeps its sense for specific (simple) observables also in $ 3+1 $
  dimensions. For instance, from the late time behaviour of $ \pi^2 $,
  $ \phi^2, \; \phi^4 $ and $ \left( {\nabla \phi}\right)^2 $ one can read off
  an effective temperature which monotonically decreases with time
  approximately as $ \sim t^{-\frac13} $. This effective temperature
  reaches the proper nonzero limit in the lattice for very late times
  showing that true thermalization has been achieved. In the continuum this
  effective temperature would always vanish for infinite time.

\item Although all our present work (and in ref.\cite{uno}) deals with
  classical field theory, some of the main features as the ultraviolet
  cascade and the slow thermalization dynamics of the infrared modes should
  be relevant in quantum field theory for large occupation numbers and
  small coupling.

\end{itemize}

The main message to early cosmology and ultrarelativistic heavy ion
collision physics from the present work is that thermalization proceeds
very slowly in classical field theory (at least for unbroken symmetry in
scalar models).  However, a state of effective equilibration is soon
reached and is characterized by a steady, smooth and relatively simple
cascade of energy flowing towards the ultraviolet. The natural
interpretation is that the ``fast'' degrees of freedom have indeed reached
some sort of statistical equilibrium defined by few macroscopic parameters
(such as the temperature, but not the temperature alone) which are slowly
varying in time. On the continuum, without any cutoff effect for arbitrary
long time, this effective temperature should become the only relevant
slow--varying quantity for late enough time; it would eventually vanish for
infinite time in classical theory, while in quantum theory it should reach
a finite nonzero value necessarily involving $\hbar$.

Thermalization has been reached in quantum field theories by numerical
studies in $3+1$ and $2+1$ dimensions in refs. \cite{ber,berR} and \cite{gre},
respectively.  In these works the 2PI expansion to lowest order and the
analogous Kadanoff-Baym approximation are used, respectively. The couplings
considered are quite strong and do not allow a direct connection with the
classical dynamics investigated in the present paper.  The numerical
studies previously reported on the dynamics of classical and quantum field
theories \cite{marcelo}-\cite{ast} have not yet focused on studying the
mechanism of energy transfer from long to short wavelengths as a function
of time.  In ref.\cite{marcelo} the emergence of spatio-temporal structures
is reported for a symmetry breaking $ \phi^4 $ model. Refs. \cite{ABW} 
study the evolution of the classical $1+1$-dimensional $\phi^4$ model.
For further work in this domain see refs.\cite{A,B} and \cite{lisref}. 

It would indeed be interesting to study if the universal cascade found in
the classical theory\cite{uno} remains at least during some early and
intermediate stages in the quantum theory.

Ref. \cite{mic} investigates the classical in the $3+1$-dimensional
massless $ \phi^4 $ model. The ultraviolet cascade is present in their
numerical results.  In addition, an analytic scenario for the scaling
behaviour is proposed following the transport treatment of weak wave
turbulence \cite{zakh}. However, this heuristic treatment does not fully
capture the thermalization dynamics found in $1 + 1$ dimensions in
\cite{uno} as well as in $3+1$ dimensions in the present paper.

\bigskip

This paper is organized as follows: in sec. II we present the classical
$\phi^4$ model in the continuum and in the lattice and our way to perform
suitable coarse-grainings. Sec. III discuss the classical $\phi^4$ model in
thermodynamic equilibrium and its properties both from perturbative as well
as Monte Carlo calculations.  Sec. IV is the core of the article where the
dynamics of thermalization is presented. We study the time evolution of
local observables as well as the correlators whose Fourier transforms yield
the power spectra of the field $\phi$ and its conjugate momentum $\pi$. The
early virialization and the late thermalization of the infrared modes 
is discussed.  Sec. V contains discussions and conclusions while
four appendices deal with more specific topics.

\section{The model in the continuum and on the lattice}
Besides defining the classical $\phi^4$ model, we present in this section
its lattice version simultaneously discretizing space and time in
such a way that an exactly conserved lattice energy can be defined. The
average procedure over the basic physical observables is presented too.

\subsection{Basic definitions and notations}

The Lagrangian density of the $(3+1)-$dimensional $\varphi^4$ field theory
reads
\begin{equation*}
   {\cal L}_{m,\lambda}\,(\varphi) = \frac12
    \dm\varphi \, \udm\varphi - \frac{m^2}2\,\varphi^2 -
   \frac\lambda4\,\varphi^4
\end{equation*}
where $ \dm \equiv \partial/\partial x^\mu$, $\mu=0,1,2,3$.
This leads to the classical equation of motion
\begin{equation}\label{eqnofmotion}
    \dm\udm \varphi + m^2\,\varphi + \lambda\,\varphi^3 = 0
\end{equation}
In $3+1$ space-time dimensions and standard units $\hbar=c=1$, the coupling
$\lambda$ is dimensionless while $\varphi$ has the dimensions of a mass.

At the classical level one can always rescale the coordinates and the field
using some reference mass $M$ and absorb the coupling $\lambda$ in the
field. Thus setting
\begin{equation}\label{adim}
  \varphi(x) = \frac{M}{\sqrt{\lambda}} \; \phi(Mx)
\end{equation}
and renaming $ (M \, t, \, M \, \vx) $ as $ (t,\vx) $, yields for the 
dimensionless field $\phi$ the equation
\begin{equation*}
{\ddot \phi} - \nabla^2\phi + \gamma\,\phi + \phi^3 = 0 
\end{equation*}
where ${\dot\phi} \equiv {\partial\phi}/{\partial t } $, $\nabla_j\phi
\equiv {\partial\phi}/{\partial x_j }$ and $\gamma\equiv m^2/M^2$. If
$m^2>0$, then we can choose $M=m$ and study the parameter--free equation
\begin{equation}\label{eqnofmot}
{\ddot \phi} - \nabla^2\phi + \phi + \phi^3 = 0
\end{equation}
This is our choice, having assumed a massive $\varphi$. At any rate, the
important point here is that, unlike in QFT, 
the notion of coupling constant at the classical level is not absolute as
it can be scaled out.

In terms of $\phi$ and its canonical conjugate momentum $\pi = \dot \phi$,
the Hamiltonian is the standard sum of a kinetic plus a potential term
\begin{equation}\label{hamilt}
  H[\pi,\phi] = \mathcal{T}[\pi]+ {\mathcal V}[\phi] \;, \quad
  \mathcal{T}[\pi] =  \frac12 \int d^3x \,{\pi}^2 \;, \quad 
  {\mathcal V}[\phi] = \frac12 \int d^3x \left[
    \left({\nabla \phi}\right)^2 + {\phi}^2 + \frac12 \;
    {\phi}^4 \right] 
\end{equation}
This Hamiltonian is dimensionless; the energy of the original field
$\varphi$ in the standard dimension-full coordinates is given by
$(m/\lambda) \, H[\pi,\phi] $.

We consider the model restricted to a finite volume, which we take to be
the cube of side $L$ (in units of $m^{-1}$) and volume $V=L^3$; we assume
periodic boundary conditions (p.b.c.), namely $\phi(\vx+L\vn,t)=
\phi(\vx,t)$ for any $\vn = (n_1,n_2,n_3) \in {\mathbb Z}^3$ .

In terms of standard Fourier mode amplitudes,
\begin{equation}\label{fourier}
  {\tilde\pi}_\vk = \int_V \frac{d^3x}{V^{1/2}}\, 
  e^{-i\vk\cdot\vx} \,\pi(\vx) = {\tilde\pi}_{-\vk}^{\,\ast} \;, \quad
  {\tilde\phi}_\vk = \int_V \frac{d^3x}{V^{1/2}}\,
  e^{-i\vk\cdot\vx} \,\phi(\vx) = {\tilde\phi}_{-\vk}^{\,\ast} \;, \quad
  \vk=\frac{2\pi}L \, \vn \;,\quad \vn\in {\mathbb Z}^3
\end{equation}
the Hamiltonian reads
\begin{equation*}
  H[\phi,\pi] = \frac12 \sum_\vk \Big[ |{\tilde\pi}_\vk|^2 +
    (1+\vk^2)|{\tilde\phi}_\vk|^2 +   \frac1{2 \, V}
    \sum_{{\bds q}\,{\bds q}^{\,\prime} }
    {\tilde\phi}_{\bds q} \; {\tilde\phi}_{{\bds q}^{\,\prime}} \;
    {\tilde\phi}_\vk \;
    {\tilde\phi}_{-\bds q-{\bds q}^{\,\prime}-\vk}  \Big] \; .
\end{equation*}
The wavenumbers $\vk$ are dimensionless; their dimension-full counterpart 
are given by $ m \,\vk$.

\subsection{Regularization on a finite space lattice}\label{finitelat}

For any numerical treatment it is necessary to discretize space and time
over some lattice. This obviously introduce an ultraviolet cutoff $\Lambda$
over the wavenumbers of the order of the inverse of the lattice spacing.
To discretize the space we consider the cubic lattice ${(2a\,\mathbb
  Z})^3$, $a$ being half the lattice spacing. Taking into account the
finite size $L$, we obtain regularized fields
\begin{equation*}
  \phi_{\vn} = \phi(\bds x) \;,\quad 
  \pi_{\vn} = \pi(\bds x) \;,\quad \bds x = 2\,a\,\vn\,,
  \quad \vn\in C_N \; ,
\end{equation*}
where $C_N$ is the cubic subset of ${\mathbb Z}^3$ formed by integer
triples $(n_1,n_2,n_3)$ satisfying $-N/2+1 \le n_j\le N/2, \; j=1,2,3$.
Here $N=L/(2a)$ is assumed even and $N^3$ is clearly the total number of
degrees of freedom of the regularized fields.

Owing to p.b.c., to the cube $2a\,C_N$ in $x-$space there 
corresponds the dual cube $(2\pi/L)\,C_N$ (the so-called first Brillouin
zone) of wavenumbers $\vk$ fulfilling
\begin{equation*}
  \vk = \frac{2\pi}L \, \vn\;,\quad -N/2+1\le n_j\le N/2
  \; , \; j=1,2,3 \; .
\end{equation*}
The largest value of each component of $\vk$,
$\Lambda=(2\pi/L)(N/2)=\pi/(2a)$, is the UV cutoff.

The lattice form of the Hamiltonian is not unique, being restricted solely
by the requirement to formally reduce to the continuum expression
eq.~(\ref{hamilt}) in the $a\to0$ limit. For all ultralocal terms, with
fields at coincident points one could assume the simplest
discretization as sums of one--site terms. Thus the kinetic energy
$\mathcal{T}$ would read on the lattice
\begin{equation}\label{eq:Tlat}
  \mathcal{T}[\pi] = \frac12 \sum_{\vn\in C_N} (2a)^3 \,\pi_\vn^2
\end{equation}
and similarly for the space integrals of $\phi^2$ and $\phi^4$. The
integral of the gradient term $({\nabla \phi})^2$ is converted, through
integration by parts, to the integral of $-\phi\,\nabla^2\phi$, with
$\nabla^2$ the Laplacian. Then, in order to define the theory on the
lattice,  $\nabla^2$ is replaced by a discretized  Laplacian.
The simplest choice is the nearest-neighbor form that leads to the well known 
replacement of the spectrum $k^2$ on the continuum
\begin{equation*}
  k^2 \longrightarrow {\hat k}^2 \;,\quad 
  {\hat k} \equiv \frac{\sin ka}a
\end{equation*}
However, our choice of discretization is different, as we show in the next
subsection.

\subsection{Discretized dynamics on a spacetime lattice}\label{ddlc}

In order to solve numerically the evolution equations for the $\phi^4$
theory, it is necessary to discretize time besides space.  In most
approaches, space and time discretization are performed separately.  As
discussed in the previous section, space discretization turns the field
theory into a classical dynamical problem with finitely many `coordinates'
and `momenta'. Their evolution is governed by ordinary differential
equations which eventually require some discretization of time to be solved
numerically. Here we proceed in a different way, treating space and time
in a symmetric way. This generalizes the scheme introduced in $1 + 1$
dimensions \cite{uno,cono}.

In our approach space and time are simultaneously discretized, with the same
lattice spacing $2a$, in a staggered fashion over the lattice $\mathbb
Z^{D+1} \cup (\mathbb Z+1/2)^{D+1}$ (for the sake of generality, we
consider here a space of generic dimensionality $D$). In other words the
discretized space-time points are 
\begin{equation*}
  (\vx,t) = a\,(\vn,s)  
\end{equation*}
where the integer components of $(\vn,s)$ are either all even or all odd.
This allows, as we shall see, to protect to a large extent the relativistic
invariance of the continuum field equation (\ref{eqnofmot}) and to provide
an exactly conserved energy on the lattice.

Let us first of all define the averages over local cubes of field powers
\begin{equation}\label{eq:PHI}
 \Phi^{(p)}(\vx,t) =  \frac1{2^D} \sum_\vs\, [\phi(\vx+a\vs,t)]^p 
 \;,\quad p=0,1,2,\dots
\end{equation}
where $\vs = (\sigma_1, \sigma_2,\ldots, \sigma_D)$, $\sigma_i = \pm$,
$i=1,2,\ldots,D$. We then construct two lattice versions of relevant
continuum observables  through the correspondences
\begin{equation}\label{lattcont}
  \begin{split}
    \dot\phi(\vx,t) \equiv \pi(\vx,t) &\longleftarrow \pi_\pm(\vx,t)
    \equiv \pm\, a^{-1}\big[\phi(\vx,t\pm a) - \Phi^{(1)}(\vx,t)\big] \\
    \phi^2(\vx,t) &\longleftarrow \phi^2_\pm(\vx,t) 
    \equiv \tfrac12 \big[\phi^2(\vx,t\pm a) + \Phi^{(2)}(\vx,t)\big] \\
    \phi^4(\vx,t) &\longleftarrow \phi^4_\pm(\vx,t)
    \equiv \phi^2(\vx,t\pm a)\,\Phi^{(2)}(\vx,t) \\
    (\nabla\phi)^2(\vx,t) &\longleftarrow (\mathrm{D}\phi)^2(\vx,t)
    \equiv a^{-2}\big\{ \Phi^{(2)}(\vx,t) - [\Phi^{(1)}(\vx,t)]^2\big\} \; .
  \end{split}
\end{equation} 
By construction, the lattice quantities on the r.h.s. tend to the
continuum expressions in the limit $a\to0$. Therefore 
the lattice energy densities
\begin{equation*}
  \E_\pm(\vx,t)= \frac12 (2a)^D \big[ \pi_\pm^2
  + (\mathrm{D}\phi)^2 + \phi_\pm^2 + \tfrac12 \phi_\pm^{\,4} \big]
\end{equation*}
both tend to the continuum energy density as $a\to0$
\begin{equation*}
 \E_\pm \simeq \E \equiv \frac12  \,d^D x  \left[  {\dot \phi}^2 + \left(
 {\nabla \phi}\right)^2 + \phi^2 + \frac12 \;  \phi^4 \right]
\end{equation*}
upon the natural identification $(2a)^D \simeq d^Dx$. Notice that 
$\E_\pm$ can be explicitly written as
\begin{equation}\label{densE}
\begin{split}
  \E_\pm(\vx,t) &= \tfrac12 a^{D-2} \sum_\vs \Big[ \phi(\vx,t\pm a) 
  - \phi(\vx+a\vs,t) \Big]^2 \\ &+ \tfrac14 (2a)^D \Big[1 + 
  \phi^2(\vx,t\pm a) \Big] \bigg[ 1 + \frac1{2^D} 
  \sum_\vs \phi^2(\vx+a\vs,t) \bigg] - \tfrac14 (2a)^D
\end{split}
\end{equation}
which shows that only diagonal, space-time symmetric finite differences
are present.

To higher orders in $a , \; \E $ and $ \E_\pm $ do differ; in fact
\begin{equation*}
  \E_+(\vx,t) - \E_-(\vx,t) =  4\,(2a)^{D-2} \; \left[ \phi(\vx,t+a) -
  \phi(\vx,t-a) \right] \; Q(\vx,t) 
\end{equation*}
where
\begin{equation*}
  Q(\vx,t) =\tfrac12 \, \big[\phi(\vx,t+a) + \phi(\vx,t-a)\big]  \Big\{
  1 + \tfrac12 a^2 \big[1 + \Phi_2(\vx,t) \big] \Big\} - \Phi_1(\vx,t)
\end{equation*}
Hence, if $Q(\vx,t) = 0$, then $\E_+(\vx,t) = \E_-(\vx,t)$
also on the lattice. In this case the total lattice energy
\begin{equation}\label{emas}
  E = \sum_{\vx \in 2a {\mathbb Z}^D}  \E_+(\vx,t)
\end{equation}
is exactly conserved in time, since it can also be written
\begin{equation}\label{emenos}
  E = \sum_{\vx \in 2a \left({\mathbb Z} + 1/2 \right)^D} \E_-(\vx,t+a) \;.
\end{equation}
All this holds exactly on infinite space. If space is restricted to the
(hyper)cube of side $L$, suitable boundary conditions on $\phi(\vx,t)$ are
necessary; p.b.c. are of this type if $L=2Na$ with $N$ an integer.

In conclusion, we may regard $Q(\vx,t) = 0$ as a discrete field equation
which conserves the total energy $E$. Explicitly, $Q(\vx,t) = 0$ reads
\begin{equation}\label{recursion}
  \phi(\vx,t+a) + \phi(\vx,t-a) =  
  \frac{2^{1-D} \sum_\vs\,\phi(\vx+a\vs,t)} {1 + \tfrac12{a^2} 
    \Big[1 + 2^{-D}\sum_\vs\,\phi^2(\vx+a\vs,t) \Big]}
\end{equation}
which evidently allows to evolve in time any configuration known on two
consecutive time slices, say $t=0$ and $t=a$.

It is easy to check that $Q(\vx,t)=0$ indeed becomes eq.~(\ref{eqnofmot})
in the continuum $ a \to 0 $ limit. The order $a^0$ is trivially satisfied,
odd powers of $a$ vanish identically as a consequence of the symmetry of
eq.~(\ref{recursion}) under $a\to-a$, while the order $a^2$ produces
eq.~(\ref{eqnofmot}).

Keeping up to $\mathcal{O}(a^4)$ in eq.(\ref{recursion}) yields, 
\begin{equation}\label{correca2} 
  \begin{split}
    &{\ddot \phi}-\nabla^2\phi + \phi + \phi^3 = a^2 \,Q_2 + {\cal O}(a^4)\\
    &Q_2 = -\frac{1}{2} (1+\phi^2)\nabla^2\phi - \frac{1}{12}{\ddddot \phi} +
    \frac14\left[\nabla^2\nabla^2\phi - \frac23
      \sum_{i=1}^{D}\frac{\partial^4\phi}{ \partial x_i^4}\right]
    - \; \phi\left[ (\nabla\phi)^2 + \phi \;
      \nabla^2\phi\right] \; .  
  \end{split}
\end{equation}
To cast eq.\eqref{recursion} in a form suitable for numerical
calculations, we define the lattice arrays $F(\vn,s)$ and $G(\vn,s)$ as
\begin{equation}\label{eq:FG}
 F(\vn,s) \equiv \phi(2\vn a,2sa) \; , \quad
 G(\vn,s)\equiv \phi(2\vn a -\vs_+a,\,2sa+a)\;,
\end{equation}
where $\vn \in {\mathbb Z}^D$, $s\in{\mathbb Z}$ and $\vs_+
\equiv (1,1,,\ldots,1)$.

We then obtain the iterative system
\begin{eqnarray}\label{evod}
  && F(\vn,s+1) = -  F(\vn,s)+ \frac{ \sum_{\vt} G(\vn + \vt,s)}
  {2^{D-1} + \frac14{a^2} \left[ 2^D + \sum_{\vt} 
    G^2(\vn + \vt,s) \right] } \cr \cr 
  && G(\vn,s+1) = - G(\vn,s)+ \frac{ \sum_{\vt} F(\vn-\vt,s+1)}
  {2^{D-1} + \frac14{a^2} \left[ 2^D + \sum_{\vt}
  F^2(\vn-\vt,s+1)\right] } \; . 
\end{eqnarray}
where $\vt =(\tau_1,\tau_2,\ldots,\tau_D)$, with $\tau_i = 0,1$ and,
according to the p.b.c, $F(\vn+N\vt,s)=F(\vn,s)$ and 
$G(\vn+N\vt,s)=G(\vn,s)$ for any $\vt$.

As initial conditions we have to specify $ F(\vn,0) $ and $ G(\vn,0) $ for
$0 \leq n_i \leq N-1, \; i=1,2,\ldots,D $. Once these values of the fields are
provided, the iteration rules eqs.~(\ref{evod}) uniquely define $ F(\vn,s) $
and $ G(\vn,s) $ for all $ s > 0 $. A comparison of this discretized
dynamics with other more traditional numerical treatments of hyperbolic
partial differential equations was performed in $1+1$ dimensions
\cite{zanlungo}. Here we only notice that this approach is particularly
efficient, stable and accurate, specially when the continuum limit $a\to0$
and very long evolution times are of interest.

All observables of the continuum can be rewritten on the lattice in terms
of the basic fields $F(\vn,s)$ and $G(\vn,s)$, according to the
correspondences rules eqs.~(\ref{lattcont}) and the identification
eq.~(\ref{eq:FG}). In the sequel, while referring to observables
discretized as above, whenever possible we shall keep using the notation
corresponding to continuum observables for simplicity.  
Particular care must be taken for $\phi^4$, since its lattice definition in
eqs.~(\ref{lattcont}) uses a product of fields over different lattice
sizes. This is harmless in the continuum limit for smooth fields, but 
makes a difference for fields with large ultraviolet support, as we shall
see shortly. For this reason, we shall keep the notation
$\phi^4_\pm(\vx,t)$ given in eqs.~(\ref{lattcont}) well distinct from the
ultralocal definition $\phi^4(\vx,t) \equiv [\phi(\vx,t)]^4$.

\subsection{Averaged observables}\label{sec:avgobs}

The key observables in our investigation are the basic quantities
\begin{equation} \label{lista}
  \phi, \,\phi^2, \,\phi^4, \,{\dot \phi}^2, \,({\nabla \phi})^2 \; ,
\end{equation}
as well as the power spectra of $\phi$ and $\pi$, that is
$|{\tilde\phi}_\vk(t)|^2$ and $|{\tilde\pi}_\vk(t)|^2$, where $
{\tilde\phi}_\vk(t) $ and $ {\tilde\pi}_\vk(t) $ are given by
eq.(\ref{fourier}). Let us recall again that we are using here the continuum
notation also for the Fourier transforms, although they actually are
discrete Fourier transforms.

The fluctuations of all these observables do not vanish upon time
evolution. Hence for generic initial conditions they do not have any limit
as $t\to\infty$. These are fine--grained or microscopic observables.
Typically there are several spatio-temporal scales, the microscopic scales
correspond to very fast oscillations and short distance variations that are
of no relevance to a thermodynamic description. We are interested in
longer, macroscopic scales that describe the relaxation of observables
towards a state of equilibrium.

In particular, the ergodic postulate states that ensemble averages must be
identified with \emph{long} time averages as well as spatial averages over
macroscopic-sized regions. Hence to make contact between the time evolution
and the thermal averages we need to properly average the microscopic
fluctuations.

First of all, for local quantities such as those in
eq.\eqref{lista} we take the spatial average. Secondly, we take
suitable time averages of all key observables in the following way
\begin{equation}\label{promT}
  \overline{\phi^2}(t) \equiv
  \frac1\tau\int_{t-\tau}^t dt' \; \;\frac1{V}\; \int_V d^3x\,
                \phi^2(\vx,t') \;.
\end{equation}
where $\tau\gg a $ fixes a limit to our resolution power in time.  $\tau$
need not be constant throughout the time evolution. For instance, we find
that a practical and efficient choice is
\begin{equation} \label{taut}
  \tau(t) = \tau(0) + C \; t
\end{equation}
where $C$ is small and positive with typical values $ \sim 0.1 $. In this
way we still keep $t \gg \tau(t)$ while the dependence on the initial
values becomes negligible for practically accessible times. This method is
quite effective in revealing general features of the (logarithmic) time
evolution such as the presence of distinct stages characterized by well
separated macroscopic time scales.

Besides the time average, for more detailed observables such us the power
spectra we performed an average over the discrete directions in space as
discussed in Appendix~\ref{app:A}. Moreover, we sometimes averaged also
over several initial conditions. All together, we denote the results of all
these coarse-grainings simply with an overbar to avoid cluttering of notation. 
For example, assuming equiprobable initial conditions
\begin{equation*}
  \overline{\phi^2}(t) = \frac{1}{M} \sum_{i=1}^M \;
  \frac1\tau \; \int_{t-\tau}^t dt' \; \frac1{V} \; \int_V d^3x \;
                [\phi^2(\vx,t')]^{(i)} \;.
\end{equation*}
where the superscript $^{(i)}$ labels the $M$ different choices of
smooth initial conditions, all with the same energy density $E/V$.
Likewise, recalling the wave-vectors quantization $\vk=(2\pi/L) \; \vn$, we have
\begin{equation}\label{promcondi}
  \overline{|{\tilde\pi}_k|^2}(t) =  \frac{1}{M} \sum_{i=1}^M \;
  \frac1{S_n} \sum_\vn \theta(n\le|\vn|<n+1) \; 
  \frac1\tau \int_{t-\tau}^t dt' \; |{\tilde\pi^{(i)}}_\vk(t')|^2
\end{equation}
where the discrete radial wavenumber $k$ is the average $|\vk|$ over the
shell of all $S_n$ wave-vectors $\vk=(2\pi/L) \; \vn$ satisfying $n\le|\vn|<n+1$
[see Appendix~\ref{app:A} for more detail]. 
Analogous expressions hold for $\phi, \; \phi^2, \;
\phi^4, \; \pi^2, \; (\nabla \phi)^2 , \;  |{\tilde\pi}_\vk|^2 $ 
and $|{\tilde\phi}_\vk|^2$.

In particular, due to the linearity of these averages, we have the
sum rules:
\begin{equation}\label{sumrules}
  \int_{-\Lambda}^{+\Lambda} \frac{d^3k}{(2\pi)^3} \;\phivk2 = 
  \overline{\phi^2}(t) \quad ,\quad
  \int_{-\Lambda}^{+\Lambda} \frac{d^3k}{(2\pi)^3} \; \pivk2 = 
  \overline{\pi^2}(t) \quad , \quad
\end{equation}
where $\Lambda=\pi/(2a)$ is the UV cutoff on the lattice and we 
write the wavenumbers as continuous although they are discrete in all actual 
calculations.

Finally, let us recall that power spectra are just the Fourier transforms of
space-averaged correlation functions. For instance:
\begin{equation}\label{phiphi}
  \overline{|{\tilde\phi_\vk}|^2}(t) = \int_V d^3x \; 
  e^{-i \vk\cdot\vx} \;   \overline{\phi\phi}(\vx,t) \; ,
\end{equation}
where, according to our general rules,
\begin{equation*}
  \overline{\phi\phi}(\vx,t) =
  \frac1M \sum_{i=1}^M \frac1\tau \int_{t-\tau}^t dt' \;
  \frac1V\int_V d^3y \; \phi^{(i)}(\bds y,t') \; 
  \phi^{(i)}(\vx+\bds y,t') \; .
\end{equation*}

\section{Thermal equilibrium}\label{sec:Thereq}

We present in this section the basic properties of the $\phi^4$ theory in 
thermal equilibrium. These results will be compared in the subsequent sections
with the time averages in order to asses whether and how thermalization is 
approached.

\subsection{General aspects}\label{eqgenral}
The thermal average of any physical quantity $ \Theta =
\Theta[\phi,\pi]$ in the canonical ensemble is written as
\begin{equation}\label{prog}
  \avg{\Theta [\pi,\phi]} = \frac{\int\!\int D\phi \; D\pi \;
    e^{- \beta H[\pi,\phi]} \; \Theta[\phi,\pi]}
      {\int\!\int D\phi \; D\pi  \; e^{- \beta H[\pi,\phi]} } \; ,
\end{equation}
where $\int\!\int D\phi \; D\pi $ stands for functional integration over
the classical phase space and $ \beta \equiv 1/T $ is the inverse
(dimensionless) temperature in the dimensionless variables. In terms of the
\emph{physical} temperature, here defined as $ T_p, \; T $ is given by
\begin{equation}\label{temp}
  T = \frac{1}{\beta}\equiv \frac{\lambda}{m} \; T_p \; .
\end{equation}
As a consequence of the field redefinition available in the classical
theory, the relevant variable for equilibrium thermodynamics is $T$.
Therefore, for a fixed physical temperature $T_p$ we see from
eq.~(\ref{temp}) that the low temperature limit $T\ll 1$ corresponds to the
weak coupling limit and/or $ T_p \ll m $. This will be relevant in the
analysis below.

Translation invariance (which is preserved by p.b.c.) implies that averages
of local observables $\Theta(\vx)$ which depend on $\pi$ and $\phi$ only at
one point $\vx$, do not depend on $\vx$, that is
$\avg{\Theta(\vx)}=\avg{\Theta(0)}\equiv\avg{\Theta}$.

Furthermore, the fact that the Hamiltonian is the sum of a kinetic
and a potential term [see Eq. (\ref{hamilt})]
entails that the average of observables which are of the form
$$
\Theta[\phi,\pi]=\Theta_1[\phi] \; \Theta_2[\pi] \; ,
$$
factorize as
\begin{equation*}
  \langle \Theta[\phi,\pi] \rangle =
  \langle \Theta_1[\phi]\rangle_{\phi} \;  \langle
  \Theta_2[\pi]\rangle_{\pi}
\end{equation*}
with
\begin{equation*}
  \langle\Theta_1[\phi]\rangle_{\phi}  =  \frac{\int D\phi \;
    e^{- \beta{\mathcal V}[\phi]} \; \Theta_{\phi}[\phi]}
  {\int D\phi   \; e^{- \beta{\mathcal V}[\phi]} } \quad  ,\quad
  \langle \Theta_2[\pi]\rangle_{\pi}  =  \frac{\int D\pi \;
    e^{- \beta \mathcal{T}[\pi]} \; \Theta_{\pi}[\pi]}
  {\int D\pi   \; e^{- \beta T[\pi]} } \; .
\end{equation*}
Moreover, since the $\pi-$integration is Gaussian and ultralocal, it can be
performed quite easily in most cases, leaving the  configurational
integral over $\phi=\phi(\vx)$ for $\vx \in V $ to be computed.

It follows from eqs.(\ref{hamilt}) and (\ref{prog}) that the two-point
correlation function of the canonical momentum $\pi(\vx)$ in the classical
theory in equilibrium is given by
\begin{equation}\label{corpi}
 \avg{\pi(\vx)\,\pi({\vx}^{\; \prime})} = T \,\delta(\vx-\vx') \;.
\end{equation}
which leads to a flat power spectrum for $\pi$
\begin{equation}\label{Tflat}
  \avg{|{\tilde\pi}_{\vk}|^2} = T \; ,
\end{equation}
where we used the Fourier transform eq.~(\ref{fourier}).  This is of course 
a consequence of equipartition, and gives a criterion to identify the
temperature: the height of the flat region in the power spectrum of $\pi$
if such flat region shows up.

Eq.~(\ref{Tflat}) implies that $\avg{\pi^2}=T\,\delta(\bds 0)$ where
$\delta(\bds 0)$ is to be understood as made finite by some UV cutoff
procedure. If a rotationally invariant sharp cutoff $\bar\Lambda$ is used,
then $\delta(\bds 0)={\bar\Lambda}^3/(6\pi^2)$. In case of the lattice
regularization of section~\ref{finitelat}, with $\pi(\vx)$ entering only in the
kinetic energy converted to a sum as in eq.~(\ref{eq:Tlat}), we have
instead $\delta(\bds 0)=1/(2a)^3 $. Thus,
\begin{equation}\label{pi2}
  \avg{\pi^2}^{(\mathrm{sharp}\,\bar\Lambda)} = 
  \frac{T\,\bar\Lambda^3}{6\,\pi^2} \quad,\qquad 
  \avg{\pi^2}^{(\mathrm{lattice,}\,2a)} =  \frac{T}{(2a)^3} 
\end{equation}
Equating these two regularized averages would yield the the identification
$\bar\Lambda=(6/\pi)^{1/3}(2\pi/a)=(6/\pi)^{1/3}\Lambda$; this connection 
between cutoffs is however
not universal; it is valid only for $\avg{\pi^2}$ and in general different
for other UV-divergent quantities. Moreover, we will see below that the
discretized dynamics described in section \ref{ddlc} provides a slightly
different UV  regularization even for $\epik2$.

\medskip
At equilibrium, the \emph{classical} virial theorem takes the form
\begin{equation}\label{virial2}
  \avg{\dot\phi^2} = \avg{(\nabla \phi)^2} + \avg{\phi^2} + \avg{\phi^4} \;.
\end{equation}
where we have trivially generalized to three spatial dimensions the
derivation in ref.~\cite{uno}. When combined with the energy functional
$H[\pi,\phi]$ given in eq.(\ref{hamilt}), it yields
\begin{equation}\label{enerdens}
  \frac{\avg{H[\pi,\phi]}}V = \frac{E}{V}= \avg{\pi^2} - \tfrac14
  \avg{\phi^4} \;,
\end{equation}
where ${E}$ is the average energy in the canonical ensemble. Since
$N=V/(2a)^3$ is the total number of degrees of freedom in the simplest
lattice regularization, we find using eq.(\ref{pi2}) that the temperature
$T$ is related to the (average) energy per degree of freedom as
\begin{equation}\label{TtoE}
  T = \frac{E}{N} + 2 \, a^3 \, \avg{\phi^4} \;.
\end{equation}
Therefore, for $ a \ll 1 $, close to the continuum limit where one finds
$a^2\avg{\phi^4}$ to be finite as $a\to0$ (see below), the temperature is
identified with the energy per site.  Furthermore,
\begin{equation}\label{TV}
  T =  (2a)^3 \;  \left[ \rho + \tfrac14  \avg{\phi^4}\right] \;,
\end{equation}
where $\rho \equiv E/V$ is the energy density, while 
the pressure $p$ is given in general (not necessarily at thermal
equilibrium) by 
\begin{equation}\label{pres} 
  p = \tfrac12 \, \left[ {\dot \phi}^2 -  
    \tfrac13
    \left({\nabla \phi}\right)^2 - {\phi}^2 - \tfrac12 \,{\phi}^4 \right]\;.
\end{equation}
At thermal equilibrium we find from eqs.(\ref{virial2}), (\ref{TV}) and
(\ref{pres}) the equation of state
\begin{equation}\label{eces}
  \avg{p} = \tfrac13 \,\rho -\tfrac13 \avg{\phi^2}
\end{equation}
that is a radiation--dominated equation of
state $\avg{p}\simeq\rho/3$, since $\avg{\phi^2}$ is of order $a^{-1}$,
much smaller than $\avg{p}$ and $\rho$ which are both of order $a^{-3}$
for fixed $T$.  

\subsection{Perturbation theory, Hartree resummation and beyond }
\label{sec:pert}

The continuum configurational functional integrals 
\begin{equation}\label{int1}
\langle\Theta[\phi]\rangle  =  \frac{\int D\phi \;
    e^{- \beta V[\phi]} \; \Theta[\phi]}
  {\int D\phi   \; e^{- \beta V[\phi]} }
\end{equation}
can be computed in a perturbative expansion in powers of $T$. In order to
do that one changes the functional integration variable $\phi(\vx)$ to
$$
\chi(\vx) \equiv \frac1{\sqrt{T}} \; \phi(\vx) \; .
$$
Eq.(\ref{int1}) takes thus the form
\begin{equation}\label{int2}
\langle\Theta[\phi]\rangle  =  \frac{\int D\chi \;
    e^{- \frac12 \int d^3x \left[
    \left({\nabla \chi}\right)^2 + {\chi}^2 + \frac{T}2 \;
    {\chi}^4 \right] } \; \Theta[\sqrt{T} \; \chi]}
  {\int D\chi   \; e^{- \frac12 \int d^3x \left[
    \left({\nabla \chi}\right)^2 + {\chi}^2 + \frac{T}2 \;
    {\chi}^4 \right] }}
\end{equation}
We read from eq.(\ref{int2}) the associate Feynman rules. The Euclidean 
$\chi$ propagator in three space dimensions takes in momentum space the form 
$$
\Delta_{\chi} = \frac1{k^2 + 1}
$$
and the quadrilinear $\chi$ vertex has $ (- 6 \; T) $ as coefficient. 

As already recalled above, this defines a superrenormalizable field theory
with only two divergent diagrams: the one-loop tadpole and the two-loops
sunset.  Recall that in classical statistical mechanics the regularized
(bare) theory is the physical one. The ultraviolet divergences are physical
and cannot be eliminated by counter-terms as in quantum field theory. 

The two-point function $\avg{\phi(\vx)\phi(\vx')}$ can be written as
\be\label{defSi}
 \avg{\phi(\vx)\phi(\vx')} = \int\frac{d^3k}{(2\pi)^3} \, \ephik2
  \, e^{i \vk \cdot(\vx-\vx')} \quad  , \quad \ephik2 = 
\frac{T}{k^2+1+\Sigma(k^2)} 
\ee
where $ \Sigma(k^2) $ stands for the self-energy.

To lowest order in $T$ the tadpole takes then the form
\begin{equation}\label{tad1}
  \avg{\phi^2} = T \, \int_{k<{\bar\Lambda}} 
  \frac{d^3k}{(2\pi)^3} \frac1{k^2 + 1}  \buildrel{\bar\Lambda\gg 1 }\over= 
\frac{T\,{\bar\Lambda}}{2\pi^2} \;
  [ 1 + {\cal O}(\,{\bar\Lambda}^{-1})]
\end{equation}
while the self-energy reads to lowest order
\begin{equation*}
  \Sigma = 3\,\avg{\phi^2}   \buildrel{\bar\Lambda\gg 1 }\over= 
\frac{3\,T\,{\bar\Lambda}}{2\pi^2} \;
  [ 1 + {\cal O}(\,{\bar\Lambda}^{-1})]
\end{equation*}
Here we are using the sharp spherically symmetric UV cutoff, to be compared
later on to the lattice regularization method with cutoff $\Lambda=\pi/(2a)$.

To next order $\Sigma$ contains the two-loops sunset diagram, which is only
logarithmically divergent; therefore the tadpole dominates the effective
mass in the limit ${\bar\Lambda}\to\infty$. Hence, in this limit the
two-point function is dominated by the Hartree approximation 
(the sum of all daisy diagrams)
\begin{equation}\label{eq:2phartree}
\ephik2 \simeq
  \frac{T}{k^2 + 1 + 3\,\avg{\phi^2}}   
\end{equation}
where the self-consistent $\avg{\phi^2}$ fulfills
\begin{equation}\label{eq:selfc}
  \avg{\phi^2} = \int\frac{d^3k}{(2\pi)^3}\;\ephik2 \;=\; 
  \int\frac{d^3k}{(2\pi)^3} \frac{T}{k^2 + 1 
    +3\,\avg{\phi^2}} \simeq \frac{T}{2\pi^2} \Big[{\bar\Lambda}
  - \sqrt{\tfrac38 T{\bar\Lambda}} + \ldots \Big]
\end{equation}
We analogously find for the gradient squared $ \left({\nabla \phi}\right)^2 $,
at lowest order
\begin{equation}\label{grad}
  \langle  \left({\nabla \phi}\right)^2\rangle = 
  \frac{T}{12\pi} \left({\bar\Lambda}^2  - 
    \frac{6}{\pi^2}\bar\Lambda\right) \,+\, {\cal O}({\bar\Lambda}^0) \;,
\end{equation}
and for  $\avg{\phi^4}$ using Wick's theorem,
\begin{equation}\label{tres}
  \avg{\phi^4} = 3 \,\avg{\phi^2}^2\,\left[1+{\cal O}(\bar\Lambda^{-1}) 
  \right] = \frac{3\,T^2}{4\pi^4} \left[{\bar\Lambda}^2 
  +{\cal O}({\bar\Lambda}) \right] \; .
\end{equation}
By construction, the Hartree resummation does not change the first order
result $\avg{\phi^4}=3 \, \avg{\phi^2}^2$. One may also check that it
provides the correct next-to-leading term of order $\sqrt{{\bar\Lambda}}$
to $\avg{\phi^4}/{\bar\Lambda}^2$, since the first, three-loop divergent
diagram not of daisy type is only linearly divergent.

Beyond the Hartree approximation the self-energy $\Sigma(k^2)$
receives $k-$dependent UV finite corrections which can be embodied into a
finite $k-$dependent wave-function renormalization $Z(k^2)$ by writing
\begin{equation}\label{eq:Zparam}
  \ephik2 = \frac{T\,Z(k^2)}{k^2 + 1 + 3\,\avg{\phi^2}}  
\end{equation}
where $ \avg{\phi^2} $ is now the exact expectation value and $Z(k^2)$ is of
course a function also of $T$ and of the UV cutoff $\bar\Lambda$, that is
$Z(k^2)=Z(k^2;T,\bar\Lambda)$\footnote{Here $\bar\Lambda$ need not be a sharp
  cutoff as in the lowest perturbative order or in the Hartree
  approximation; our only request is that it is the UV cutoff in some
  spherically symmetric regularization procedure. For instance, one could
  use free-field propagators smeared in the ultraviolet by some smooth
  function of $k/\bar\Lambda$ dying at infinity faster than any power}.
We have from eqs. (\ref{defSi}) and (\ref{eq:Zparam}), 
\begin{equation*}
  \frac1{Z(k^2)} = \frac{k^2+1+\Sigma(k^2)}{k^2 + 1 + 3\,\avg{\phi^2}}
  = 1 + \frac{\tilde\Sigma(k^2)} {k^2 + M^2}
\end{equation*}
where $ M^2=1 + 3\,\avg{\phi^2} $ and $ \tilde\Sigma(k^2) = \Sigma(k^2) -  
3\,\avg{\phi^2} $. Notice that $ \tilde\Sigma(k^2) $ can also be written as 
sum of Feynmann diagrams with dressed propagators $ (k^2+M^2)^{-1} $ and no 
tadpole insertion of any order. Thus, the lowest order contribution to
$ \tilde\Sigma(k^2) $ comes from the sunset diagrams with dressed
propagators.  Now, since the renormalized $\phi^4$ QFT is asymptotically
free in the ultraviolet (that is the ultraviolet is controlled by the
Gaussian fixed point), the exact $\phi$ two-point function must behave as
$1/k^2$ in the ultraviolet (even in the presence of the cutoff
$\bar\Lambda$, provided $k$ remains much smaller than $\bar\Lambda$). Thus
we expect $Z(k^2)$ to tend to unity for large $k$ (and from above, since
the two-loop sunset diagram enters in $\tilde\Sigma$ with a minus sign),
while $\avg{\phi^2}$ still diverges linearly with $\bar\Lambda$. Since the
only divergence in $\tilde\Sigma(k^2)$ comes from the sunset diagrams and
is only logarithmic, $Z(k^2;T,\bar\Lambda)$ tends to $1$ for all $k$ in the
limit $\bar\Lambda \to\infty$ to any finite order of perturbation theory.
A different behaviour in  $\bar\Lambda$ would signal a nonperturbative effect.

To know more about $Z(k^2)$, in a nonperturbative
way, we  resort to Monte Carlo simulations in
section~\ref{sec:MC}.  In preparation, we need to re-express some relevant
equilibrium quantities with the lattice regularization which corresponds to
the discrete dynamics of section~\ref{ddlc}.

\subsection{Equilibrium on the lattice}\label{sec:eqlatt}

Thermal expectation values on the lattice are written as their continuum
counterparts, eq.~(\ref{prog}), in terms of standard multiple integrals
over phase-space degrees of freedom and the statistical weight $\exp(-H/T)$
where $H$ is now the lattice Hamiltonian. As phase-space degrees of freedom
we take $\pi_+(\vx,0)\equiv\pi_+(\vx)$ and
$\phi(\vx+a\vs_+,0)\equiv\phi(\vx+a\vs_+)$ with $\vx\in 2a{\mathbb Z}^3$,
or better $\vx\in 2aC_N$ after the restriction to a finite volume (see 
section~\ref{finitelat}).  As lattice Hamiltonian we take the total
conserved energy of eqs.~(\ref{densE}) and~(\ref{emas}) with $D=3$, namely
\begin{equation}\label{eq:Hpiphi}
  \begin{split}
    H[\pi_+,\phi] &= \sum_{\vx \in 2aC_N}  \E_+(\vx,0) 
    \;=\; \frac12\sum_\vx(2a)^3 \big[ \pi_+^2
  + (\mathrm{D}\phi)^2 + \phi_+^2 + \tfrac12 \phi_+^{\,4} \big] \\
  &=\frac12 \sum_\vx (2a)^3 \Big[ \pi_+^2  + (\mathrm{D}\phi)^2 + 
  \tfrac12\Phi^{(2)}+ \tfrac12(\Phi^{(1)} + a\,\pi_+)^2 (1+\Phi^{(2)})\Big]
  \end{split}
\end{equation}
where we used the first of eqs.~(\ref{lattcont}) to express $\phi(\vx,a)$
as $\Phi^{(1)}(\vx,0) + a\,\pi_+(\vx,0)$. We see that the momentum field
$\pi_+$ enters the lattice Hamiltonian in a way definitely more involved
than in the continuum, eq.~(\ref{hamilt}), or in the naive lattice
regularization where only space is discretized and $\pi$ enters only the
kinetic energy as in eq.~(\ref{eq:Tlat}). However, $H[\pi_+,\phi]$ still
depends on $\pi_+$ only ultra-locally, so that $\pi_+$ plays the role of a
Gaussian auxiliary field which could be integrated away to produce an
effective Hamiltonian for $\phi$ alone with a non-polynomial local
self-interaction.  This self-interaction reproduces the standard continuum
$\phi^4$ potential ${\mathcal V}[\phi]$ in the limit $a\to0$. However, it
is simpler not to integrate over $\pi_+$ and keep working with the quartic
Hamiltonian eq.(\ref{eq:Hpiphi}).

\medskip 

We discuss first the Gaussian approximation to the thermal equilibrium on the
lattice. This is valid for the linearization around classical solutions (zero
field and the cnoidal) as well as for the selfconsistent Hartree approximation.

Consider the lattice equation of motion (\ref{recursion}) and the first of
the correspondence rules eq. (\ref{lattcont}) written in Fourier space, that is
\begin{equation}\label{pikphik}
  {\tilde\pi}_{\pm,\vk}(t) = \pm\,\frac1{a}\big[ {\tilde\phi}_\vk(t\pm a) - 
  {\tilde\phi}_\vk(t) \big] \;,\quad 
  C(\vk a) \equiv \prod_{j=1}^D \cos k_ja 
\end{equation}
Next we take the free-field limit of eq.~(\ref{recursion}) by linearizing
it.  Details are reported in Appendix~\ref{app:B}, where linearization is
performed more generally over the uniform time-dependent solution of the
discrete dynamics (the discrete cnoidal). Free-field dynamics is decoupled
in Fourier space and harmonic also on the lattice, so that the standard
virial theorem for harmonic oscillations (which holds true for phase-space
averages as well as for time averages) yields, as in eq.~(\ref{fvirial})
specialized to $D=3$;
\begin{equation}\label{eq:hvirial}
  \avg{|{\tilde\phi}_\vk|^2} = a^2 \Big[1 - 
  \frac{1-\tfrac12 a^2}{1+\tfrac12 a^2}\,C(\vk a)
  \Big]^{-1}\avg{|{\tilde\pi}_{+,\vk}|^2} =   \frac{a^2\,T}
  {\big[1 + \tfrac12a^2 - (1- \tfrac12a^2)\, C^2(\vk a) \big]} 
  \equiv G_0(\vk,a)
\end{equation}
having used equipartition, that is
$\avg{|{\tilde\pi}_{+,\vk}|^2}=T/(1+\frac12a^2)$, as follows from the
quadratic part of the Hamiltonian eq.(\ref{eq:Hpiphi}). $G_0(\vk,a)$ in
eq.~(\ref{eq:hvirial}) is the tree-level (Fourier transform of) the $\phi$
two-point function on the lattice and can be seen to coincide with the
result obtained by integrating $\pi_+$ away within a quadratic
approximation. Notice that $\avg{|{\tilde\phi}_{\bds 0}|^2}=1$ as its
continuum counterpart. The small $a$-dependent tree-level correction in
$\avg{|{\tilde\pi}_{+,\vk}|^2}=T/(1+\frac12a^2)$ w.r.t. the continuum
$\epik2=T$ is evidently a first effect of our specific lattice
regularization. Similarly, the finite lattice version of the tree-level
tadpole reads
\begin{equation}\label{eq:tadlat}
  \begin{split}
    \avg{\phi^2} &= \frac1{L^3}\sum_{\vk\in (2\pi/L)C_N} G_0(\vk,a) 
    \buildrel{L\rightarrow \infty}\over = 
    \frac{T}{a}\,[1+{\cal O}(a)]\int_{-\pi/2}^{+\pi/2}\frac{d^3q}{(2\pi)^3} 
    \frac1{1-C^2(\bds q)} \\ & = 0.1741 \ldots \frac{T}{a}\,[1+{\cal O}(a)]
    =  0.1108 \ldots T\,\Lambda\;[1+{\cal O}(\Lambda^{-1})]
  \end{split}
\end{equation}
to be compared with the infinite-volume continuum expression eq.~(\ref{tad1}). 
If these two expressions are identified, one obtain the tree-level relation
${\bar \Lambda} = 2.187 \ldots \Lambda$, which differs from that 
implied by $\avg{\pi^2}$ [see section \ref{eqgenral}]. 

Another important issue about lattice effects concerns the
equilibrium expectation value of $\phi^4$. We have already commented at the
end of section \ref{ddlc} on two different lattice regularizations of $\phi^4$,
the slightly nonlocal $\phi_\pm^4$ of eqs.~(\ref{lattcont}) and the
ultralocal $\phi^4(\vx) \equiv [\phi(\vx)]^4$. 
They do not yield the same values 
due to ultraviolet effects. At equilibrium this appears
quite evident already at tree level: application of Wick theorem to the
expectation value of the ultralocal $\phi^4$ yields
\begin{equation*}
  \avg{\phi^4} = 3\,\avg{\phi^2}^2 \;,\quad \avg{\phi^2} = 
  \frac1{V}\sum_\vk G_0(\vk,a)
\end{equation*}
as in the continuum, only with a sum over wave-vectors in the first
Brillouin zone rather than a rotation invariant integral as in
section~\ref{sec:pert}. The same Wick rules on Gaussian integration plus
translation invariance give instead for $\phi_+^4$:
\begin{equation*} 
  \begin{split}
    \avg{\phi_+^4} &= \avg{(\Phi^{(1)} + a\,\pi_+)^2\Phi^{(2)}} =
    \frac{T\,\avg{\Phi^{(2)}}}{8a(1+\tfrac12a^2)} +  
    \frac{\avg{[\Phi^{(1)}]^2\Phi^{(2)}}}{(1+\tfrac12a^2)^2} \\
    &= \frac{\avg{\phi^2}}{1+\tfrac12a^2} \Big[\frac{T}{8a} + 
    \frac{I_2}{1+\tfrac12a^2} \Big] + \frac{2\,I_2^2}
    {(1+\tfrac12a^2)^2} \quad,\qquad 
    I_2 \equiv \frac1{V}\sum_\vk C^2(\vk a)\,G_0(\vk,a)
  \end{split}
\end{equation*}
A straightforward numerical integration yields the tree-level ratio
\begin{equation} \label{fi42}
  \avg{\phi_+^4}/\avg{\phi^2}^2= 1.154748\ldots
\end{equation}
for $L=12.8$ and $a=0.0125$.

\medskip 

When the interaction is turned on both $\avg{|{\tilde\pi}_{+,\vk}|^2}$ and
$\avg{|{\tilde\phi}_\vk|^2}$ gets modified. The relations between
$\avg{|{\tilde\pi}_{+,\vk}|^2}$ and the temperature $T$ gets new corrections in
terms of expectation values of $\phi$ dependent observables. A look at
eq.~(\ref{eq:Hpiphi}), with the knowledge that the Euclidean $\phi^4$ QFT is
superrenormalizable (so that UV divergences can be fully assessed from
perturbation theory as in the previous section), reveals that these corrections
all vanish at least linearly as $a\to0$, so that the continuum result $\epik2=T$
is recovered. At any rate, here we are only seeking a convenient parametrization
for $\avg{|{\tilde\phi}_\vk|^2}$ in terms of $\avg{|{\tilde\pi}_{+,\vk}|^2}$ and
$\avg{\phi^2}$.

Again, we may resort to the linearization of the discrete equation of
motion, this time over a uniform nonzero background $\phi(t)$, as
detailed in Appendix \ref{app:B}. First of all we learn how to obtain the
Hartree resummation on the lattice: we replace the background $\phi^2(t)$
with the equilibrium expectation value $\avg{\phi^2}$ in the
dispersion relation eq.~(\ref{disprule}), obtaining
\begin{equation}\label{eq:eqdisp}
  \cos \left[ \omega_\vk a \right] = B(\avg{\phi^2})\, C(\vk a) \quad , \quad
    B(u) = \frac{1 + \tfrac12 \; {a^2}  \; (1-u)} 
  {\bigl[1 + \tfrac12 \; {a^2} \;  (1+u) \bigr]^2}
\end{equation}
and invoke the virial theorem (the Hartree approximation is Gaussian, that
is describes harmonic oscillations) to write 
\begin{equation}\label{eq:eqhartree}
  \avg{|{\tilde\phi}_\vk|^2} = 
  \frac{a^2 \; \avg{|{\tilde\pi}_{+,\vk}|^2}}
  {1 - (1-\tfrac14 \; a^4)\cos^2\omega_\vk a} \simeq
  \frac{a^2 \; \avg{|{\tilde\pi}_{+,\vk}|^2}} 
  {1 - [B(\avg{\phi^2}) \; C(\vk a)]^2}
\end{equation}
We have neglected the term $a^4\cos^2\omega_\vk$ in the last step since it
is of order $a^4$ for {\bf all} $\vk$. 
Beyond the Hartree approximation, we promote eq.~(\ref{eq:eqhartree}) to an exact
relation by introducing the $k-$dependent lattice wave-function
renormalization $Z_\vk=Z_\vk(T,a)$
\begin{equation}\label{eq:Zparamlatt}
  \avg{|{\tilde\phi}_\vk|^2} = 
  \frac{a^2 \; Z_\vk \; \avg{|{\tilde\pi}_{+,\vk}|^2}} 
  {1 - [B(\avg{\phi^2}) \; C(\vk a)]^2}
\end{equation}
$Z_\vk(T,a)$ is by construction as much as possible free from lattice
effects (such as the deformation of free-field propagator at large $|\vk|$
due to the replacement of continuum derivatives with finite differences)
and, once averaged over discrete directions to $Z_k(T,a)$, it should
provide a rather accurate representation of its continuum counterpart
$Z(k^2;T,\bar\Lambda)$, if a universal relation between $\bar\Lambda$ and
$\Lambda=\pi/(2a)$ could be determined. As we have seen above, in the case
of UV divergent quantities like $\avg{\pi^2}$ and $\avg{\phi^2}$ this is
not possible, but should be possible for renormalized or UV finite
observables. Perturbation theory suggests this to be the case for
$Z(k^2;T,\bar\Lambda)$. In the next section we shall follow a different,
nonperturbative strategy, based on the simple observation that the
agreement of lattice calculations with their continuum
$\bar\Lambda-$cutoffed counterparts is guaranteed for all observables in a
different limit, that is $\Lambda\to\infty$ first, at fixed large
$\bar\Lambda$, rather than $\Lambda,\,\bar\Lambda\to\infty$ with
$\Lambda={\cal O}(\bar\Lambda)$. For $\vk-$dependent quantities this means
agreement for $|\vk|\lesssim\bar\Lambda\ll\Lambda$.

\subsection{Monte Carlo simulations}\label{sec:MC}

Monte Carlo simulations allow to compute thermodynamical averages beyond
perturbation theory and Hartree resummation. We performed Metropolis
simulations for the $\phi^4$ model discretized on the lattice as in section
\ref{ddlc}.

\bigskip

\noindent
Consider the lattice Hamiltonian $H[\pi_+,\phi]$ of eq.~(\ref{eq:Hpiphi}).
In this context it is more convenient to revert to
the notation with the single field $\phi$, using both $\phi(\vx,0)$ and 
$\phi(\vx,a)$. We may drop the distinction between the two time slices 
by considering the degrees of freedom attached to a cell-centered cubic 
lattice.
There are therefore the points of $2a {\mathbb Z}^3$, labeled by
$ \vx=2a\vn , \; \vn$ having integer coordinates, and the points of
$ 2a({\mathbb Z+1/2})^3 $ labeled $\vx+a\vs_+$, where $\vs_+=(1,1,1)$.
The Hamiltonian can then be written as
\begin{equation}\label{Hto}
  H = 2\,a ( 2 + a^2)  \sum_{\vx \in 2a {\mathbb Z}^3} \left[\phi^2(\vx)
    + \phi^2(\vx + a\vs_+) \right]
  - a \sum_{\vx \in 2a {\mathbb Z}^3}\sum_\vs \phi(\vx) \; \phi(\vx+a\vs)
  \; \Big[ 1 - \tfrac14 a^2\; \phi(\vx) \; \phi(\vx+a\vs) \Big] \; .
\end{equation}
This Hamiltonian contains local terms and couplings between
nearest-neighbors along the main diagonals.  The quartic term contribute to
these couplings and there is no quartic local term.  However, to perform
Monte Carlo simulations is more convenient to add and subtract a quartic
local piece. In this way one can change the local integration variables
$-\infty < \phi(\vx) < +\infty $ in Eq.  (\ref{prog}) to a new variable 
$C(\vx) $ ranging from zero to one as in \cite{bin}.

We therefore split our Hamiltonian eq.~(\ref{Hto}) as
\begin{eqnarray}
  && H = H_1 + H_2 \; , \cr \cr
  &&H_1 =  \sum_{\vx \in 2a {\mathbb Z}^3} 
  \left\{h_1[\phi(\vx)] +h_1[\phi(\vx+a\vs_+)] \right\}  \quad,\qquad
  \quad  h_1[\phi] \equiv
  4 \, a \, ( 2 + a^2)\;  \phi^2 + 2 \; a^3 \; \phi^4 , \cr \cr
  && H_2 = - a \,  \sum_{\vx \in 2a {\mathbb Z}^3}\sum_\vs \phi(\vx) 
  \; \phi(\vx+a\vs) + \frac{a^3}4 \sum_{\vx \in 2a {\mathbb Z}^3}
  \sum_\vs \phi^2(\vx) \left[ \phi^2(\vx+a\vs) - \phi^2(\vx) \right] \; .
\label{ising}
\end{eqnarray}
The functional integral in  Eq. (\ref{prog}) can be written as,
\begin{eqnarray}
&&\avg{\Theta} =
\frac{\int_{ \infty}^{+\infty} \cdots \int_{-\infty}^{+\infty}
\left[ \prod_\vx d \phi(\vx) \; e^{-\beta \; 
h_1[\phi(\vx)]}  \right] 
 \; \Theta[\phi(.)] \; e^{ -\beta  H_2[\phi(.)]}}
{\int_{ \infty}^{+\infty} \cdots \int_{-\infty}^{+\infty}
\left[ \prod_\vx d \phi(\vx) \; e^{-\beta \; 
h_1[\phi(\vx)]}  \right] \; e^{ -\beta  H_2[\phi(.)]}}   \cr \cr
&&= \frac{\int_{0}^{1} \cdots \int_{0}^{1}\left[ \prod_\vx 
d C(\vx)  \right]  \; \Theta[C(.)] \; e^{ -\beta  H_2[C(.)]}}{\int_{0}^{1} 
\cdots \int_{0}^{1}\left[ \prod_\vx 
d C(\vx)  \right]\; e^{ -\beta  H_2[C(.)]}}
\end{eqnarray}
where the new field $ C $ is defined as
\begin{equation}\label{cfi}
C(\phi) \equiv = \frac{\int_{-\infty}^{\phi} dx \; e^{-\beta \; 
h_1[x]}}{\int_{-\infty}^{+\infty} dx \; e^{-\beta \; h_1[x]}}\; .
\end{equation}
In this way the classical $ \phi^4 $ model is mapped into a classical
continuous Ising-like model where the dynamical variables $ C(\vx) $ run
between zero and one.  The Ising-like Hamiltonian is given by $H_2$ [Eq.
(\ref{ising})].

Following the Metropolis method we generate a sequence of configurations
for the whole cell-centered cubic lattice, as follows. We start by choosing
random values at each point of the lattice for the variables $ C(\vx) $.
Then, we choose one point in the lattice $ \vx $ at random and consider a
new value $ C' $ for it picked at random between zero and one. We then have
to go back from the variables $ C(\vx) $ to the variables $ \phi(\vx) $ in
order to compute the energy of the old and new configurations. Notice that the
inverse function $ \phi = \phi(C) $ is unique since $ dC/d\phi > 0
$ as one sees from Eq. (\ref{cfi}). Their energy change is given by
$$
\Delta \equiv H_2(\mathrm{new}) - H_2(\mathrm{old}) = 2 \; a \;  \left[\phi' - 
\phi(\vx) \right]  \left\{ \sum_\vs \phi^2(\vx+a\vs) - 4 \left[  \phi'^2 + 
 \phi^2(\vx) \right]\right\}
$$
where we used Eq.(\ref{ising}) and $ \phi' \equiv \phi(C') $

We now follow the standard Metropolis procedure. That is, we compare $ e^{-
  \beta \; \Delta } $ with a random number $r$ with $ 0 < r < 1 $.  If $ r
< e^{- \beta \; \Delta } $ we pick the new configuration. Otherwise, for $
r > e^{- \beta \; \Delta } $ we keep the old one.  We repeat this process
many times producing in this way a sequence of configurations on which we
can compute the values of any observable $ \Theta $.

It can be shown \cite{bin} that the average of these values of $ \Theta $
converges to the thermodynamical phase average in the limit when the number 
of iterations $ {\cal N} $ goes to infinity 
The expectation value is then given by
$$
\avg{\Theta} = \frac1 {\cal N} \sum_{n=1}^{\cal N} \Theta_n \; .
$$
In the continuous limit $ a \to 0 $ we see from eq.~(\ref{ising}) that the 
quartic terms
become negligible and hence all quantities will be functions of
$ \beta \; a = a/T $. In addition, gradients of the fields scale as $ 1/a $.

\begin{figure}[htbp]
  \centering
  \psfrag{<phi^2> vs. T/a}{$\avg{\phi^2}~\mathrm{vs.}~T/a$} 
  \psfrag{a^2 <(grad phi)^2> vs. T/a}[r][r]
  {$a^2\avg{(\nabla\phi)^2}~\mathrm{vs.}~T/a$} 
  \psfrag{<phi^4>/[<phi^2>]^2 vs. T/a}[r][r]
  {$\dfrac{\avg{\phi^4}}{\avg{\phi^2}^2}~\mathrm{vs.}~T/a$} 
  \includegraphics[width=10cm,height=12.72cm,angle=270]{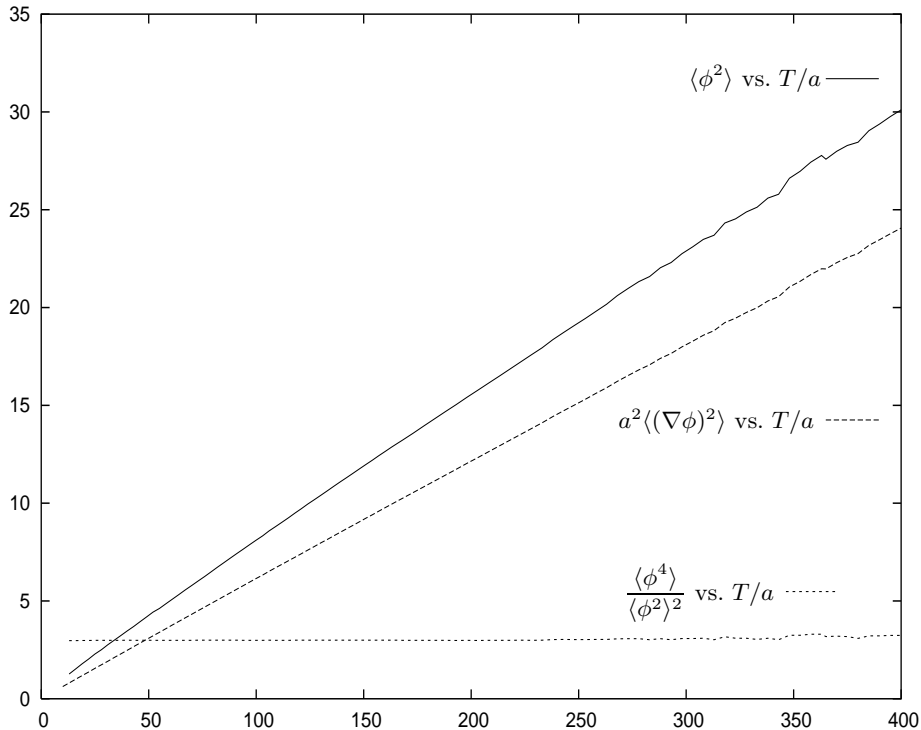}
  \caption{ $ \avg{\phi^2} , \;  \avg{(\nabla \phi)^2} $ 
    and the ratio $ \avg{\phi^4}/\avg{\phi^2}^2 $ as a function of $ T/a
    $ in thermal equilibrium from Monte Carlo simulations.}
  \label{fi2gradtreMC}
\end{figure}

In fig. \ref{fi2gradtreMC} we plot $ \avg{\phi^2}, \; \avg{(\nabla \phi)^2}
$ and the ratio $ \avg{\phi^4}/[\avg{\phi^2}]^2 $ as functions of $ T/a $
from Monte Carlo simulations. We see that $ \avg{\phi^2} $ and $
\avg{(\nabla \phi)^2} $ follow the linear behaviour predicted by low
temperature perturbation theory eq.~(\ref{tad1}) while the ratio $
\avg{\phi^4}/[\avg{\phi^2}]^2 $ takes the value $3$ predicted by Wick's
theorem in low temperature perturbation theory eq.~(\ref{tres}) or in
Hartree approximation. In the Monte Carlo simulations we used in the
lattice the ultralocal definition for $ \phi^4 $. For the definition $
\phi_+^4 $, we find from the Monte Carlo simulations $
\avg{\phi_+^4}/[\avg{\phi^2}]^2 \simeq 1.1\ldots $ for $ a=0.05 $ and $ L =
3.2 $ in agreement with eq.(\ref{fi42}).

As a matter of fact, by measuring also the two-point function $
\avg{\phi(\vx)\phi(\vx')}$ we verified that the Hartree approximation is
indeed quite accurate for all wavelengths, except for the largest ones, and
for a wide range of temperatures. We recall that this fact depends on the
classical statistical mechanics interpretation of the ultraviolet
divergences, which are physical and cannot be eliminated by counter-terms as
in quantum field theory. Thermal equilibrium at nonzero temperature really
makes sense for the classical $\phi^4$ theory in three space dimensions only
with a UV cutoff present, that is on a lattice. As $ a \to 0 $, the only
way to keep things finite is to decrease $T\to0$, making the Hartree
approximation, which resumes all $T/a$ corrections, more and more accurate.

\section{Dynamics of Thermalization}

We present here the exact evolution of the lattice $\phi^4$
theory defined in sec. \ref{ddlc} and the development of the ultraviolet
cascade. We considered three system sizes, $L=6.4,\,12.8$ and $25.6$ and
several lattice spacings, $a=0.1,\,0.064,\,0.05,\,0.025, \,0.0125$ and
$0.00625$. The largest system contains $1024^3$ lattice sites. We worked
with the largest volume $25.6^2$ only for $a=0.05$ and $E/V=89.5$, verifying
that the change from $L=12.8$ to $L=25.6$ does not change any observable in
a significant way. Therefore $L=12.8$ has been our preferred choice.

\subsection{Initial Conditions}\label{incond}
We used a large variety of initial conditions in our calculations. In
these studies the initial power is concentrated in the infrared; that is,
$|{\tilde\phi}_\vk|^2(0)$ and $|{\tilde\pi}_\vk|^2(0)$ are significant only
for wavenumbers well below the cutoff $\Lambda = \frac{\pi}{2a} $. In
particular, we considered superpositions of infrared plane waves,
when the initial fields have the form
\begin{equation}\label{ondp}
  \phi(\vx,0) = \phi_0 + A\,\sum_{i=1}^K c_i \; \cos({\vk}_i \cdot {\vx}+2
  \, \pi \, \gamma_i) \quad , 
  \quad    \pi(\vx,0) = B\,\sum_{i=1}^K d_i \;  \cos({\vk}_i \cdot {\vx}
  +2 \, \pi \, \delta_i) \;.
\end{equation}
as well as superpositions of localized wave packets of the form
\begin{equation} \label{campa}
  \phi(\vx,0) = \phi_0 + A\sum_i c_i \; w(\vx-\vx_i) \;, \quad
  \pi(\vx,0) = B\sum_i d_i  \; w(\vx-\vx'_i) \;.
\end{equation}
where 
\begin{equation*}
  w(\vx) = \sum_{\vn \in {\mathbb Z}^3} w_0(k_{max}[\vx-L\; \vn]) 
\end{equation*}
and $ w_0(\vx) $ is either the Gaussian, $ w_0(\vx)=e^{-\vx^2}$, or the 
Lorentzian, $w_0(\vx)=(1+\vx^2)^{-1}$. The sum over $\vn$ in the last
equation is needed by p.b.c, but in practice, with our choice 
$L\sim 10$, only few terms in the sum are needed. These fields have
support throughout Fourier space, but peaked as Gaussians or simple
exponentials at low wavenumbers $ k \lesssim k_{\rm max}$. 

In eqs.~(\ref{ondp}) and (\ref{campa}), $\phi_0$ is a uniform background,
or homogeneous condensate, while the wave-vectors $\vk_i=2\pi {\vn}_i/L$ in
eq.~(\ref{ondp}) have nonzero modulus $|\vk| \le k_{\rm max} \lesssim
30\pi/L = 15\pi/(N \, a) \ll \pi/(2a)$. The number $K$ in eq.(\ref{ondp}) 
is chosen within the
range $10-100$, proportionally to $k_{\rm max}$, and the specific $\vk_i$
in eq.~(\ref{ondp}) or the $\vx_i$ and $\vx_i$ in in eq.~(\ref{campa}) are
chosen at random. The phases $\gamma_i$, $\delta_i$ in eq.~(\ref{ondp}) and
the relative amplitudes $c_i\,,d_i$ in both cases are chosen at random in
the interval $[0,1]$. Finally, for any given realization of these random
numbers, the background $\phi_0$ and the overall amplitudes $A$ and $B$ are
constrained in such a way that the energy density $E/V$ takes a given,
predefined value. Typically, we further restricted the remaining freedom by
choosing either $B=0$ or $B=A$, so that the `macroscopic' part of the
initial state is entirely characterized by the values of $E/V$, of the
condensate $\phi_0$ and of $k_{\rm max}$. In principle, one should then
regard the randomly chosen numbers as `microscopic', or fine-grained
properties providing the initial entropy; one should then average all
measured observables over these random choices, as described in
section~\ref{sec:avgobs}, to reduce the fluctuations in the measures.
Actually, we performed such average (over 20 to 40 initial conditions) only
rarely, since fluctuations were kept under control already by the 3D space
average, by the sliding time average and/or by the average over discrete
directions of Appendix~\ref{app:A}.

It should be noticed that all our initial configurations have vanishing
total momentum $ \int d^3x \; \pi(\vx) \; \nabla \phi(\vx) $, which is a
conserved quantity for p.b.c..

\subsection{Time evolution of local observables}\label{basico}

Here we show the time evolution of the basic local observables of
eq.~(\ref{lista}) averaged in space and time according to section
\ref{sec:avgobs}. As mentioned above, we considered many different initial
conditions, with several values of the energy density ranging from
$E/V=0.1$ up to $E/V=5000$.

\begin{figure}[h]
  \centering
  \psfrag{L6.4}{$L=6.4$} 
  \psfrag{a0.1}{$a=0.1$} 
  \psfrag{EoV1000 }{$E/V=1000$} 
  \psfrag{logt}[tc][tc]{$\log t$} 
  \psfrag{logphi2 }{$\log\overline{\phi^2}$}
  \psfrag{logphi4 }{$\log\overline{\phi_+^4}$}
  \psfrag{logdphi2 }{$\log\overline{(\nabla\phi)^2}$}
  \psfrag{logpi2 }{$\log\overline{\pi^2}$}
  \includegraphics[width=11.66cm,height=10cm]{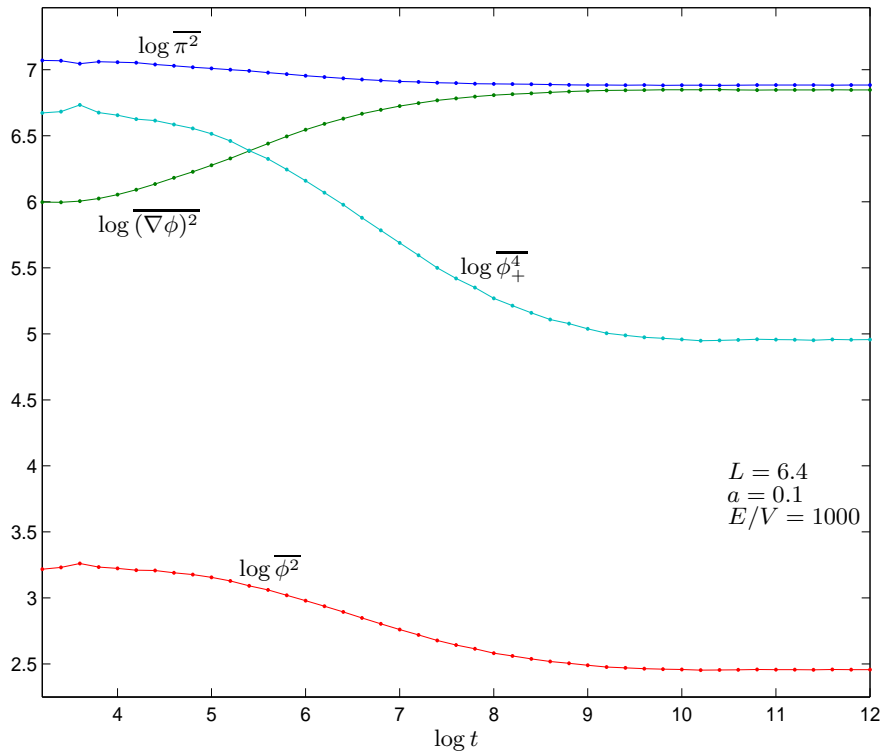}
  \caption{Log-log plot of the time evolution of local observables with
    sliding time average. Initial conditions are localized wave packets of
    Lorentzian shape without any zero-mode condensate. No average is
    performed over initial packet amplitudes or positions.}
  \label{logopf4}
\end{figure}

\begin{figure}[p]
  \centering
  \psfrag{L12.8}{$L=12.8$} 
  \psfrag{a0025}{$a=0.025$} 
  \psfrag{EoV419.5}{$E/V=419.5$} 
  \psfrag{EzoE0.98}{$E_0/E=0.98$ }
  \psfrag{logt}[tc][tc]{$\log t$} 
  \psfrag{logphi2}{$\log\overline{\phi^2}$}
  \psfrag{logphi4}{$\log\overline{\phi_+^4}$}
  \psfrag{logdphi2}{$\log\overline{(\nabla\phi)^2}$}
  \psfrag{logpi2}{$\log\overline{\pi^2}$}
  \includegraphics[width=11.66cm,height=10cm]{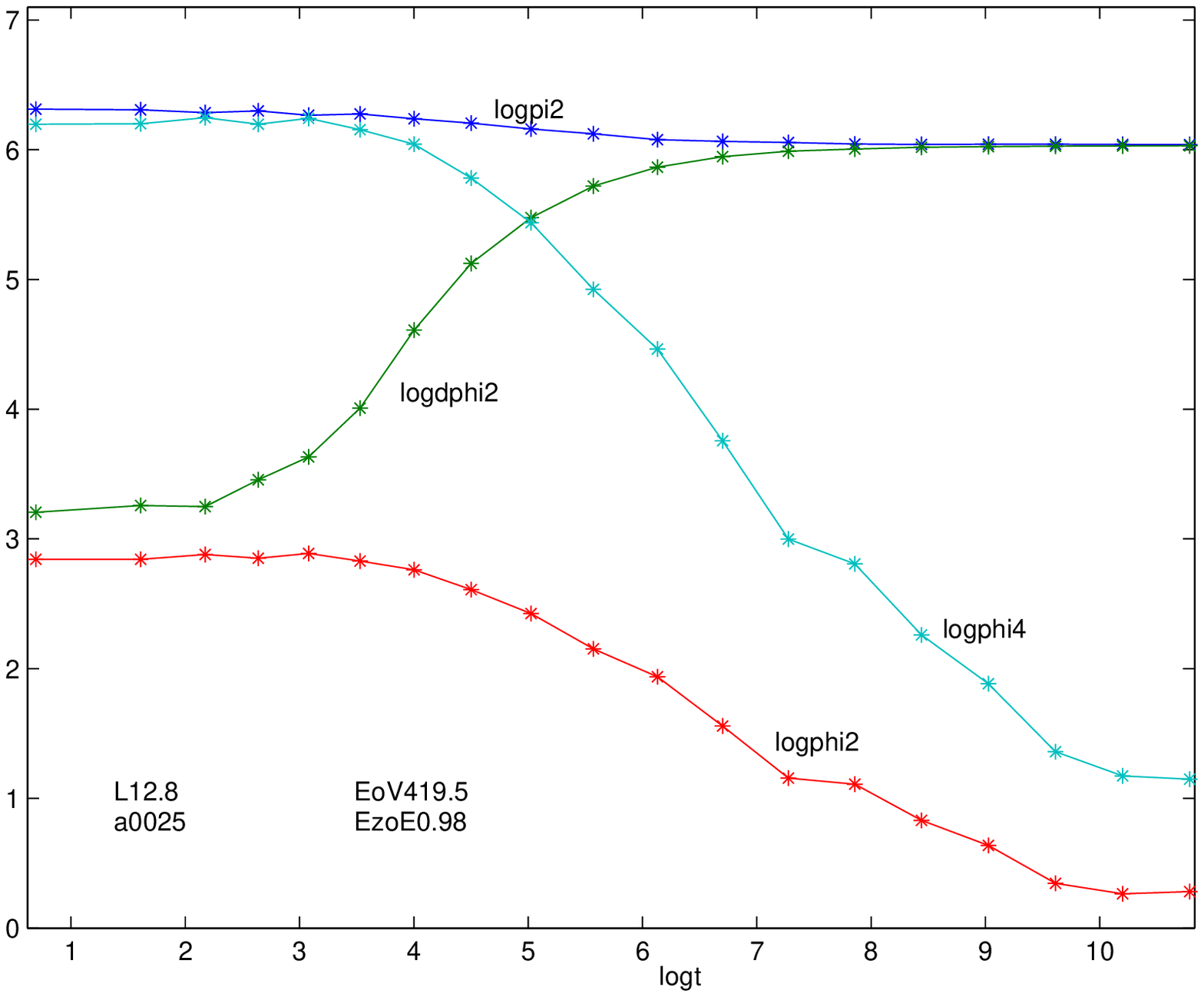}
  \caption{Log-log plot of the time evolution of local observables.
    The time averaging interval is quite small, $\tau=2$, and the initial
    conditions are plane waves with a large zero-mode condensate. No
    average is performed over initial amplitudes or phases.}
  \label{logopf3}
\end{figure}

\begin{figure}[p]
  \centering
  \psfrag{L12.8}{$L=12.8$} 
  \psfrag{a0.025}{$a=0.025$} 
  \psfrag{EoV12.5}{$E/V=12.5$} 
  \psfrag{EzoE0.98}{$E_0/E=0.98$ }
  \psfrag{logt}[tc][tc]{$\log t$} 
  \psfrag{logphi2 }{$\log\overline{\phi^2}$}
  \psfrag{logphi4 }{$\log\overline{\phi_+^4}$}
  \psfrag{logdphi2 }{$\log\overline{(\nabla\phi)^2}$}
  \psfrag{logpi2 }{$\log\overline{\pi^2}$}
  \includegraphics[width=11.66cm,height=10cm]{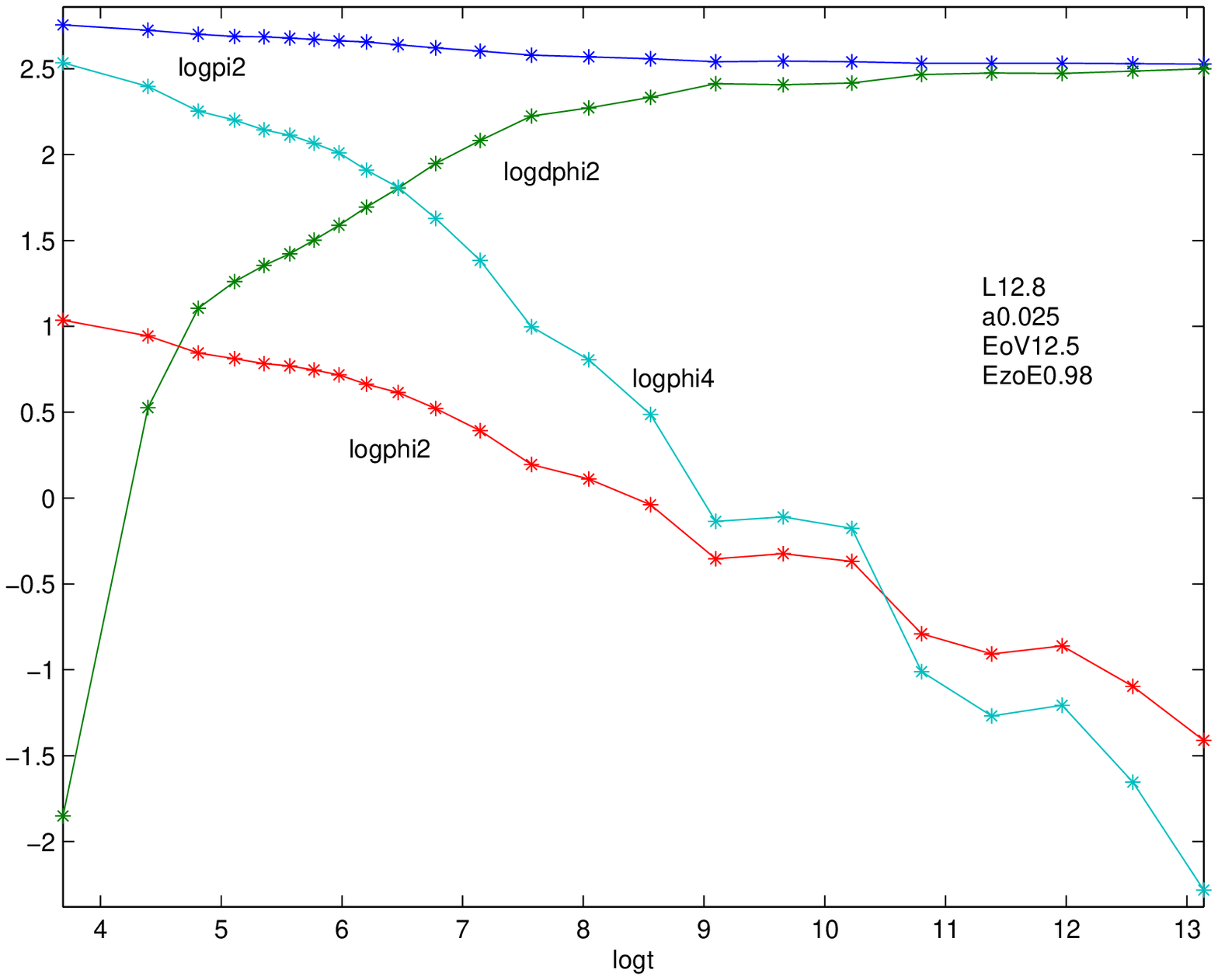}
  \caption{Log-log plot of the time evolution of local observables. 
    The time averaging interval is $\tau=40$ and the initial conditions are
    plane waves with a large zero-mode condensate. No average is performed over
    initial amplitudes or phases.}
  \label{logopf1}
\end{figure}

\begin{figure}[ht] 
  \centering
  \psfrag{L12.8}{$L=12.8$} 
  \psfrag{a0.064}{$a=0.064$} 
  \psfrag{EoV9.5}{$E/V=9.5$} 
  \psfrag{logt}[tc][tc]{$\log t$} 
  \psfrag{logphi2 }{$\log\overline{\phi^2}$}
  \psfrag{logphi4 }{$\log\overline{\phi_+^4}$}
  \psfrag{logdphi2 }{$\log\overline{(\nabla\phi)^2}$}
  \psfrag{logpi2 }{$\log\overline{\pi^2}$}
  \includegraphics[width=11.66cm,height=10cm]{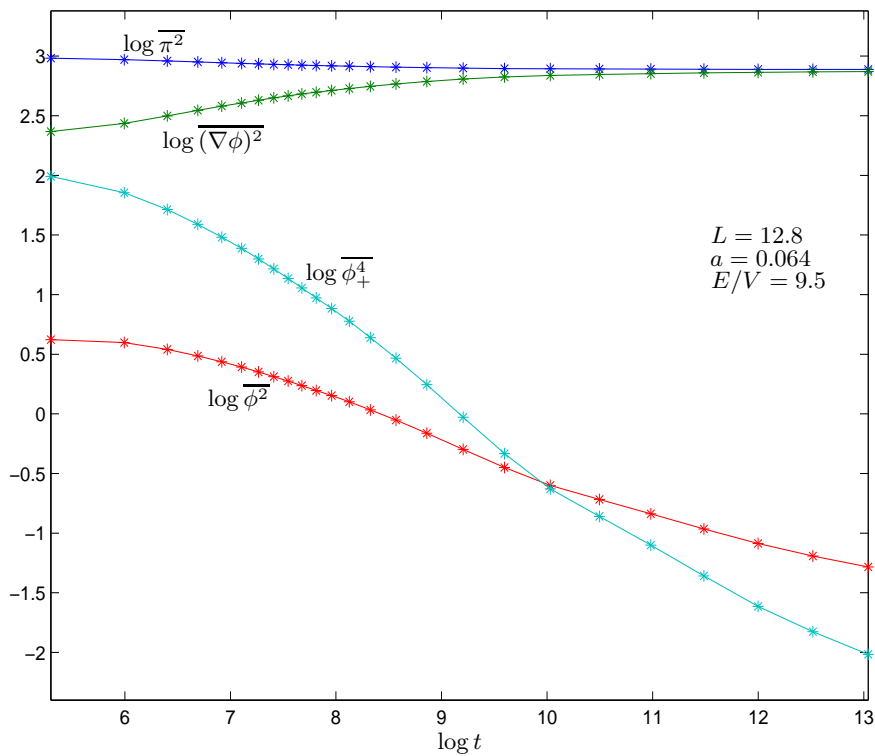}
  \caption{Log-log plot of the time evolution of local observables. 
    Initial conditions are plane waves without zero-mode condensate. The
    time averaging interval is $\tau=200$ and an extra average is performed
    over 40 random choices of initial amplitudes and phases. This explains
    why the curves in this picture are smoother than those in
    fig.~\ref{logopf3} and   \ref{logopf1}.}
  \label{logopf2}
\end{figure}

Figs.~\ref{logopf4} to \ref{logopf2}, display the local observables
$ \overline{\pi^2}(t), \; \overline{(\nabla\phi)^2}(t), \; 
\overline{\phi^2}(t) $ and $ \overline{\phi_+^4}(t) $, as functions of time for
different values of the relevant parameters. When an initial zero-mode
condensate is present [ $ \phi_0 $ in Eqs.(\ref{ondp}) and  (\ref{campa}) ], 
we quantify its weight with the ratio $E_0/E$ between the
energy built solely from the zero-mode and the total energy.

These plots clearly reveal the main qualitative feature of the
thermalization process in $ 3 + 1$ dimensions. Initially
$\overline{(\nabla\phi)^2}(t)$ is small reflecting the fact that the
initial conditions determine a power spectrum localized at wave-vectors with $
k \ll 1/a $.  The mode mixing entailed by the interaction is transferring
power to larger wave-vectors, thus effectively transferring energy from the
interaction and $ \pi^2 $ terms, which decrease, to the spatial gradient
term which increases. As is clear from these figures, $\overline{\pi^2}(t)$
and $\overline{(\nabla\phi)^2}(t)$ tend to a limit for late time.
$\overline{\phi^2}(t)$ and $\overline{\phi_+^4}(t)$ show a clear limiting 
behaviour only for $E/V$ large enough. 

The steady growth of $\overline{(\nabla\phi)^2}(t)$ at the expense of the
interaction term ${\overline {\phi_+^4}}(t)$, the kinetic energy 
$\overline{\pi^2}(t)$ and the mass term $\overline{\phi^2}(t)$ 
shows that the basic mechanism leading towards
thermalization is a rather uniform flow of energy towards larger $k$ modes,
namely the ultraviolet cascade. We can associate a time scale $t_0$ to the
onset of this ultraviolet cascade by looking at the point when
$\overline{(\nabla\phi)^2}(t)$ overtakes ${\overline {\phi_+^4}}(t)$, if such
point exists. For $E/V$ too small, implying small field amplitudes and very
small initial $\phi_+^4(0)$, $\overline{(\nabla\phi)^2}(t)$ might be larger
than ${\overline {\phi_+^4}}(t)$ from the beginning. In this case we can take
the point where the rise of $\overline{(\nabla\phi)^2}(t)$ is the steepest.
>From figs.~\ref{logopf4} to \ref{logopf2}, as well as from the rest of our
data, we see that $t_0$ does depend on $E/V$ and on the condensate, but stays
within the range $100 \lesssim t_0 \lesssim 400$ for $E/V$ in the range 
from $10$ to $5000$, with $t_0$ decreasing as $E/V$ increases at fixed
value of the condensate. 

\medskip

\begin{figure}[hp]
  \centering
  \psfrag{|grad phi|^2 vs. ln t}[r][r]
  {$\avg{(\nabla\phi)^2}~\mathrm{vs.}~\log t$} 
  \psfrag{phi^4 vs. ln t}[r][r]{$\avg{\phi_+^4}~\mathrm{vs.}~\log t$}
  \psfrag{phi^2 vs. ln t}[r][r]{$\avg{\phi^2}~\mathrm{vs.}~\log t$}
  \psfrag{pi^2 vs. ln t}[r][r]{$\avg{\pi^2}~\mathrm{vs.}~\log t$}
  \psfrag{a = 0.1, N=32^3, L=6.4}{$\!\!\!\!a = 0.1,\;N=32^3,\;L=6.4$}
  \psfrag{E/V = 0.1}{$E/V = 0.1$}
  \includegraphics[width=10cm,height=12.72cm,angle=270]{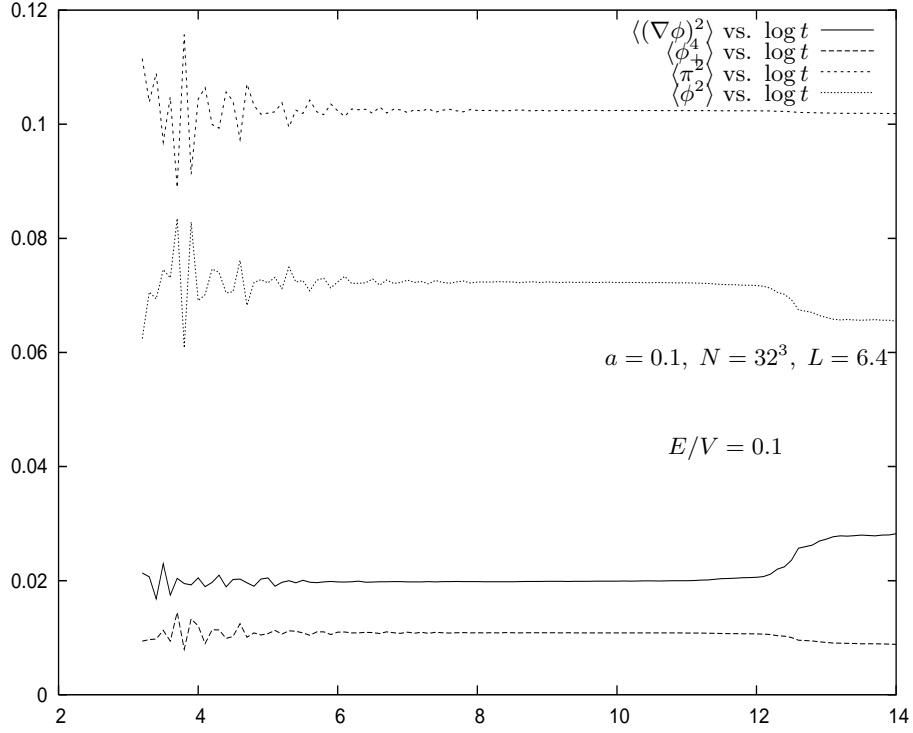}
  \caption{Log-log plot of the time evolution of local observables with
    sliding time average and very small energy density. Initial conditions
    are of the same type as in fig.~\ref{logopf4}. No average is performed 
over initial packet amplitudes or positions. Time averaging is performed here 
as by eqs.(\ref{promT})-(\ref{taut}).}
\label{e01a01n32}
\end{figure}

We are interested in somehow large energy densities $E/V$ which is the
relevant regime both for the early universe and the ultrarelativistic
heavy ion collisions. However, it is physically interesting to also
study the low energy density regime. 
We find that thermalization happens for small values of $E/V$. 
No threshold to a non-ergodic behaviour was found. 
However, the thermalization dynamics slows down
dramatically when $E/V$ is  reduced well below a value $ \sim 10 $ 
and  $t_0$ grows substantially. We depict in fig. \ref{e01a01n32} 
 $\overline{\left( {\nabla \phi}\right)^2}(t) , \; {\overline
    {{\dot \phi}^2}}(t), \; {\overline{ \phi^2}}(t) $ and $
  {\overline{\phi_+^4}}(t)$ for $ E/V = 0.1 , \; a=0.1, \; L=6.4 $.
We see that the ultraviolet cascade only starts in this case 
for much later times $ \ln t \simeq 11, \; t  \simeq 50000 $.

\bigskip

\begin{figure}[htbp]
  \centering
  \psfrag{E/V = 1000}{$E/V=1000$} 
  \psfrag{E/V = 10}{$E/V=10$} 
  \psfrag{E/V = 0.1}{$E/V=0.1$}
  \psfrag{ln |phi| vs. ln t}{$\log|\phi|~\mathrm{vs.}~\log t$}
  \psfrag{a = 0.1, N=32^3, L=6.4}{$a = 0.1, N=32^3, L=6.4$}
  \includegraphics[width=10cm,height=12cm,angle=270]{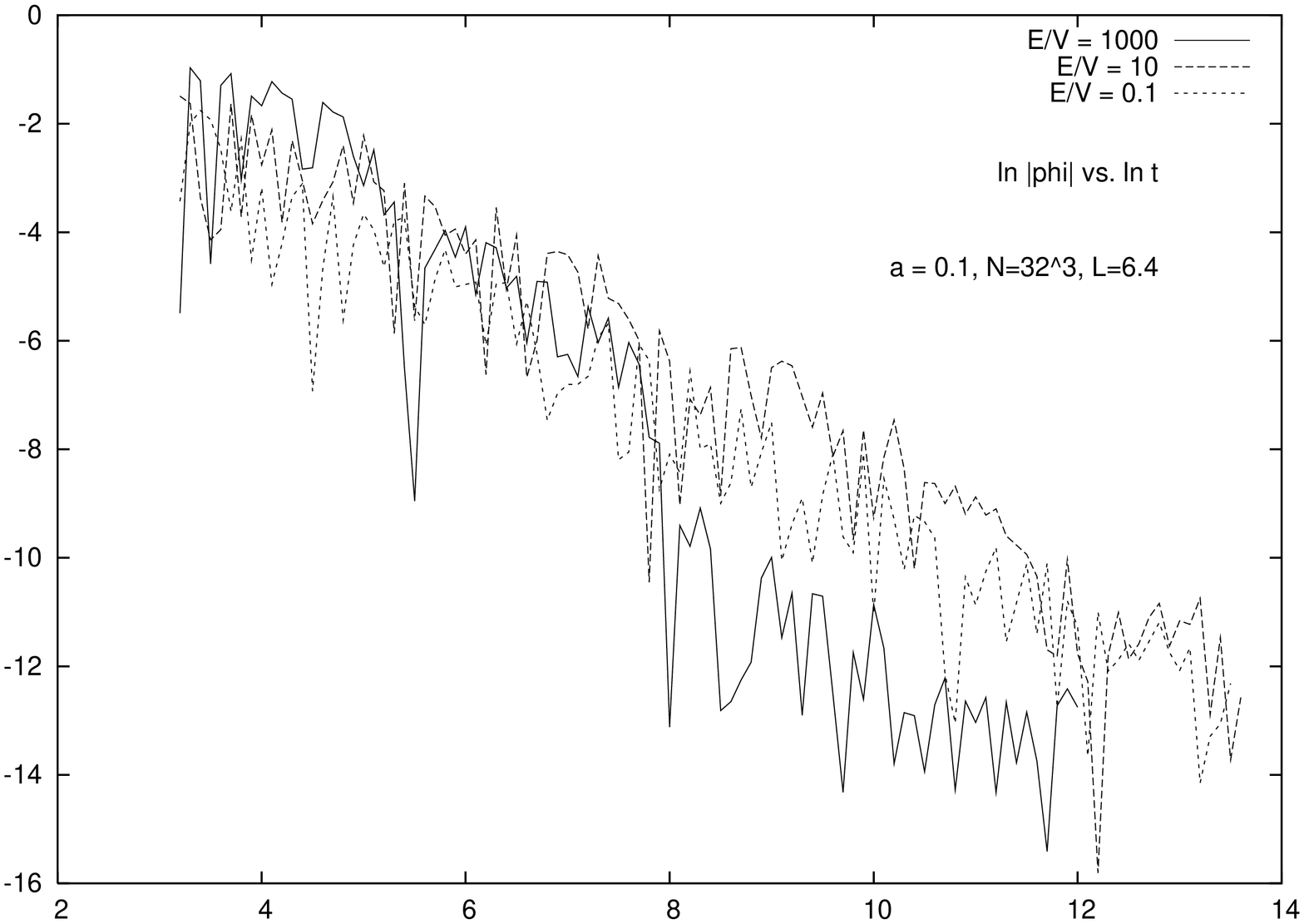}
\caption{$\log|{\overline {\phi} }(t)|$ as a function of the logarithm of the 
time $ t $ for $ E/V = 1000 $ and $ E/V = 10 $ with $ L = 6.4 $ and 
$ a = 0.1 $. No average is performed over initial packet amplitudes 
or positions. Time averaging is performed here as by 
eqs.(\ref{promT})-(\ref{taut}).}
\label{fi}
\end{figure}

Fig. \ref{fi} displays $\log|{\overline {\phi}}(t)|$ as a function of the
logarithm of time for  $ E/V= 1000, \; 10  $ and $
E/V = 0.1 $ for $  L = 6.4 $ and $ a = 0.1 $, respectively.
We see that the relaxation of $ \phi $ towards its thermal equilibrium
value ($\phi=0$) is different from the other physical quantities
previously discussed. This is due to the fact that the vanishing
of ${\overline {\phi}}$ is connected to a symmetry of the model.
We find that ${\overline {\phi}}(t)$ relaxes as  $\sim 1/t$ for $  \ln
t > \ln t_0 \simeq 6, \; t > t_0 \simeq 500 $ in the universal stage.
This can be seen from fig. \ref{fi}.

\bigskip


\begin{figure}[p]
  \centering
  \psfrag{"tfi24e1000.dat"}
  {$\avg{\phi_+^4}~\mathrm{vs.}~\avg{\phi^2}~
    \mathrm{from~time~averages}~E/V=1000,\;N=32^3,\;a=0.1$}
  \psfrag{"tfi24e500.dat"}[r][r]{$E/V=500,\;N=32^3,\;a=0.1$}
  \psfrag{"tfi24e100.dat"}[r][r]{$E/V=100,\;N=32^3,\;a=0.1$}
  \psfrag{"Xfi24MC.dat"}
  {$\avg{\phi_+^4}~\mathrm{vs.}~\avg{\phi^2}~\mathrm{from~Monte~Carlo}~
    N=32^3,\;a=0.1$}
  \includegraphics[width=10cm,height=12.72cm,angle=270]{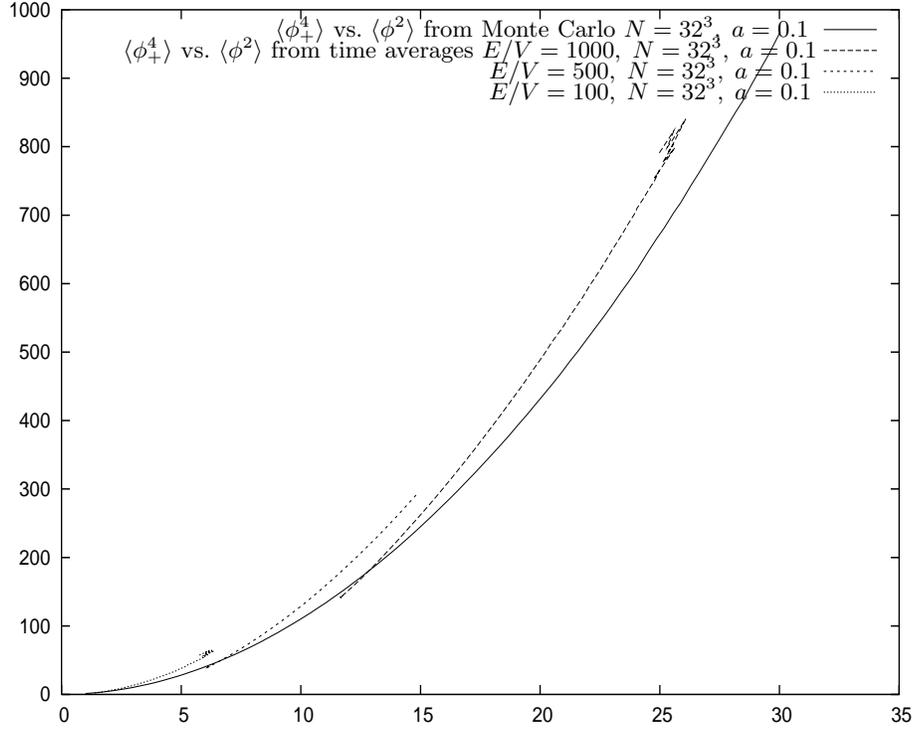}
  \caption{$ \avg{\phi_+^4} $ vs.  $ \avg{\phi^2} $ in thermal
    equilibrium (obtained from Monte Carlo simulations) and $
    {\overline{\phi_+^4}}(t) $ vs. $ {\overline{\phi^2}}(t) $ from time
    averages. Both $ {\overline{\phi_+^4}}(t) $ and $ {\overline{\phi^2}}(t)
    $ decrease for increasing time.}
\label{ptvsmc}
\end{figure}

We plot in fig. \ref{ptvsmc} $ \avg{\phi_+^4} $ vs.  $ \avg{\phi^2} $ 
in thermal equilibrium (obtained from Monte Carlo simulations) as well as 
$ {\overline{\phi_+^4}}(t) $ vs. $ {\overline{\phi^2}}(t) $ from time
averages. 
We see that the curve obtained from the lattice Monte Carlo calculations is
in agreement with the ones from time averages.  Since both $
{\overline{\phi_+^4}}(t) $ and $ {\overline{\phi^2}}(t) $ vary with time,
the agreement with the thermal curve implies that, at least for these
observables, we are in a situation of effective thermalization with a time
depending temperature. As in ref. \cite{uno} the effective temperature
decreases with time.  The small disagreement (less than $10\%$) between the
thermal equilibrium curve (from Monte Carlo) and the time averages in fig.
\ref{ptvsmc} comes from the low $k$ modes contribution. As it will be
discussed below, infrared modes are much slower to thermalize than the
modes with $ k^2 > {\overline{\phi^2}}(t) $. Actually, the disagreement
decreases with increasing time [decreasing $ {\overline{\phi_+^4}}(t) $ and
$ {\overline{\phi^2}}(t) $] showing that the IR modes are getting
thermalized equilibrating with the modes with $ k^2 >
{\overline{\phi^2}}(t) $.

\subsection{Virialization and Equation of State}

\begin{figure}[p]
  \centering
  \psfrag{"vire01n32a01.dat"}[r][r]{$E/V=0.1$}
  \psfrag{"vire1n32a01.dat"}[r][r]{$E/V=1$}
  \psfrag{"vire10n32a01.dat"}[r][r]{$E/V=10$}
  \psfrag{"vire100n32a01.dat"}[r][r]{$E/V=100$}
  \psfrag{"vire1000n32a01.dat"}[r][r]{$E/V=1000$}
  \includegraphics[width=10cm,height=12.72cm,angle=270]{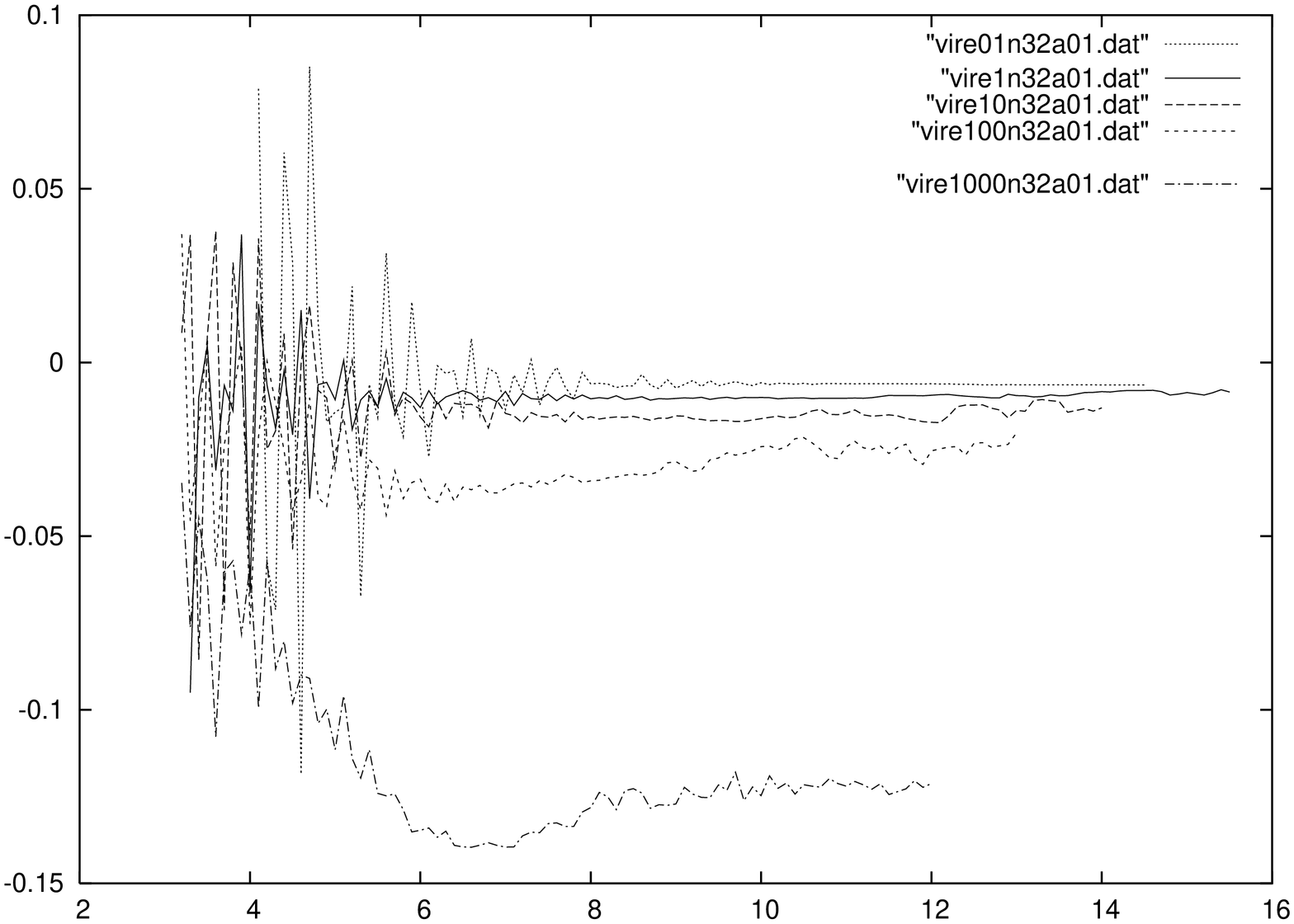}
  \caption{ The normalized l.h.s. of the virial theorem $\Delta(t)$ vs. the 
    logarithm of the time $ t $ for $ E/V=\rho = 1, \; 10, \; 100 $ and $ 1000
    $ for $ a=0.1 $ and $ L = 6.4 $ [see eq.(\ref{Dvir})]. No average is 
performed over initial packet amplitudes or positions. Time averaging is 
performed here as by eqs.(\ref{promT})-(\ref{taut}).}
\label{vir}
\end{figure}

We depict in fig. \ref{vir} the quantity,
\begin{equation}\label{Dvir}
  \Delta(t) \equiv\frac{\avg{\dot\phi^2}(t) - \avg{(\nabla\phi)^2}(t) -
    \avg{\phi^2}(t) - \avg{\phi_+^4}(t)}{\avg{\dot\phi^2}(t)}\; .
\end{equation}
This quantity vanishes when the virial theorem is fulfilled [see
  eq.(\ref{virial2})]. It turns out to be negative for finite times
and nonzero $a$. We see from fig. \ref{vir} that $|\Delta(t)|$
starts to decrease at times earlier than $ t_0 $. Therefore, the
model starts to virialize {\bf before} it starts to thermalize.
$|\Delta(t)|$ keeps decreasing with time and tends  to a nonzero
value which is of the order $ {\cal O} (a^2) $ for $ t \to \infty
$. This is to be expected since eq.(\ref{virial2}) only holds in
the continuum limit and receives corrections in the lattice 
$ {\cal O} (a^2) $ as shown in eq.(\ref{correca2}).

We computed the pressure as a function of time from eq.(\ref{pres}) as
\begin{equation*}
\overline{p} = \frac12 \, \left[\overline{ {\dot \phi}^2} - 
  \frac13 \,\overline{\left({\nabla \phi}\right)^2}  
  -\overline{{\phi}^2} - \frac12 \;\overline{  {\phi}_+^4} \right]\;.
\end{equation*}
Notice that we are {\bf not} using the virial theorem.

\begin{figure}[h]
  \centering
  \psfrag{pe100n64a005}[r][r]{$E/V=100,\;a=0.05$}
  \psfrag{pe10n64a005}[r][r]{$E/V=10,\;a=0.05$}
  \psfrag{pe5n32a01}[r][r]{$E/V=5,\;a=0.1$}
  \psfrag{pe01n32a01}[r][r]{$E/V=0.1,\;a=0.1$}
  \includegraphics[width=10cm,height=12.72cm,angle=270]{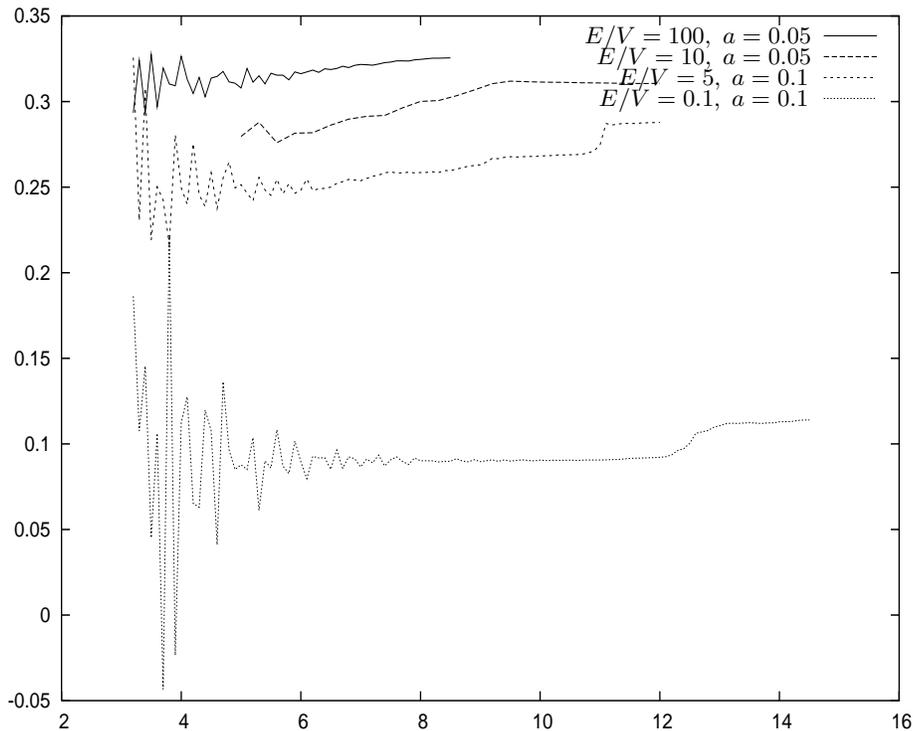}
  \caption{ The equation of state $\overline{p}/\rho$
    vs. the logarithm of the time $ t $ for $ \rho = 100 $ and $ 10 $ with
    $ a=0.05 $ and for $ \rho = 5 $ and $ 0.1 $ with and $ a = 0.1 $. We
    have in all four cases $ L = 6.4 $. No average is performed over initial 
packet amplitudes or positions. Time averaging is performed here as by 
eqs.(\ref{promT})-(\ref{taut}).}
\label{p}
\end{figure}

We depict in fig. \ref{p} $ \overline{p}/\rho $ as a function of time.
We see that 
\begin{equation*}
\frac{\overline{p}}{\rho}\lesssim \frac13 \quad,\quad t\rightarrow \infty \;,
\end{equation*}
for the whole range of $ \rho $ considered. This inequality is in agreement
with eq.(\ref{eces}). That is, fig. \ref{p} shows that the equation of state
approaches approximately the radiation--dominated equation of state unless
$ \rho $ is too small.

Notice from figs. \ref{logopf4}-\ref{logopf2} and fig. \ref{p}
that the approach to the radiation--dominated equation of state
parallels the decrease of $\overline{ {\phi}_+^4}$ since both are governed
by the UV cascade.

We see that the virial as well as the equation of state reach stationary
values despite thermalization is not achieved. As already noticed in ref.
\cite{Ngran} the equation of state follows by taking time average over the
period for the cnoidal solution. Moreover, as we show in the Appendix, the
virial theorem is exactly fulfilled by the cnoidal solution averaging over
one period.  These type of phenomena has been recently highlighted in ref.
\cite{preter}.

\subsection{Evolution of power spectra and UV cascade}\label{correti}

We now turn our attention to the study of correlation functions.

According to the Fourier transform relationship eq.~\eqref{phiphi} between
the power spectrum $\phik2$ and the equal--time correlation function
$\overline{\phi\phi}(x,t)$, there are two approaches to the numerical
evolution of such quantities (we specialize now on $\pi$ but the
discussions applies equally well to $\phi$). We extract the field $\pi(\vx,
t)$ from the lattice fields $F(\vn,s)$ and $G(\vn,s)$ [see first line in
eq.~(\ref{lattcont}) and eq. (\ref{eq:FG})], Fourier--transform it to
${\tilde\pi_\vk}(t)$ and then perform all needed averages on
$|{\tilde\pi}_\vk(t)|^2$. Or we directly compute averages of the
correlations of $F(\vn,s)$ and $G(\vn,s)$ and extract from them the
correlations $\overline{\pi\pi}(\vx,t)$.  We found that both methods yield
the same results.

Moreover, when using the approach with growing time averages as in
eq.\eqref{taut} with a unique initial condition, one realizes that
the simply time--averaged correlation
\begin{equation}\label{corfi}
  \frac1\tau \; \int_{t-\tau}^t dt' \; \pi(\vx,t') \; \pi(\vx',t') 
\end{equation}
very soon (in the logarithm of time) becomes translation
invariant, that is a function only on the distance $|\vx-\vx'|$, 
making the time consuming space average unnecessary.

\begin{figure}[htbp]
  \centering
  \psfrag{L12.8}{$L=12.8$} 
  \psfrag{a0.025}{$a=0.025$} 
  \psfrag{EoV89.5 }{$E/V=89.5$}
  \psfrag{Lambda }{$\Lambda=62.832$} 
  \psfrag{t0 }{$t=0$} 
  \psfrag{times }{$t=0,2,5,8.5,12.75$} 
  \psfrag{logtimes }{$\log t=2.54,2.89,3.22,3.54,$}
  \psfrag{row1}{$3.86,4.2,4.54,4.89,$}
  \psfrag{row2}{$5.26,5.63,6.01,6.4,$}
  \psfrag{row3}{$6.79,7.18,7.58,7.98$}
  \psfrag{t--> }{$t \longrightarrow$} 
  \psfrag{kvar}[tc][tc]{$k$} 
  \psfrag{pipower}{$4\pi k^2\pik2$} 
  \includegraphics[width=12.25cm,height=10cm]{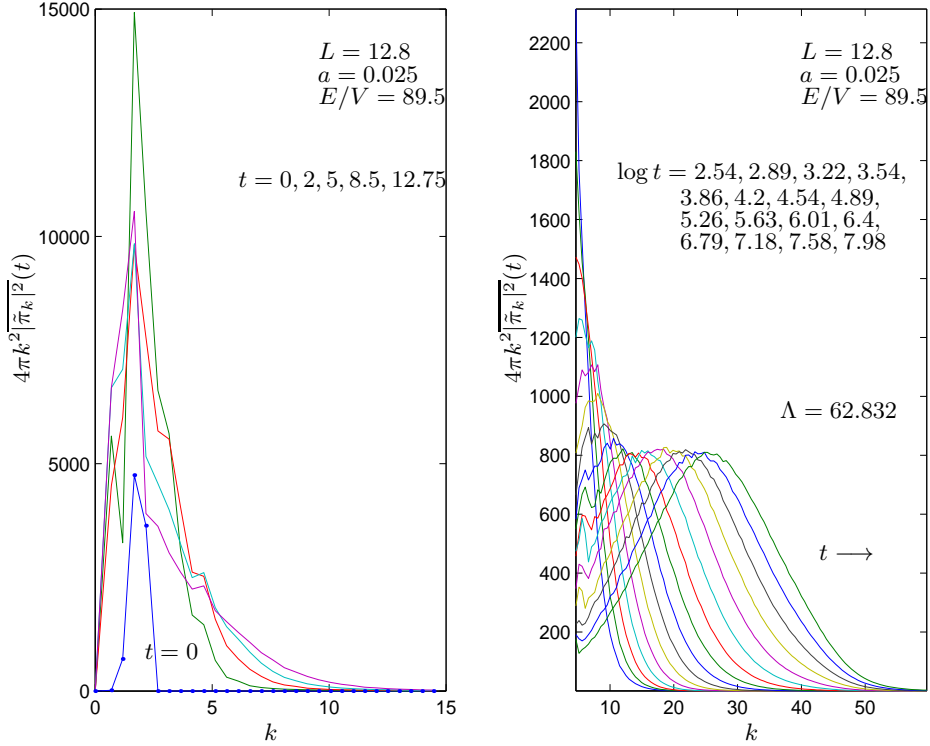}
  \caption{The power $4\pi k^2\pik2$ vs. $k=|\bds k|$ at 20 different times
    ranging, in an approximately exponential way, from $t=0$ to $t=12.75$
    (left) and from $t=12.75$ to $t=2948$ (right).  The time averaging
    interval is $\tau=2$ and the initial conditions are infrared
    random plane waves, as apparent from the IR peaks at early times. The
    front of the UV cascade arrives close to the cutoff $\Lambda$ for the 
latest times.}
  \label{pw2a025e89}
\end{figure}

\begin{figure}[htbp]
  \centering
  \psfrag{L12.8}{$L=12.8$ } 
  \psfrag{a0.1}{$a=0.1$ }
  \psfrag{a0.05}{$a=0.05$ }
  \psfrag{a0.0125}{$a=0.0125$ }
  \psfrag{a0.00625}{$a=0.00625$ }
  \psfrag{Lambda1}{$\!\!\!\!\longleftarrow\Lambda=15.708$}
  \psfrag{Lambda2}{$\Lambda=31.416$}
  \psfrag{Lambda3}{$\!\!\!\Lambda=125.664$}
  \psfrag{Lambda4}{$\Lambda=251.328$}
  \includegraphics[width=12.25cm,height=10cm]{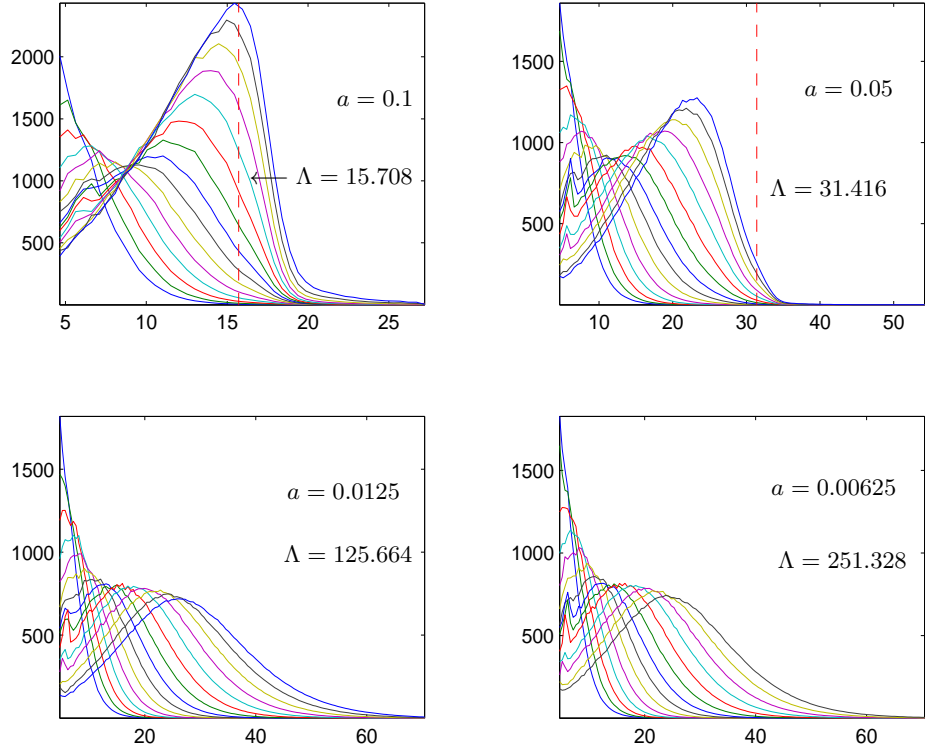}
  \caption{The power $4\pi \, k^2  \; \overline{|{\tilde\pi}_k|^2}(t)$
vs. $k$ as in the right of Fig.~\ref{pw2a025e89} but
at values of the UV cutoff $\Lambda$ scaled by $1/4$, $1/2$, $2$ and
$4$. When $\Lambda=251.328$ the last time $t=2978$ is missing.
The cascades shown in the lower panels are cut-off independent since they are 
evolving well below the cut-off $ \Lambda $.}
\label{fourpw2}
\end{figure}

Figs. \ref{pw2a025e89} and \ref{fourpw2} show the average power
$\overline{|{\tilde\pi}_k|^2}(t)$, multiplied by the spherical 
measure  $4\pi k^2$, for five values of the lattice spacing and one given
choice of all other parameters.  The region where the power is significant is
spreading towards the UV cutoff.

The chosen initial conditions eqs.(\ref{ondp})--(\ref{campa}) are
such that the power is concentrated in long wavelength modes with $k$
well below the ultraviolet cutoff $\Lambda = \pi/(2a)$. Therefore,
$\overline{|{\tilde\pi_k}|^2}(0) $
is concentrated on small $k$. During the time evolution the
non-linearity  gradually transfers energy off to higher $k-$modes
leading to the ultraviolet cascade as discussed above.

It is important to observe the effects of the finite UV cutoff $\Lambda$.
At the given value $E/V=89.5$ of the energy density, when $\Lambda=62.832$
(Fig. \ref{pw2a025e89}),
which corresponds to a lattice with $256^3$ sites, the shape as a function
of $k$ and the time evolution of the power spectrum are still markedly
distorted by the presence of the cutoff, 
although much less than for the smaller values $\Lambda=15.708$
and $\Lambda=31.416$ (Fig. \ref{fourpw2}). 
Only when doubling the cutoff from $125.664$
to $251.328$ (corresponding resp. to $512^3$ and $1024^3$ lattices) the
 effect of the cutoff in $\pik2$ does not appear significant.

That is, the cascades shown in the lower panels of fig. \ref{fourpw2} are
cut-off independent since they are evolving well below the cut-off $ \Lambda $.

\medskip

\begin{figure}[htbp]
  \centering
  \psfrag{xxxxxxL12.8}{$L=12.8$ } 
  \psfrag{xxxxxxa0.0125}{$a=0.0125$ }
  \psfrag{xxxxxxEoV89.5}{$E/V=89.5$}
  \psfrag{kzero }{$k=0$}
  \psfrag{k0.695}{$k=0.695$}
  \psfrag{k1.178}{$k=1.178$}
  \psfrag{k1.675}{$k=1.675$}
  \psfrag{k2.174}{$k=2.174$}
  \psfrag{kkkk smaller}{$k~\mathrm{smaller}$}
  \psfrag{kkkk larger}{$k~\mathrm{larger}$}
  \psfrag{logt}{$\log t$}
  \psfrag{logphik2}{$\log\phik2$}
  \includegraphics[width=12.25cm,height=10cm]{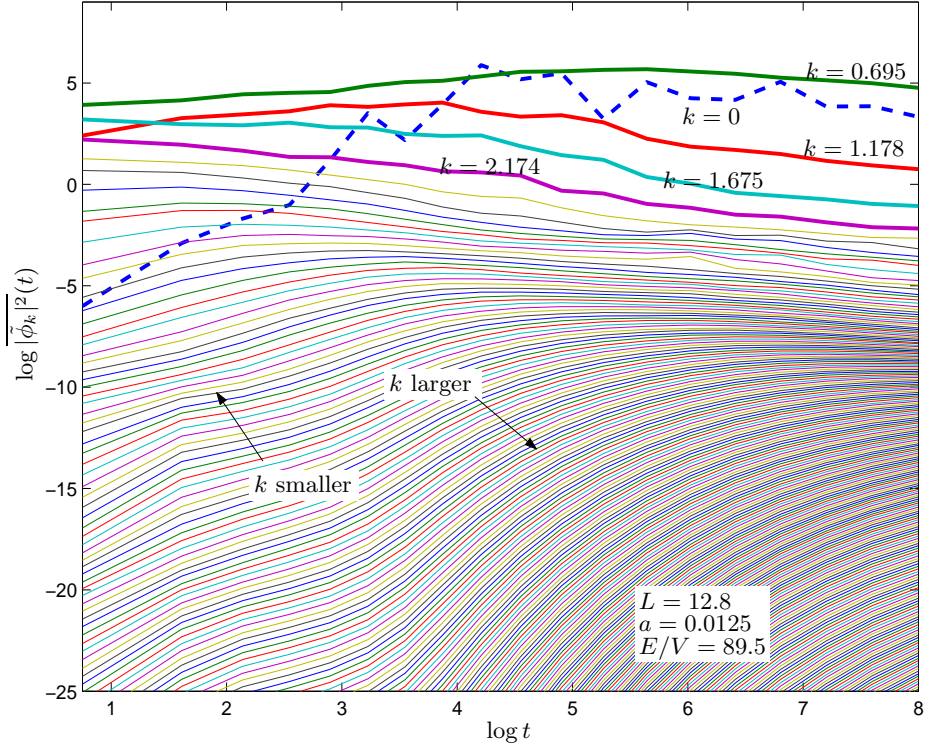}
\caption{Log-log plot of $\phik2$ vs. time. Initial conditions are as in 
fig.~\ref{pw2a025e89}. The four thicker lines correspond to the modes initially
filled (after averaging over directions). The zero-mode (dashed line) was not
filled at $t=0$. Notice the peculiar behaviour of the lowest $k$-modes compared
with the rest of the $k$-modes.}
\label{rpw1ta0125}
\end{figure}

\begin{figure}[htbp]
  \centering
  \psfrag{xxxxxxL12.8}{$L=12.8$ } 
  \psfrag{xxxxxxa0.0125}{$a=0.0125$ }
  \psfrag{xxxxxxEoV89.5 }{$E/V=89.5$}
  \psfrag{kzero }{$k=0$}
  \psfrag{k0.695}{$k=0.695$}
  \psfrag{k1.178}{$k=1.178$}
  \psfrag{k1.675}{$k=1.675$}
  \psfrag{k2.174 }{$k=2.174$}
  \psfrag{kkkk smaller}{$k~\mathrm{smaller}$}
  \psfrag{kkkk larger}{$k~\mathrm{larger}$}
  \psfrag{logt}{$\log t$}
  \psfrag{logpik2}{$\log\pik2$}
  \includegraphics[width=12.25cm,height=10cm]{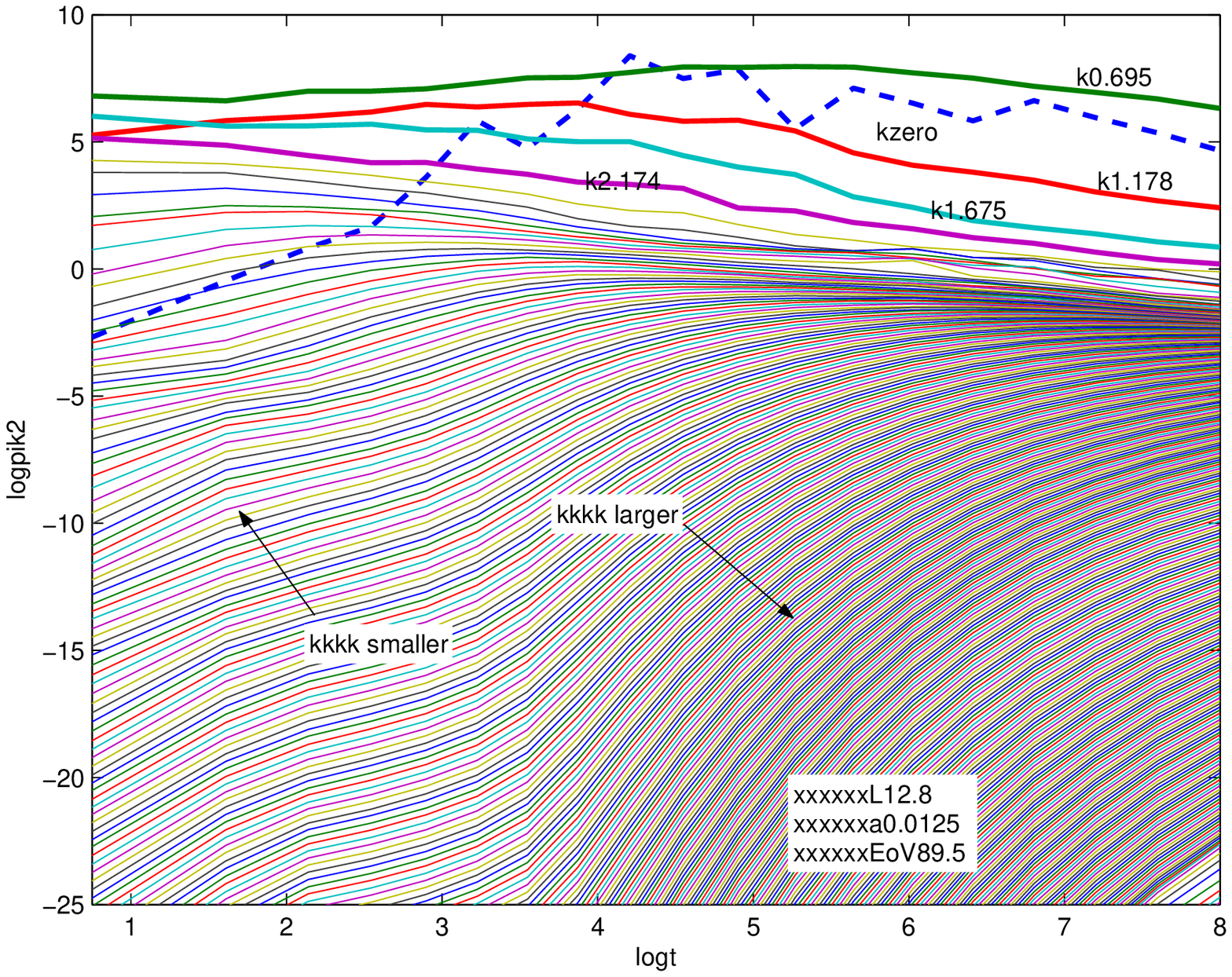}
\caption{As in fig.~\ref{rpw1ta0125}, but for $\pik2$ vs. time. Notice the 
peculiar behaviour of the lowest $k$-modes compared with the rest of the 
$k$-modes.}
\label{rpw2ta0125}
\end{figure}

Plots analogous to figs.~\ref{pw2a025e89} and \ref{fourpw2} can be drawn
for the field power $ k^2 \; \phik2$. However, since unlike $ k^2 \; \pik2 $,
$ k^2 \; \phik2 $ never grows with $k$ [at thermal equilibrium and in the 
continuum
$\phik2$ goes like $k^{-2}$, see section \ref{sec:pert}], the UV cascade
for $ k^2 \; \phik2$ is not as evident as for $ k^2 \; \pik2$.  The time
evolution of both powers can be better appreciated in figs.~\ref{rpw1ta0125}
and \ref{rpw2ta0125}. One can see a rather complicated behaviour in the
strongly interacting infrared, while the rest of the modes exhibit a rather
orderly evolution which indicate a weakly interacting dynamics.

One sees from figs.~\ref{rpw1ta0125} and~\ref{rpw2ta0125} that modes with
low $k$ {\bf decrease} in amplitude with time except for $k=0$ and the first 
two non-zero modes. Modes with larger $k$ grow in amplitude
monotonically with time showing the existence of the smooth UV cascade. 
Since total energy is conserved the
larger $k$ modes grow in amplitude at the expense of the lower $k$ modes. 
Notice that only a restricted
number of low $k$ modes have a significant initial amplitude. They feed the
growth of a large number of modes with larger $k$ which start with very
small initial amplitudes. Later, for $ t \gtrsim t_0 $ all IR modes
decrease with time.

\medskip

\noindent 
As said before, the $k=0$ mode and the first two non-zero modes in
figs.~\ref{rpw1ta0125} and \ref{rpw2ta0125} behave differently to the rest
of the modes. They start by {\bf growing} with time and they stay {\bf
  larger} than all the other modes for a while. Actually, as we shall see
below in sec. \ref{conde} , the modes with $ 0 \leq k \lesssim \sqrt{
  \overline{\phi^2}}(t) $ behave differently to the rest keeping a
significant coupling among them while modes with $ k \gtrsim \sqrt{
  \overline{\phi^2}}(t) $ exhibit weak nonlinearities for late times in the
lattice model. More precisely, these latter modes obeys the Hartree
approximation and exhibit effective equilibration much earlier than the
infrared modes.

\medskip

A single parameter that efficiently measures the UV cascade for $\pik2$ is
the average wavenumber $\bk(t)$, defined [in continuum notation and
recalling the sum-rules eq.~(\ref{sumrules})] as
\begin{equation}\label{defikb1}
\frac1{\overline{\pi^2}(t)} \int_{-\Lambda}^\Lambda \dk3  \;|k|\, \pivk2 \;.
\end{equation}
However, it will be more appropriate to exclude from the integration in
eq.(\ref{defikb1}) the infrared modes. More precisely, since for the
evolution time considered, a large portion of energy lingers on the IR
modes filled at $t=0$ (see fig.~\ref{rpw2ta0125}), one should restrict the
averaging over the modes not filled at $t=0$ and define more properly,
after averaging over discrete directions,
\begin{equation}\label{defkbar}
   \bk(t) =  \frac1{P}\int_{k_0}^\Lambda 
     \frac{k^2\,dk}{2\,\pi^2} \;k\, \pik2 \;,\quad P= 
     \int_{k_0}^\Lambda \frac{k^2\,dk}{2\,\pi^2} \,\pik2 \;.
\end{equation}
where $k_0$ is the smallest unfilled radial wavenumber larger than all
wavenumbers of the modes filled at $t=0$. This procedure applies directly
when the initial conditions are superposition of IR plane waves. In case of
superposition of localized wave packets,  $k_0$ may be defined as the value of
$k$ at which $|{\tilde\pi}_k|^2(0)$ drops below some fixed small value.
  
\begin{figure}[htbp]
  \centering
  \psfrag{L12.8 }{$L=12.8$} 
  \psfrag{EoV89.5}{$E/V=89.5$}
  \psfrag{a0000000.1logkbeq2.714}
  {$~~a=0.1,\;\log\bk_{\mathrm{eq}}=2.714$}
  \psfrag{a0000000.05logkbeq3.407}
  {$~~a=0.05,\;\log\bk_{\mathrm{eq}}=3.407$}
  \psfrag{a0000000.025logkbeq4.100}
  {$~~a=0.025,\;\log\bk_{\mathrm{eq}}=4.100$}
  \psfrag{a0000000.0125logkbeq4.793}
  {$~~a=0.0125,\;\log\bk_{\mathrm{eq}}=4.793$}
  \psfrag{a0000000.00625logkbeq5.586}
  {$~~a=0.00625,\;\log\bk_{\mathrm{eq}}=5.486$}
  \psfrag{logt}[tc][tc]{$\log t$} 
  \psfrag{logkbar}{$\log \bar k$}
  \includegraphics[width=11.66cm,height=10cm]{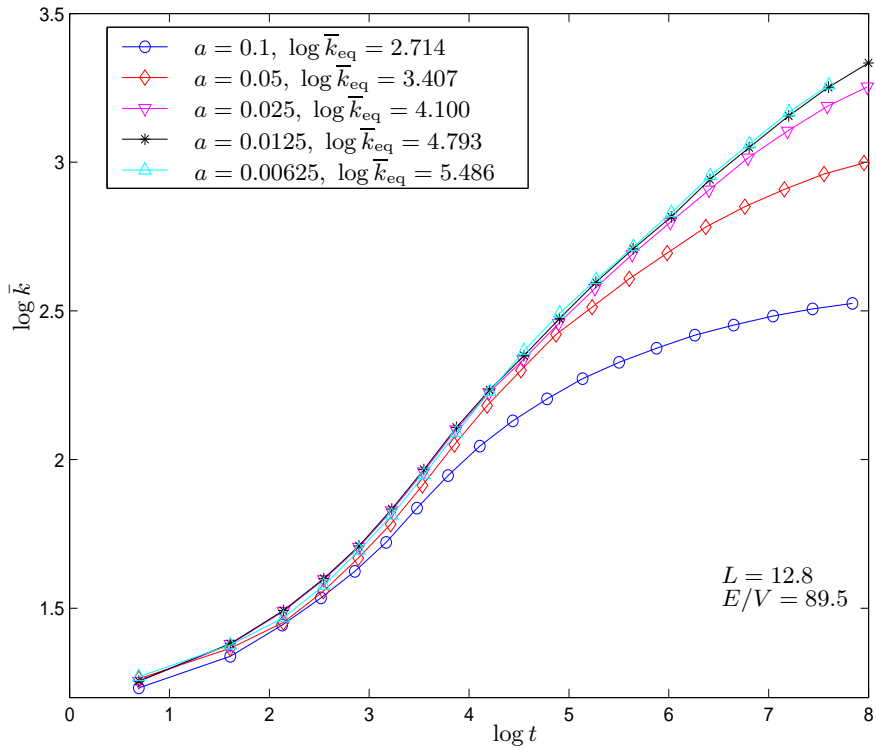}
  \caption{Log-log plot of the time evolution of average wavenumber
    $\bk(t)$, for five values of the lattice spacing $2a$, with 
    initial conditions as in figs. \ref{pw2a025e89} and \ref{fourpw2}.}
\label{kbar1red}
\end{figure}

We plot $\bk(t)$ in fig.~\ref{kbar1red} for the same context of
figs.~\ref{pw2a025e89} and \ref{fourpw2}. The cutoff effects at larger
values of $a$ and the saturation in the continuum limit $a\to0$ are quite
evident. Still, in spite of the fact that the UV cascade has time to fully
develop, as evident from figs. \ref{pw2a025e89} and \ref{fourpw2},
fig.~\ref{kbar1red} shows that the growth of $\bk(t)$ with time is not a
simple power, not even for the latest times considered. However, 
\begin{equation}\label{expk} 
  \bk(t) \sim k_0 \; t^{1/3}
\end{equation}
provides a rough estimate. 

\begin{figure}[htbp]
  \centering
  \psfrag{kvar}{$k$}
  \psfrag{logradpw2}{$\log[\pik2]$}
  \psfrag{L12.8}{$L=12.8$ } 
  \psfrag{a0.0125}{$a=0.0125$ }
  \psfrag{EoV89.5}{$E/V=89.5$}
  \psfrag{time=}{$t=$}
  \psfrag{time}{$t$}
  \psfrag{Lambda}{$\Lambda$}
  \includegraphics[width=11.66cm,height=10cm]{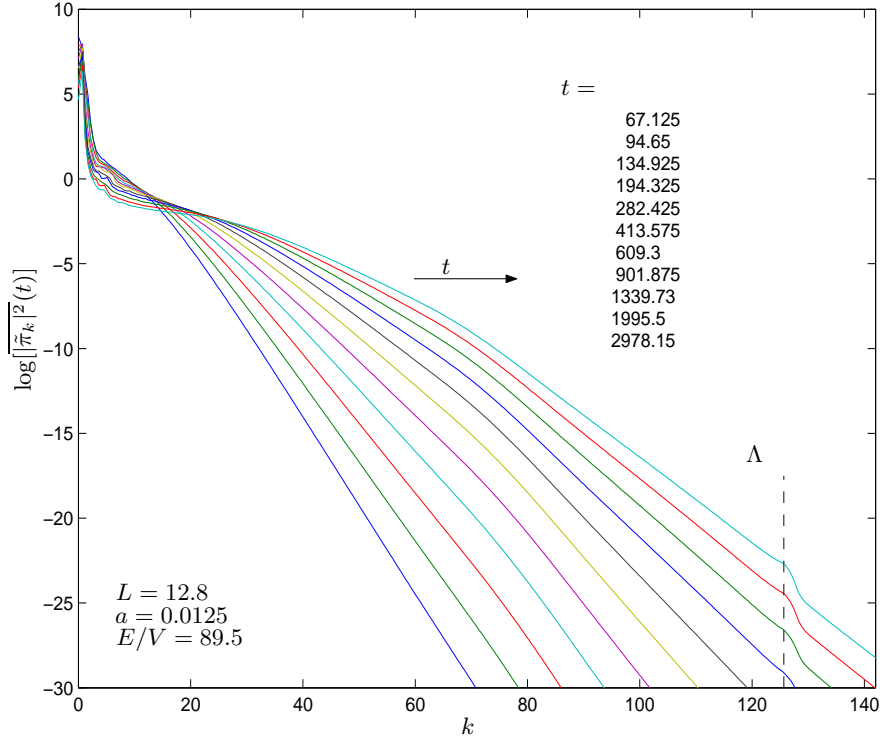}
  \caption{$ \log[\pik2] $ vs. $ k $ for several values of time. All relevant 
parameters are at the indicated values. Initial conditions were random plane 
waves. $ \pik2 $ starts decreasing with $k$ as $ \sim k^{-\alpha} $ and later
it dies exponentially for $ \Lambda > k > \bk(t) $.}
  \label{tail125}
\end{figure}

\begin{figure}[htbp]
  \centering
  \psfrag{logk}{$\log k$}
  \psfrag{logrpw2}{$\log[4\pi k^2\pik2]$}
  \psfrag{L12.8}{$L=12.8$ } 
  \psfrag{a0.025}{$a=0.025$ }
  \psfrag{EoV89.5}{$E/V=89.5$}
  \psfrag{EzoE0.98}{} 
  \psfrag{time        slope}{$~~time~~~~~~~slope$ }
  \psfrag{t812        s0.92}{$~~~812~~~~~~~~0.92$ }
  \psfrag{t1446      s1.18}{$~~1446~~~~~~~1.18$ }
  \psfrag{t2585      s1.31}{$~~2585~~~~~~~1.31$ }
  \psfrag{t4635      s1.49}{$~~4635~~~~~~~1.49$ }
  \psfrag{t8321      s1.69}{$~~8321~~~~~~~1.69$ }
  \psfrag{t14955    s1.81}{$~14955~~~~~~~1.81$ } 
  \psfrag{t26895    s1.78}{$~26895~~~~~~~1.78$ } 
  \psfrag{t48385    s1.83  }{$~48385~~~~~~~1.83$ } 
  \includegraphics[width=11.66cm,height=10cm]{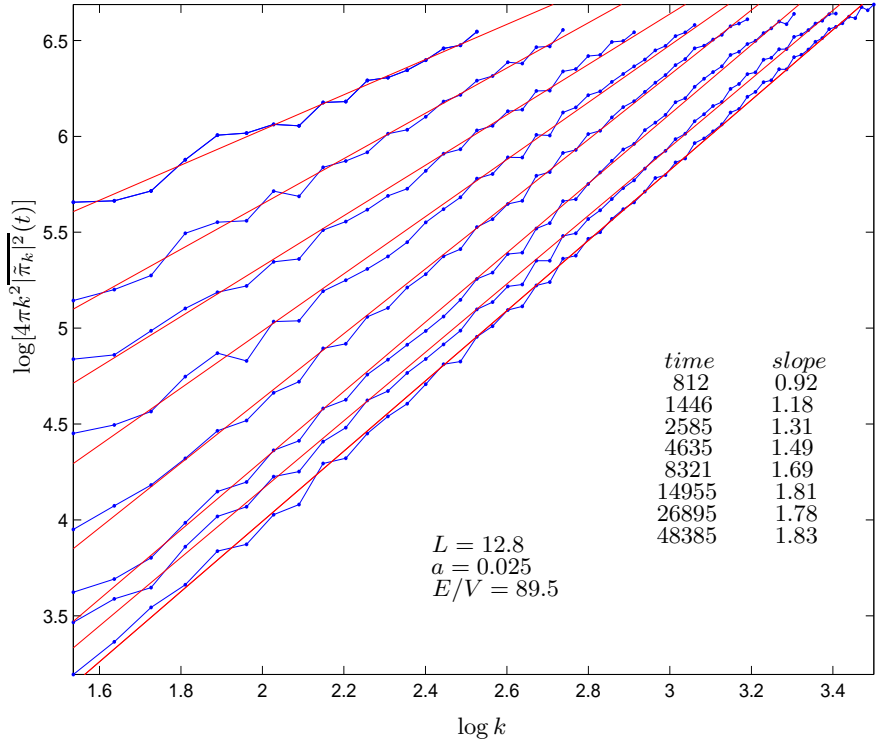}
  \caption{$ \log[4 \, \pi \; k^2\pik2] $ vs. $\log k$ in a fixed 
    wavenumber window for several values of time. All relevant parameters
    are at the indicated values.}
  \label{slopesa025co1}
\end{figure}

Let us now turn our attention to the behavior in $k$ of $ \pik2 $ during the
UV cascade. From all our data, it is clear that $\pik2$ dies exponentially
fast for $ k>\bk(t) $ as long as $\bk(t)\ll\Lambda$ [see fig.~\ref{tail125}
as an example].  This exponential behavior is clearly time-dependent. For
$k$ not too small nor too close to $\bk(t)$, $\pik2$ exhibits a decreasing
power-like behaviour, as can be seen for instance in
fig.~\ref{slopesa025co1}, where the log-log plot of $ 4 \, \pi \;  
k^2  \; \pik2) $ vs. $ k $ is shown. That is, $ \pik2 \sim k^{-\alpha} $ at 
different times, where
the  values of $-\alpha$ are obtained subtracting $2$ from the numbers 
indicated in fig.~\ref{slopesa025co1}. $\alpha$ decreases 
monotonically from $ 1.12 $ at $t=812$ to $ 0.17 $ at $t=48385$, when the
forefront of the cascade is reaching the cutoff $ \Lambda $.

\begin{figure}[htbp]
  \centering
  \psfrag{L12.8}{$L=12.8$} 
  \psfrag{a0.025}{$a=0.025$} 
  \psfrag{a0.0125}{$a=0.0125$} 
  \psfrag{EoV89.5}{$E/V=89.5$}
  \psfrag{EzoE0.98}{$E_0/E=0.98$ }
  \psfrag{Lambda1}{$\Lambda=62.832$} 
  \psfrag{Lambda2}{$\Lambda=125.664$} 
  \psfrag{time }{$t\longrightarrow$} 
  \psfrag{t= }{$t=$} 
  \psfrag{kvar}[tc][tc]{$k$} 
  \psfrag{k2pik2}{$4\pi k^2\pik2$} 
  \includegraphics[width=12.25cm,height=10cm]{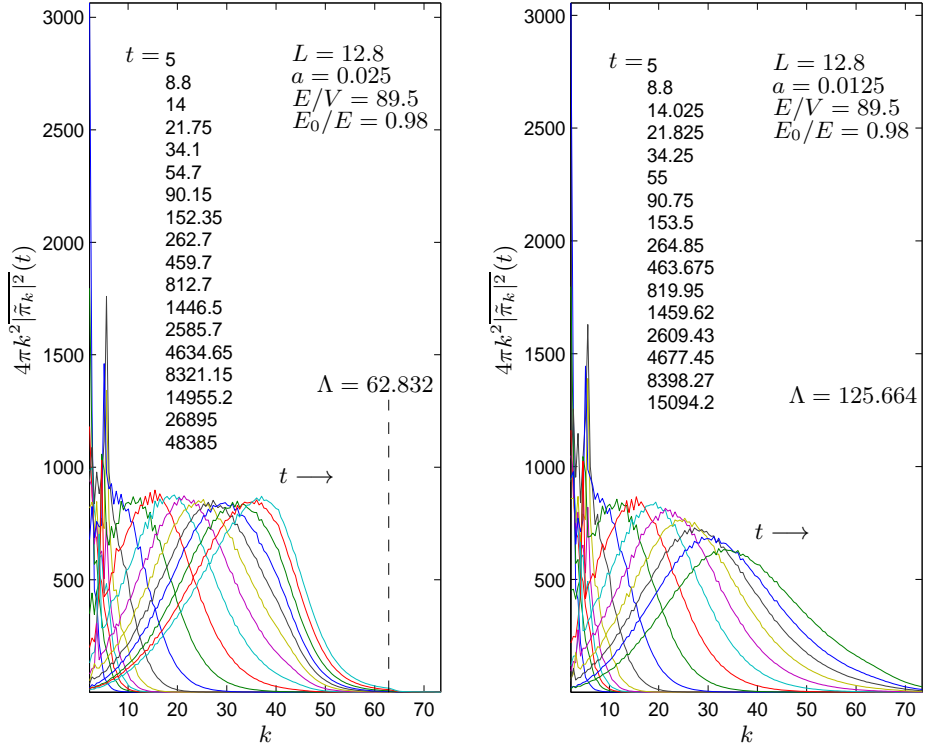}
  \caption{The power $ 4\pi \;  k^2\pik2 $ vs. $ k=|\bds k| $ at several 
times, $ E/V=89.5 $ and two lattice spacings. The initial conditions are 
infrared random plane waves with a zero-mode condensate dominating the energy.
Notice the peaks due to parametric resonance.}
  \label{pikco1}
\end{figure}

\begin{figure}[htbp]
  \centering
  \psfrag{L12.8}{$L=12.8$ } 
  \psfrag{a0.025 }{$a=0.025$ }
  \psfrag{EoV89.5}{$E/V=89.5$}
  \psfrag{EzoE0.98 }{$E_0/E=0.98$ }
  \psfrag{kzero }{$k=0$}
  \psfrag{logt}{$\log t$}
  \psfrag{logphik2}{$\log\phik2$}
  \includegraphics[width=12.25cm,height=10cm]{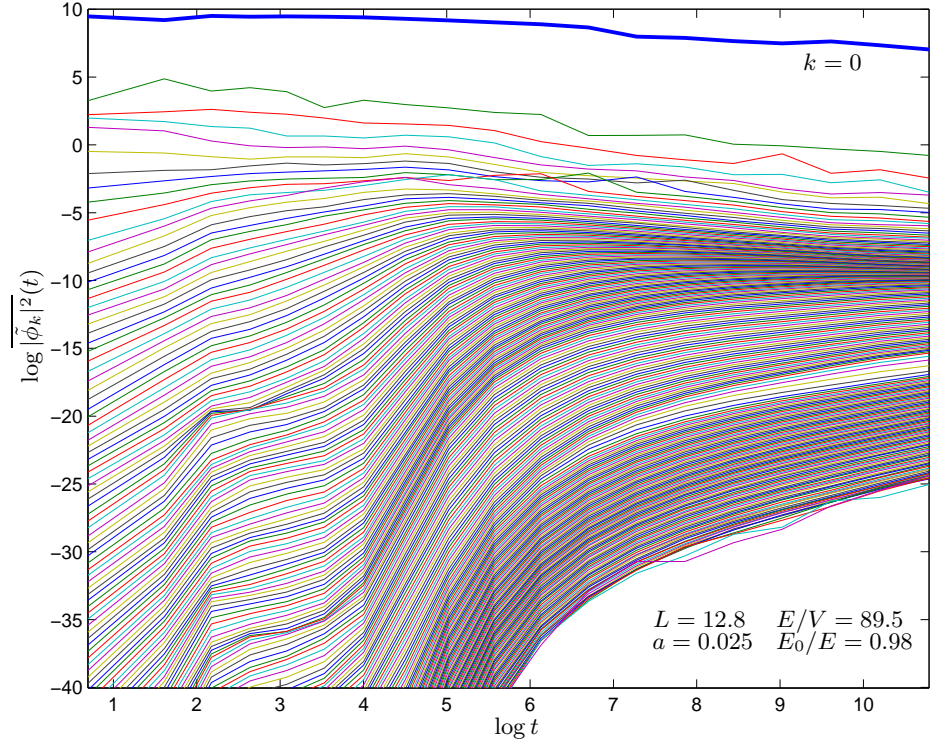}
  \caption{Log-log plot of $ \phik2 $ vs. time. Initial
    conditions are as in fig.~\ref{pikco1}. Only the zero-mode (thik line)
    was macroscopically filled at $t=0$.}
\label{rpw1ta025co1}
\end{figure}

\medskip
In figs.~\ref{pikco1} and \ref{rpw1ta025co1} we give some graphic examples
of the UV cascade in the presence of an initial zero-mode condensate [$
\phi_0 $ in Eqs.(\ref{ondp}) and (\ref{campa})]. We notice the evidence for
the parametric resonance, whose location in $k-$space well agrees with the
analytic prediction from the solution of the Lam\'e equation for the
linearized model [see eq.(\ref{banda})]. It is quite evident however, that
such parametric resonance plays practically no role in the UV cascade, as
already noticed in ref.\cite{mic}.  We also stress once more the importance
of UV cutoff effects when $a$ is doubled from $a=0.0125$ to $a=0.025$
(corresponding to a reduction of the lattice from $512^3$ to $256^3$
sites). This is particularly relevant since the zero-mode (and few lower
$k$ modes) remain quite large for all times considered (see
fig.~\ref{rpw1ta025co1}): according to the scenario of ref.\cite{mic}, the
regime should then remain that of driven turbulence until the lattice
effect become dominant; thus, for the energy density of figs.~\ref{pikco1},
\ref{rpw1ta025co1}, no cutoff-independent regime of free turbulence is
observable in our evolution when $a=0.025$ and $L=12.8$.  (Notice that our
lattices reach a size $4^3$ times larger than those in ref.\cite{mic}).

\medskip

It would be indeed very interesting to find this universal UV cascade in 
quantum theory for large occupation numbers and small couplings. 
Most of the works consider other regimes \cite{ber,berR,gre}. 
In ref. \cite{ber2} the small coupling regime in quantum theory is investigated
including the leading $1/N$ corrections. However, thermalization is not 
reached in ref. \cite{ber2} since the times considered are not long enough.
The classical evolution is compared with the quantum evolution to first 
order in $1/N$ for the $ \phi^4 $ model in $3+1$ dimensions in 
ref. \cite{ast}. It is stated there that the classical evolution is a good 
approximation to the quantum evolution for non-asymptotic times therefore
supporting the relevance for quantum field theory of the 
classical dynamics studied here.

\subsection{Thermalization of the power spectra}\label{fulltherm}

We present here the results of a long simulation yielding to a very good
approximation thermalized powers $ \pik2 $ and $ \phik2 $, on a lattice
cube $ C_N $ with $ N=100 $, $ a=0.064 $ and $ L=12.8 $. The energy density
is $ E/V=569.5 $, still relatively small compared to $ \Lambda^3\simeq
14785 $.  The initial conditions are random plane waves as in
eq.~(\ref{ondp}), with the initial powers of $ \phi $ and $ \pi $ exactly
null for $ k>3.436 $. The rather large time averaging interval $ \tau=200 $
is still negligible compared to the length $ t=915494 $ of the evolution.
To further reduce fluctuations, in this case we also performed an average
over $ 32 $ different random initial mode amplitudes and phases.

\begin{figure}[htbp]
  \centering
  \psfrag{L12.8}{$L=12.8$} 
  \psfrag{a0.064}{$a=0.064$} 
  \psfrag{E569.5 }{$E/V=569.5$} 
  \psfrag{logt}[tc][tc]{$\log t$} 
  \psfrag{logkbar }{$\log \bk$}
  \psfrag{logphi2 }{$\log\overline{\phi^2}$}
  \psfrag{logphi4}{$\log\overline{\phi_+^4}$}
  \psfrag{logdphi2 }{$\log\overline{\partial\phi^2}$}
  \psfrag{logpi2 }{$\log\overline{\pi^2}$}
  \psfrag{equilibrium }{equipartition}
  \includegraphics[width=12.72cm,height=10cm]{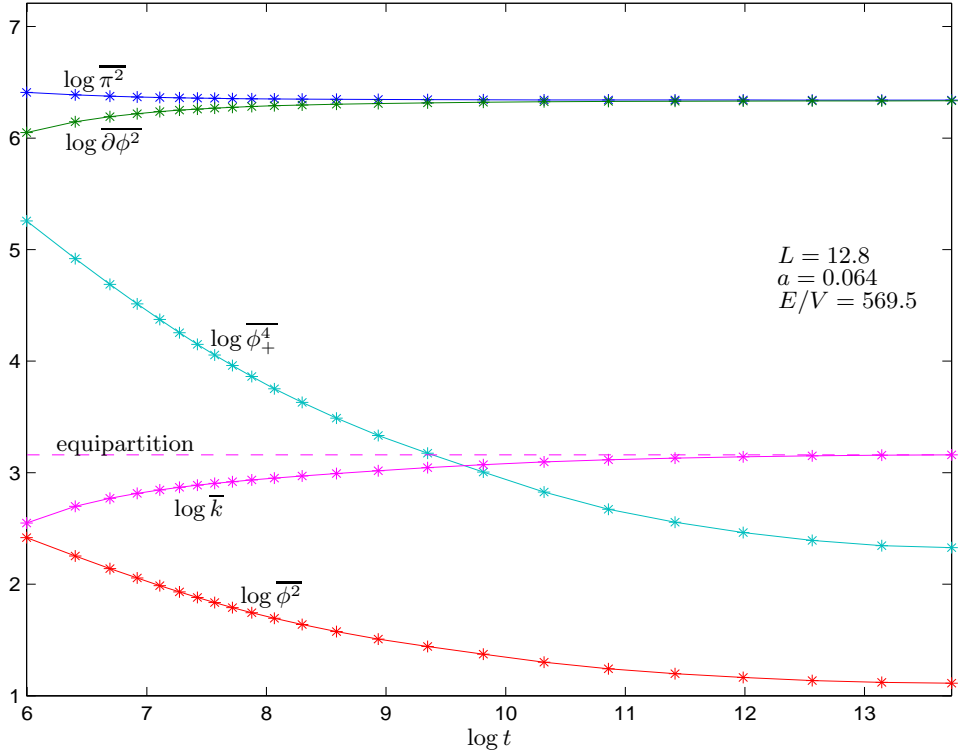}
  \caption{Log-log plot of the time evolution of space-averaged local
    observables and of the average wavenumber. The time averaging interval
    is $ \tau=200 $ and the initial conditions are infrared random plane
    waves. This plot is analogous to fig. \ref{logopf3} but with the time
    evolution lasting for significatively later times.}
  \label{opa064}
\end{figure}

\begin{figure}[htbp]
  \centering
  \psfrag{L12.8}{$L=12.8$} 
  \psfrag{a0.064}{$a=0.064$} 
  \psfrag{E569.5 }{$\!\!E/V=569.5$} 
  \psfrag{kvar}[tc][tc]{$k$} 
  \psfrag{r2radpw2}{$4\pi k^2\pik2$} 
  \psfrag{logt }{$\log t=6,6.69,7.1,7.42,7.71,8.07,$} 
  \psfrag{row2 }{$\quad8.58,9.35,10.32,11.41,12.56,13.73$} 
  \psfrag{time}[bl][bl]{$t$} 
  \psfrag{time }[tr][tr]{$t$} 
  \psfrag{Lambda }{$\Lambda$} 
  \includegraphics[width=14cm,height=10cm]{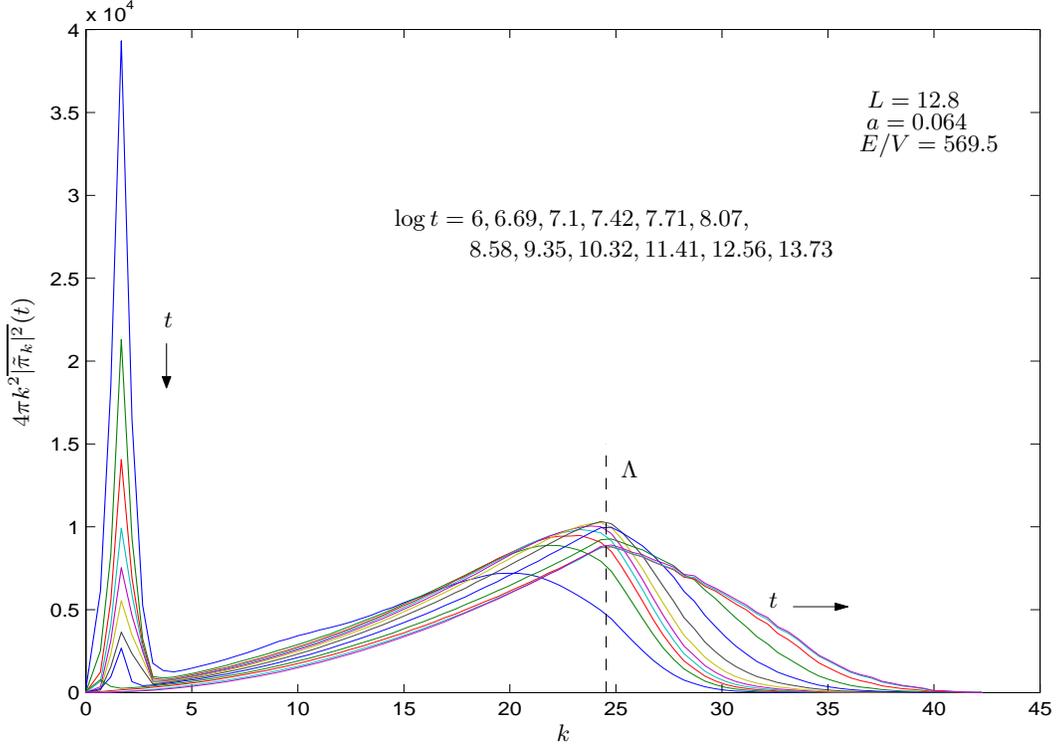}
  \caption{$4\pi k^2\pik2$ vs. $k$ at the times indicated. All parameters
    are as in Fig.~\ref{opa064}.}
  \label{rpw2a064}
\end{figure}

\begin{figure}[htbp]
  \centering
  \psfrag{logradpw2}{$\log\pik2$} 
  \psfrag{logradpw1}{$\log\phik2$} 
  \psfrag{logk}{$\log k$} 
  \psfrag{t401 }{$t=401$} 
  \psfrag{t915494 }{$t=915494$} 
  \psfrag{logLambda }[l]{$\log\Lambda$} 
  \includegraphics[width=12.72cm,height=10cm]{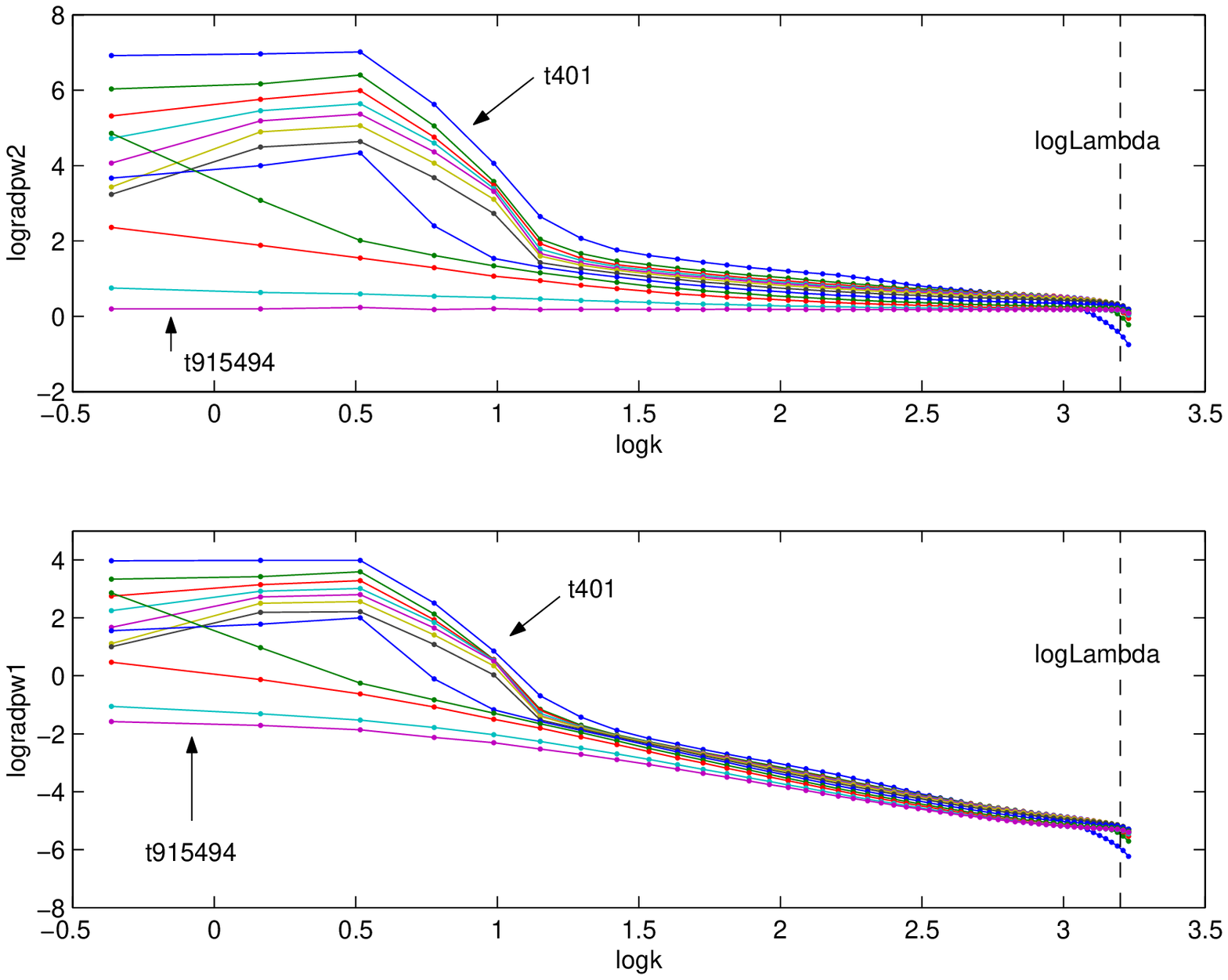}
  \caption{Log-log plot of $\pik2$ and $\phik2$ vs. $k$. Data points are
    explicitly indicated. All parameters are as in Fig.~\ref{opa064}.
Comparison with fig. \ref{opa064} shows that modes with $k^2 < 
\overline{\phi^2}(t)$ thermalize much slower than those with 
$ k^2 > \overline{\phi^2}(t)$. }
  \label{logrpwa064}
\end{figure}

\begin{figure}[htbp]
  \centering 
  \psfrag{logradpw2}{$\log\pik2$}
  \psfrag{logradpw1}{$\log\phik2$} 
  \psfrag{logt}[bl]{$\log t$}
  \psfrag{kok }[bl]{$k$}
  \includegraphics[width=12.72cm,height=10cm]{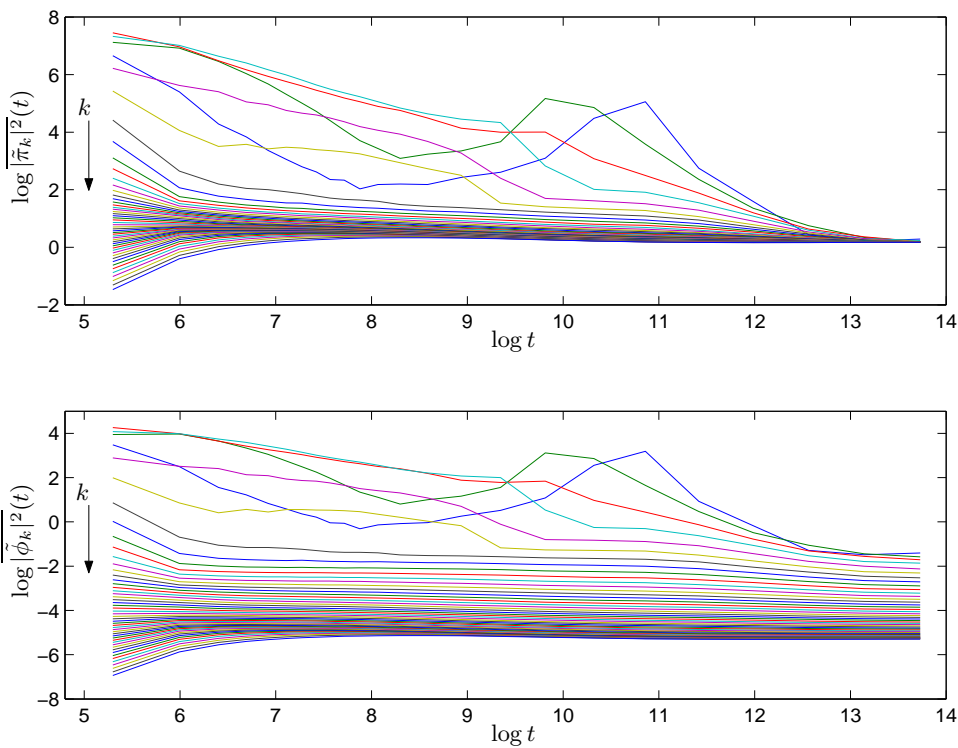}
  \caption{Log-log plot $ \pik2 $ and $ \phik2 $ vs. $ \ln t $. Only modes
    up to $ k\le\Lambda $ are shown. All parameters are as in
    Fig.~\ref{opa064}.  We see that modes with $ k^2 < \overline{\phi^2}(t)
    $ thermalize here the last.  [Compare with fig. \ref{opa064} to see $
    \log \overline{\phi^2}(t) $].}
  \label{logtrpwa064} 
\end{figure}

\begin{figure}[htbp]
  \centering 
  \psfrag{L12.8}{$L=12.8$} 
  \psfrag{a0.064}{$a=0.064$} 
  \psfrag{EoV569.5}{$E/V=569.5$} 
  \psfrag{logradpw2}{$\log\pik2$}
  \psfrag{zkt}{$Z_k(t)$} 
  \psfrag{time }{$t=$}
  \psfrag{kvar}[tc][tc]{$k$}
  \psfrag{times}{$t=$ as above}
  \includegraphics[width=12.72cm,height=10cm]{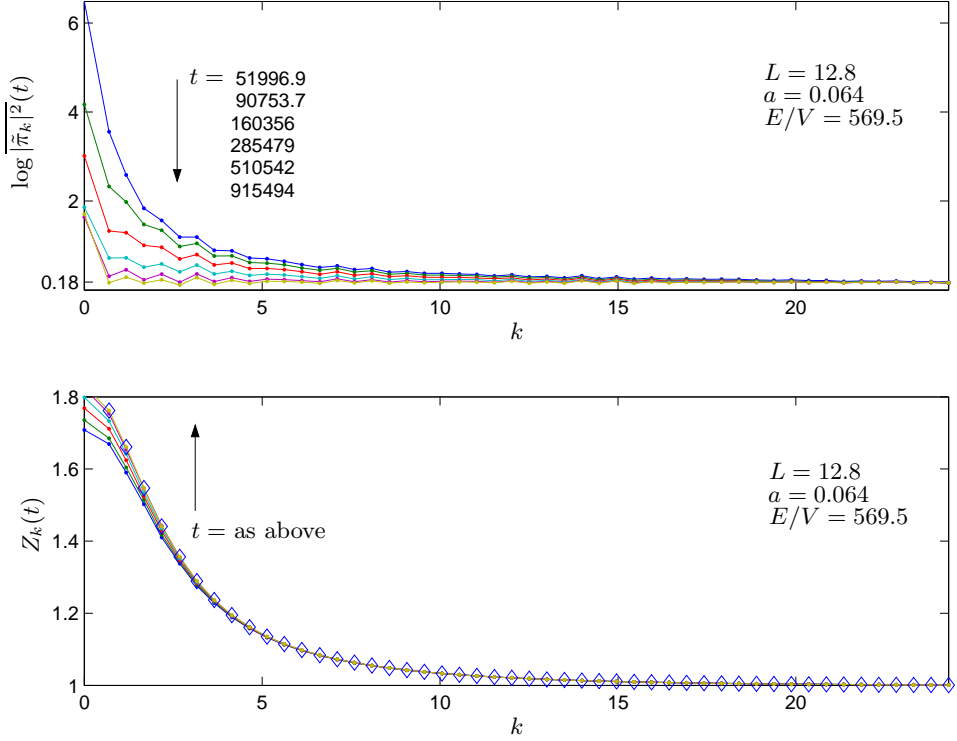}
  \caption{Comparison of $ \log\pik2 $ vs. $ Z_k(t) $. Diamonds in the
    bottom represent equilibrium values.  All parameters are as in
    Fig.~\ref{opa064}.  We see that $ Z_k(t) $ equilibrates much earlier
    than $ \pik2 $.}
  \label{zkta064} 
\end{figure}

In Fig. \ref{opa064} we plot the basic one--point observables and the
average wavenumber $ \bk(t) $ as function of time. The system clearly reaches
a time-independent stage for $ \ln t \gtrsim 12.5 $. In Fig. \ref{rpw2a064} 
we plot the power spectra $ \pik2 $, times the spherical volume 
$ 4\pi \,k^2 $, vs. the wavenumber $ k $ at
several times: the UV cascade up to the cutoff is evident. We plot in
log--log scale both $ \pik2 $ and $ \phik2 $ vs. $ k $ in Fig. \ref{logrpwa064}
and vs. time in Fig. \ref{logtrpwa064}. Again, the system shows an evident
limiting behaviour: the large initial peaks in the IR modes have almost
completely disappeared for $\log t\simeq 12.5$ and at the latest times
$ \log t\simeq 14 $ the power spectrum $ \pik2 $ is flat, except for
fluctuations and a small remnant of the infrared peaks exactly at $ k=0 $
(see top of fig.~\ref{zkta064}). From the height of $\pik2$ in the plateau
we read the temperature $ T= 1.2 $, very close to $ E/N^3=1.19432\ldots $. The
difference is almost completely accounted for by the last value
of $ \overline{\phi_+^4}= 10.264\ldots $ [see eq.~(\ref{TtoE})].

To ascertain the thermalization of $ \phik2 $ is very convenient to
consider the time-dependent analog $ Z_\vk(t) $ of the equilibrium
wave-function renormalization $ Z_\vk $ [see eq.(\ref{eq:Zparamlatt})]
replacing there the equilibrium quantities by their time-dependent
out-of-equilibrium counterparts, that is
\begin{equation}\label{eq:Zkt}
  \phivk2 = \frac{a^2\,Z_\vk(t)\, \pivk2}
  {1 - [B(\overline{\phi^2}(t))\,C(\vk a)]^2}
\end{equation}
This relation implies that the mode $k$ has indeed virialized in the 
interacting theory.

In the bottom of fig.~\ref{zkta064} we plot the (average over discrete
directions of) the equilibrium $ Z_k $ and $ Z_k(t) $ for several late
times.  It is evident that $Z_k(t)$ approaches its equilibrium value $ Z_k
$ earlier and much more closely than $ \pik2 $. In other words, when both $
\pik2 $ and $ \phik2 $ are still out of equilibrium over a wide range of
low wavenumbers, $ Z_k(t) $ is practically already at equilibrium except
for a very narrow range of very small wavenumbers. One can see that this
range is approximately given by $ k^2 < \overline{\phi^2}(t) $.  These
infrared modes are the last to effectively thermalize.  [Recall figs.
\ref{rpw1ta0125}-\ref{rpw2ta0125} and the discussion in sec. \ref{correti}
about the peculiar behaviour of these low $ k$-modes].

\medskip
In conclusion, the field {\bf does} thermalize completely in the lattice
with a time scale of the order $ 10^6 $. For such times $ \pivk2 $ is $k$
independent as seen in fig. \ref{logrpwa064}.  Contrary to the $1+1$
dimensional case\cite{uno}, the infrared modes with $ k^2 <
\overline{\phi^2}(t) $ thermalize here the last.

\subsection{Equilibration in the cascade}\label{eqcasc}

The effectiveness of the time-dependent wavefunction renormalization
$ Z_\vk(t) $ in the study of the equilibration process for very late times,
suggests to use it also at earlier times, during the universal UV cascade.
As a matter of fact, we may regard eq.~(\ref{eq:Zkt}) just as a {\em
  definition} of $ Z_\vk(t) $, which is possible for any time but might very
well turn out to yield a function of $ \vk $ quite different from its
equilibrium counterpart $ Z_\vk $. What we find, however, is something
very close to equilibrium already at times of order $ t_0 \simeq 500 $,
which are those proper of the universal cascade. To have a
cutoff-independent window wide enough, we consider here a lattice spacing
$ a=0.0125 $, much smaller that that of the the previous section, and a
smaller energy density $ E/V=89.5 $. Notice that this would correspond to an 
equilibrium temperature of order $ E/N = 0.001398\ldots $, almost negligible
as compared to that of the previous section.

\begin{figure}[htbp]
  \centering 
  \psfrag{L12.8}{$L=12.8$} 
  \psfrag{a0.0125}{$a=0.0125$} 
  \psfrag{EoV89.5 }{$E/V=89.5$} 
  \psfrag{kvar}[tc][tc]{$k$}
  \psfrag{kphivar}{$k/[\overline{\phi^2}(t)]^{1/2}$}
  \psfrag{zkvar}{$Z_k(t)$} 
  \psfrag{tttttttt135}{$t=135$} 
  \psfrag{tttttttt194}{$t=194$} 
  \psfrag{tttttttt282}{$t=282$} 
  \psfrag{tttttttt414}{$t=414$} 
  \psfrag{tttttttt609}{$t=609$} 
  \psfrag{tttttttt902}{$t=902$} 
  \psfrag{tttttttt1340}{$t=1340$} 
  \psfrag{tttttttt1995}{$t=1995$} 
  \psfrag{tttttttt2978}{$t=2978$} 
  \includegraphics[width=12.72cm,height=10cm]{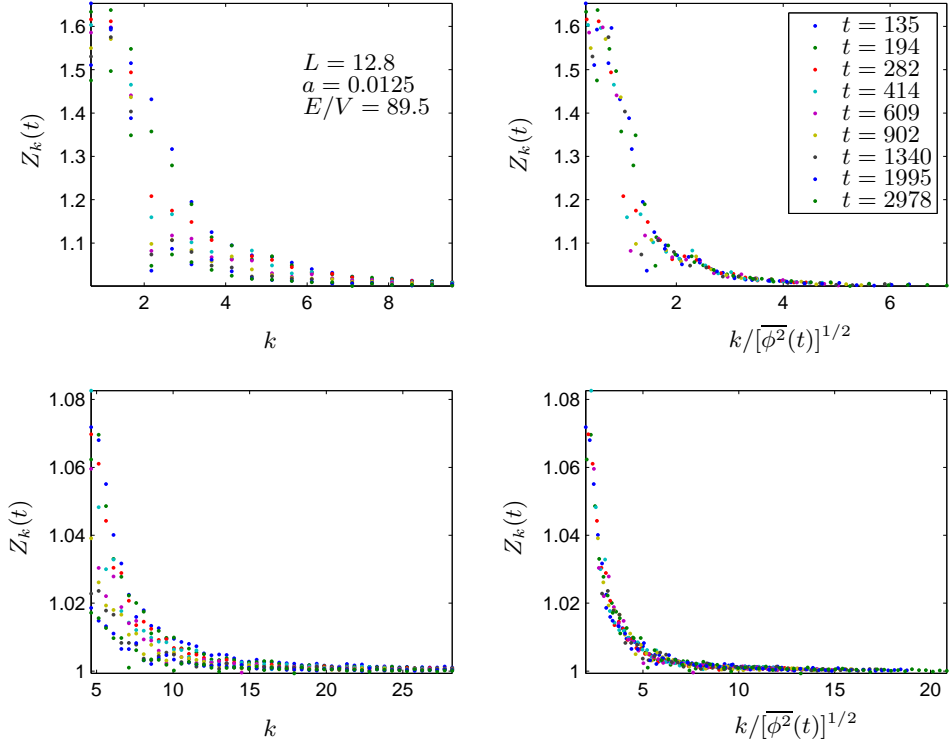}
  \caption{Scaling properties of the $k-$dependent wavefunction
    renormalization for two low--lying ranges of the radial wavenumber $k$.}
  \label{zkt} 
\end{figure}

In Fig.~\ref{zkt}, on the left column, we plot $ Z_k(t) $, the direction
averaged $Z _\vk(t) $, for several times ranging from $ t = 135$ to 
$ t = 2978 $
and for $ 0.695\ldots \le k \le 9.559\ldots $ in the top and $4.644\ldots \le
k \le 28.218\ldots$ in the bottom. On the left column, we plot $ Z_k(t) $ vs.
the scaled wavenumber $ u=k/\sqrt{\overline{\phi^2}(t)} $. The good collapse
in the bottom left shows that $ Z_k(t) $ is a almost function only of $u$ in
the range $ 3.430\ldots\le u\le 20.843\ldots $. The collapse for $ 0.513\le u
\lesssim 2.5 $ is definitely worse, in agreement with the slower
equilibration of the infrared modes. Most remarkably, by direct comparison
with fig.~\ref{zkta064}, we observe that $Z_k(t)$ is very similar,
qualitatively and quantitatively, to its equilibrium counterpart at a
temperature roughtly a thousand times larger than the temperature the
system will eventually reach when $t\to\infty$. One could say that
equilibration in the bulk of cascade has taken place with an effective
temperature of order one. The precise time dependence of this effective
temperature as well as a check on its universality (that is to say unicity
when the studied observables are changed) require a much more elaborate
analysis which is beyond the scope of the present work. Here we only stress
that, as a function of $ u , \; Z_k(t) $ resembles closely its equilibrium
counterpart and in particular it decreases to unity rather fast as $ u $
grows, that is when $ k^2 \gg \overline{\phi^2}(t) $. 
 
\subsection{Effective frequency and particle number}\label{frecefe}

The motion of each Fourier mode ${\tilde\phi}_\vk$ of the field is
characterized by fast oscillations into an envelope which varies relatively
slowly in time. In the linear approximation, the scale of fast oscillation is 
fixed by the  frequency of the free massive dispersion relation, that is 
$\omega_k = k^2+1$ in the continuum. Of course on our staggered lattice one 
should consider the lattice dispersion relation, 
which differs from the continuum 
form by power corrections in $ a^2\vk^2 $. In Appendix~\ref{app:B} we show that
the exact free dispersion relation on our lattice reads
\begin{equation}\label{freedisprule}
  \cos(\omega_\vk a) = \frac1{1+\tfrac12 a^2}\prod_{j=1}^3 \cos k_ja
\end{equation}
which correctly reproduces $k^2+1$ in the limit $a\to0$ at fixed $\vk$.

On the continuum the modes ${\tilde\phi}_\vk$ obey the Fourier transform
of the exact field equation (\ref{eqnofmot}), namely
\begin{equation}\label{eq:kfe}
  \frac{d^2}{dt^2}\, {\tilde\phi}_\vk = -\omega_k \, {\tilde\phi}_\vk
  - ({\tilde\phi^3})_\vk
\end{equation}
where $({\tilde\phi^3})_\vk$ is the Fourier transform of $\phi^3(x)$.

\begin{figure}[htbp]
  \centering
  \psfrag{Spatial average of phi(x,y,x,t) vs. t. E/V=100, L=6.4, a=0.1}{$
Spatial ~average ~of ~ \phi(x,y,x,t) ~vs.~ t, ~ E/V=100, ~ L=6.4, ~ a=0.1$}
  \includegraphics[width=10cm,height=12.72cm,angle=270]{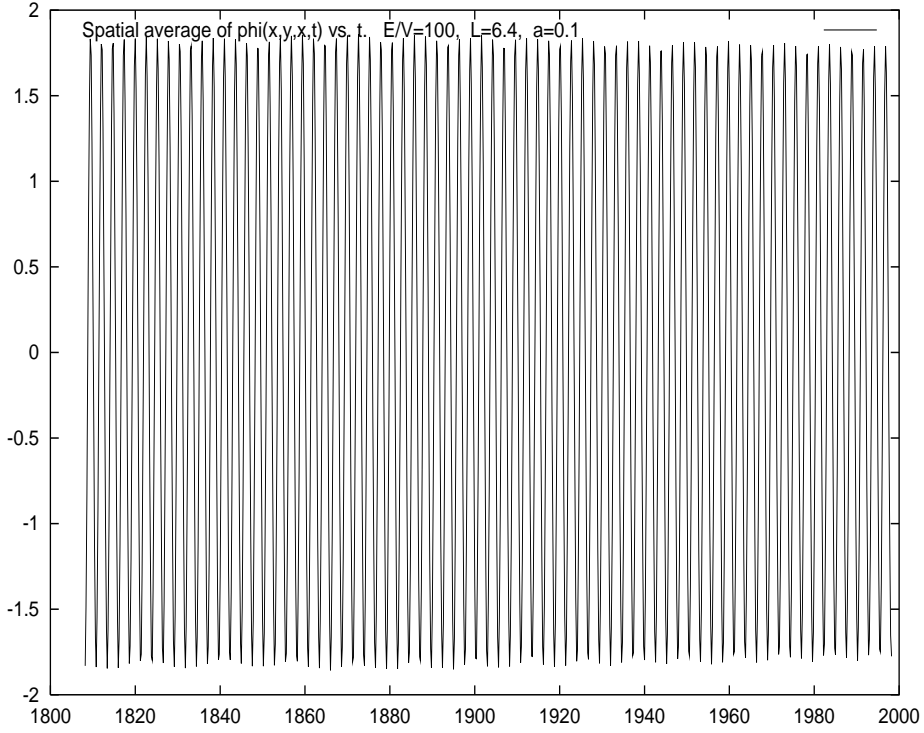}
  \caption{ The space average of $ \phi(\vx,t) $ as a function of time $ t
    $ for $ E/V = 100 $ with $ L = 6.4 $ and $ a = 0.1 $ over a sample time
    interval.  Notice the fast oscillations displayed here which are erased
    by the time averaging. Thanks to time averaging the slow dynamics
    becomes visible in figs. \ref{logopf4}-\ref{logtrpwa064}.}
\label{osce100n32a01}
\end{figure}

Now to disentangle fast and slow motions one can use time averages
over intervals large compared to the periods of the fast oscillations but
small compared to the time scale of the slow motion. This is
exactly what we did above, revealing the smooth UV cascade. To give
a quantitative description of this scenario we can use the following
simple argument. Consider the quantity
\begin{equation*} 
I_\vk = \frac{d}{dt}\left(\mathrm{Re} \,
{\tilde\phi}_{-\vk}{\tilde\pi}_\vk\right)\;.
\end{equation*}
If the motion of each mode were strictly periodic, the time average over
several periods of $I_\vk$ would vanish. But this is  also true 
if the fast and slow motions are indeed separable, at least after some
coarse graining like averaging over discrete directions. Then we may
proceed as in the standard derivation of the virial theorem and obtain, by
use of the equation of motion (\ref{eq:kfe}),
\begin{equation*}
  \overline{|{\tilde\pi_\vk}|^2} = \omega_k^2 \; 
  \overline{|{\tilde\phi_\vk}|^2} + \mathrm{Re} \left[
  \overline{{\tilde\phi}_{-\vk}({\tilde\phi^3})_\vk} \right] \; .
\end{equation*}
We may now define the effective frequency $ \Omega_\vk $ as 
\begin{equation}\label{eq:effom}
  \Omega_\vk^2 \equiv \frac{\overline{|{\tilde\pi_\vk}|^2}}{
  \overline{|{\tilde\phi_\vk}|^2}} = \omega_k^2 + \mathrm{Re}\left[\frac{
  \overline{{\tilde\phi}_{-\vk}({\tilde\phi^3})_\vk}}{
  \overline{|{\tilde\phi_\vk}|^2}}\right] \;,
\end{equation}
where, with the exception of $ \omega_k^2 $, everything depends on time, but
only according to the slow motion.

Our aim here is to derive a quasi-particle description for the classical 
evolution. Quasi-particle pictures have been succesfully
derived in the context of the Hartree approximation both classically
and quantum mechanically (see for example refs.\cite{A,B,ours2,Ngran,quasi}). 

On the staggered lattice there are purely kinematic modification to
eq.~(\ref{eq:effom}), since the fields satisfy discrete recursion relations
rather than differential equations. In Appendix~\ref{app:B} we derive the
virial theorem for the free discrete dynamics on the staggered lattice. A
comparison of eq.~(\ref{fvirial}) with eq.~(\ref{eq:effom}) the suggests
the following lattice relation for the effective frequency $\Omega_\vk$
\begin{equation}\label{eq:leffom}
   \frac{\overline{|{\tilde\pi}_\vk|^2}}{\overline{|{\tilde\phi}_\vk|^2}}
   = \frac1{a^2}\Big[1 -\big(1-\tfrac14 a^4\big)\cos^2(\Omega_\vk a)\Big]
\end{equation}
After averaging over discrete directions, we obtain the following practical
rule for calculating the lattice direction-independent effective frequency
\begin{equation}\label{eq:leffom2}
  (1-\tfrac14 a^4)\cos^2(\Omega_ka) \simeq 1- a^2\,   
  \frac{\overline{|{\tilde\pi}_k|^2}}{\overline{|{\tilde\phi}_k|^2}}
\end{equation}

\begin{figure}[htbp]
  \centering 
  \psfrag{"mefe1000n64a005.dat"}[r][r]{$E/V=1000,\;a=0.05$}
  \psfrag{"mefe100n64a005.dat"}[r][r]{$E/V=100,\;a=0.05$}
  \psfrag{"mefe10n64a005.dat"}[r][r]{$E/V=10,\;a=0.05$}
  \psfrag{"mefe1n64a005.dat"}[r][r]{$E/V=1,\;a=0.05$}
  \psfrag{"mefe01n64a005.dat"}[r][r]{$E/V=0.1,\;a=0.05$}
  \includegraphics[width=10cm,height=12.72cm,angle=270]{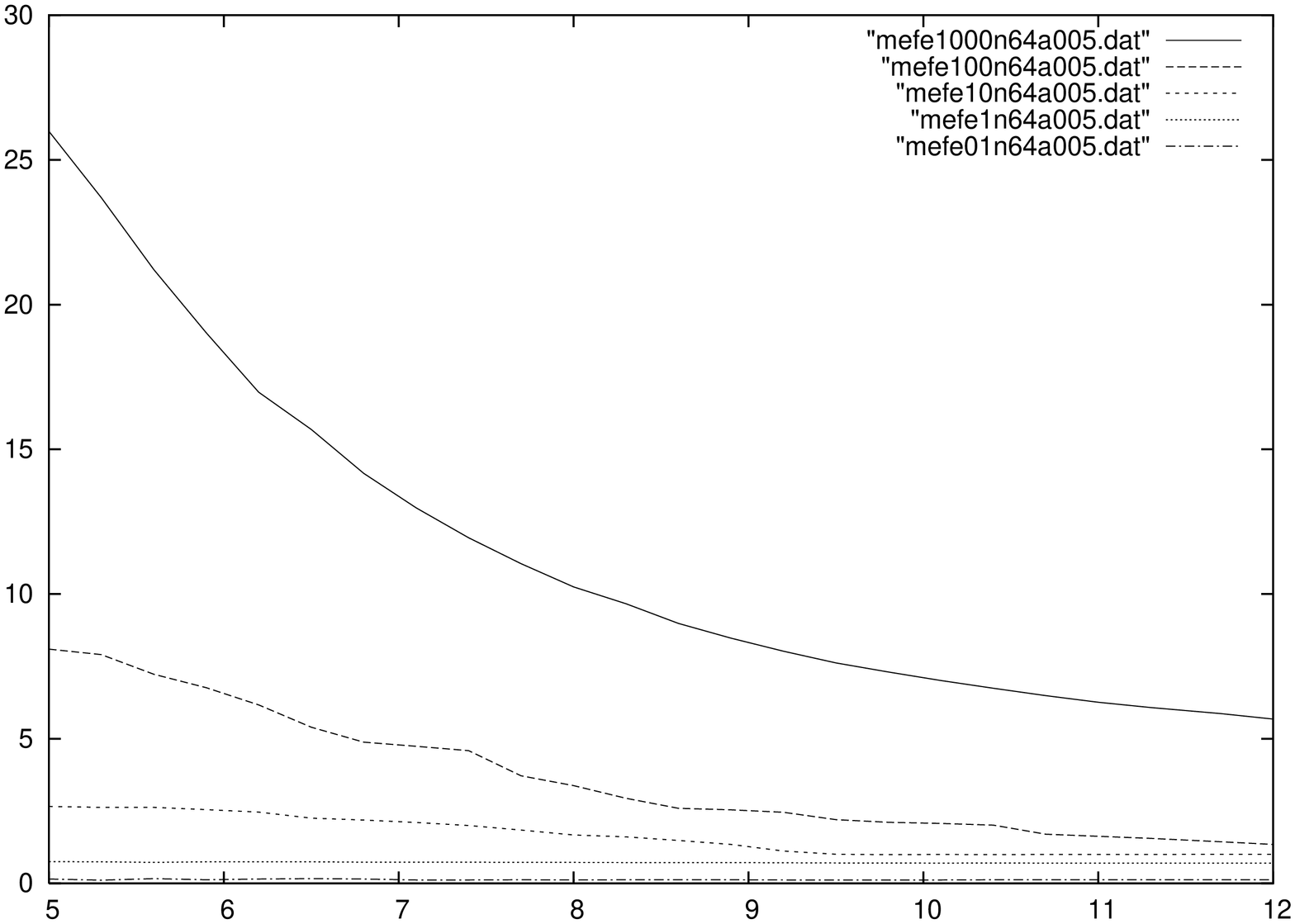}
  \caption{$M_{\rm eff}^2(t) - 1$ as a function of the logarithm of the
    time $ t $ for $ E/V = 0.1, \; 10, \; 100 $ and $ 1000 $ with $ L = 6.4
    $ and $ a = 0.1 $. No average is performed over initial packet
    amplitudes or positions.  Time averaging is performed here as by
    eqs.(\ref{promT})-(\ref{taut}).}
\label{mef}
\end{figure}

\begin{figure}[htbp]
  \centering
  \psfrag{"re1000n64a005.dat"}[r][r]{$E/V=1000,\;a=0.05$}
  \psfrag{"re100n64a005.dat"}[r][r]{$E/V=100,\;a=0.05$}
  \psfrag{"re10n64a005.dat"}[r][r]{$E/V=10,\;a=0.05$}
  \psfrag{"re1n64a005.dat"}[r][r]{$E/V=1,\;a=0.05$}
  \includegraphics[width=10cm,height=12.72cm,angle=270]{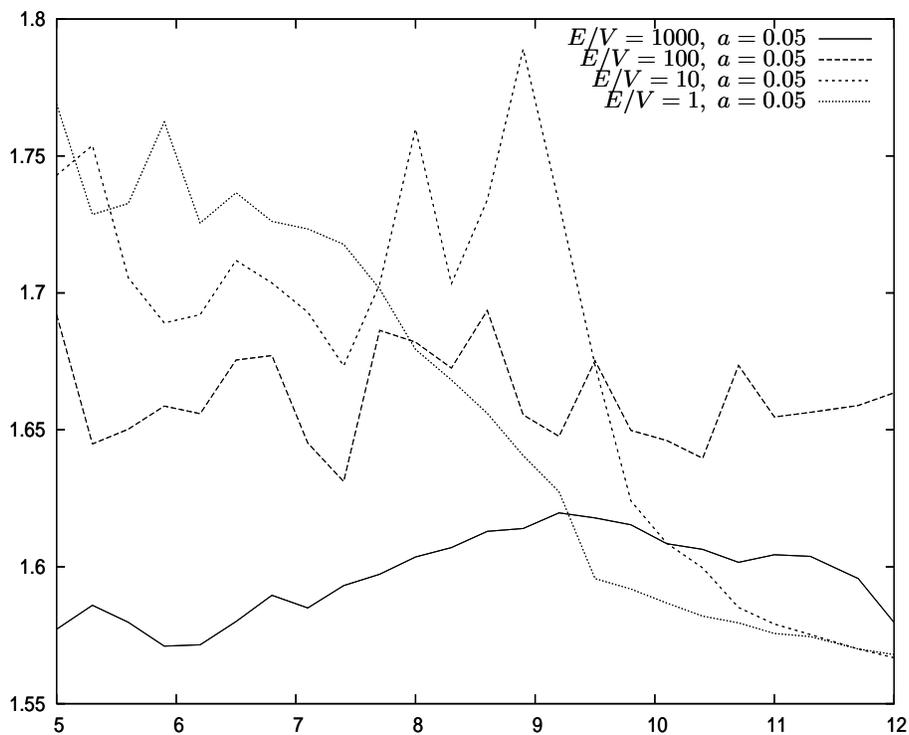}
  \caption{$R \equiv [M_{\rm eff}^2(t) - 1]/ {\overline{\phi^2}}(t)$ as a
    function of the logarithm of the time $ t $ for $ E/V = 0.1, \; 10, \;
    100 $ and $ E/V = 1000 $ with $ L = 6.4 $ and $ a = 0.1 $.  Notice that
    $R$ varies in a much narrower interval than $ M_{\rm eff}^2(t) $
    depicted in fig. \ref{mef}. No average is performed over initial packet
    amplitudes or positions. Time averaging is performed here as by
    eqs.(\ref{promT})-(\ref{taut}).}
\label{r}
\end{figure}

Eq.~(\ref{eq:leffom2}) suggests also a way to parametrize the equilibrium
two-point function $\ephik2$ on the lattice in terms of an equilibrium
effective frequency $\Omega^\mathrm{eq}_\vk$: replacing
$\overline{|{\tilde\pi}_k|^2}$ by its equilibrium counterpart $\epik2=T$
and $\overline{|{\tilde\phi}_k|^2}$ by $\ephik2$ yields
\begin{equation}\label{eq:eqom}
  \ephik2 = \frac{a^2\,T}{1 -\big(1-\tfrac14 a^4\big)
    \cos^2(\Omega^\mathrm{eq}_\vk a)}  \; .
\end{equation}
Going back to the continuum, we recall that the Hartree approximation for
$\Omega_\vk$ follows from the reduction of
\begin{equation*}
  ({\tilde\phi^3})_\vk = \frac1{V}
    \sum_{{\bds q}\,{\bds q}^{\,\prime} } \; 
    {\tilde\phi}_{\bds q} \; {\tilde\phi}_{{\bds q}^{\,\prime}} \; 
    {\tilde\phi}_{\vk-\bds q-{\bds q}^{\,\prime}}
\end{equation*} 
to the sum of all terms proportional to $ {\tilde\phi}_\vk $, that is
\begin{equation*}
  ({\tilde\phi^3})^\mathrm{H}_\vk = \frac3{V}
  \sum_{\bds q}|{\tilde\phi}_{\bds q}|^2 \; {\tilde\phi}_\vk
  = \frac3{V} \; {\tilde\phi}_\vk \, \int_V d^3x \; \phi^2(x) \;.
\end{equation*} 
Using eq.(\ref{eq:effom}) this yields the Hartree frequency,
\begin{equation}\label{eq:hartree}
  (\Omega^\mathrm{H}_\vk)^2 = \omega_k^2 + 3 \; \overline{\phi^2}(t) \; .
\end{equation} 
It must be noticed that the same effective frequency follows by using
the Whitham approach\cite{whi}. That is, considering  a multi-wave 
configuration as in eq.(\ref{ondp}), the first nonlinear correction to the 
frequency in the Whitham approach can be shown after some work to coincide 
with the Hartree formula eq.(\ref{eq:hartree}).

Taking into account the definition of time--dependent wavefunction
renormalization $Z_k(t)$ as defined by eq.(\ref{eq:Zkt}), we now have
\begin{equation} \label{Z}
  Z_k(t) \equiv 
  \frac{\overline{|{\tilde\phi}_k|^2}}{a^2\,\overline{|{\tilde\pi}_k|^2}}
  \Big[1 -\big(1-\tfrac14 a^4\big)\cos^2(a\,\Omega^{\mathrm{H}}_k)\Big]
  = \frac{(\Omega^{\mathrm{H}}_k)^2}{\Omega^2_k(t)}
\end{equation}
Hence $1/Z_k(t)$ is the renormalization which turns the approximated
Hartree frequency into the exact effective frequency. In the previous
section we have provided numerical evidence showing that $Z_k(t)$ takes a
form quite close to the equilibrium one (at some time--dependent effective
temperature) for $t\gtrsim t_0$ and in particular that it approaches unity
rather fast as $k^2 > \overline{\phi^2}(t)$. As shown in
section~\ref{fulltherm} and in more detail in appendix~\ref{sec:hart}, this
holds for very late times of order $10^4 t_0$, close to complete lattice
thermalization (necessarily cutoff--dependent), with $Z_k(t)$ almost
constant in time much earlier than $\overline{\phi^2}(t)$ or the power
spectra and the effective temperature very close to the final equilibrium
temperature. But it also holds for much shorter times, of order $t_0$ only,
as soon as the universal cascade has set in, with $Z_k(t)$ almost constant
as a function of $k[\overline{\phi^2}(t)]^{1/2}$ and some time--dependent
effective temperature much larger than the final equilibrium temperature.

We tested the universality of this picture by studying the changes of $Z_k$
(or absence thereof) upon changes of the lattice spacing, physical size,
energy density, initial condensate and microscopic details of initial
conditions in our numerical evolutions.

\medskip
Notice that we always use the exact classical time evolution for the
fields and {\bf never} the Hartree approximation to it. However,
this exact evolution of the modes with $ k^2 > \overline{\phi^2}(t) $ 
is well reproduced with an effective mass as given by the Hartree 
approximation eq.(\ref{eq:hartree}).

\bigskip 

In quantum theory, as is well known the definition of the number operator
is not unique (see, for example refs.\cite{ours,ours2}).  However, we can
introduce a classical number of modes based in the correspondence
principle. In the classical limit the number of quanta is given by the
phase space area encircled by the classical phase space trajectory divided
by $ 2 \, \pi $ (in units where $ \hbar = 1 $). In the Hartree
approximation both $ \pi_\vk(t) $ and $ \phi_\vk(t) $ oscillate with
frequency $ \Omega_\vk^H(t) $. The phase space area for a slow varying
frequency is then given by the ellipse area $ \pi \; |\pi_\vk|_{max} \;
|\phi_\vk|_{max} $ where in addition for harmonic oscillations $
|\pi_\vk|_{max} = (2\,\overline{|{\tilde\pi}_\vk|^2})^{1/2} $ and
$|\phi_\vk|_{max} = (2\,\overline{|{\tilde\phi}_\vk|^2})^{1/2} $.
Therefore, the number of modes is given by
\begin{equation}\label{eq:nk}
  \overline{n}_\vk(t) = \left[\pivk2\,\phivk2 \right]^{1/2} \; .
\end{equation}
Also, 
\begin{equation*}
\overline{n}_\vk(t) =\tfrac12 \,\frac{\sqrt{Z_k(t)}}{
\Omega_\vk^{\mathrm{H}}(t)} \; \pik2 + 
\tfrac12 \, \frac{\Omega_\vk^{\mathrm{H}}(t)}{\sqrt{Z_k(t)}} \; \phik2 \; ,
\end{equation*}
where we used the continuum limit of eq.(\ref{Z}), namely,
\be
Z_k(t) \; \pik2 = \left[\Omega_\vk^{\mathrm{H}}(t)\right]^2 \; \phik2 \; .
\ee

Although derived within the assumption of harmonic oscillations with a
slowly varying frequency, eq.~(\ref{eq:nk}) should have a more general
validity, since $ n_\vk $ provides in any case a measure of the phase space
occupied by the trajectory of the $ \vk $ mode. The only important condition
is that the different modes are weakly coupled. This is true, when the UV
cascade has fully developed, for all $\vk$ such that
$ [\overline{\phi^2}(t)]^{1/2} <|\vk|\lesssim\bk(t) $, where
$ [\overline{\phi^2}(t)]^{1/2} $ decreases with time tending to a small
equilibrium value $ \sqrt{\avg{\phi^2}} $ of order $ \sqrt{T/a} \sim a \,
\sqrt{E/V}$.  Furthermore, if $t$ is not too large so that
$\bk(t)\ll\Lambda$, then the occupied modes do not feel the lattice
discretization and have the relativistic dispersion relation. Therefore
\begin{equation}\label{eq:nk2}
  \overline{n}_\vk(t) = \frac{\sqrt{Z_k(t)}}
  {\Omega_k^{\mathrm{H}}(t)} \; \pivk2 \; ,
\end{equation}
where now $[\Omega_k^{\mathrm{H}}(t)]^2= k^2+1+3\,\overline{\phi^2}(t)$.

\begin{figure}[htb]
   \centering
   \psfrag{L12.8}{$L=12.8$}
   \psfrag{a0.0125}{$a=0.0125$}
   \psfrag{EoV = 89.5}{$E/V=89.5$}
   \psfrag{Lambda}{$\Lambda=125.664$}
   \psfrag{time}{$t\longrightarrow$}
   \psfrag{tval1}{$t=25.17,\,34.75,\,48.1,\,67.12,\,94.65,\,134.92,\,194.32,$}
   \psfrag{tval2}{$~~~~~282.4,\,413.6,\,609.3,\,901.88,\,1339.73,\,1995.5,\,2978.15$}
   \psfrag{kvar}[tc][tc]{$k$}
   \psfrag{nkvar}{$4\pi k^2\overline{n}_k(t)$}
   \includegraphics[width=12.25cm,height=10cm]{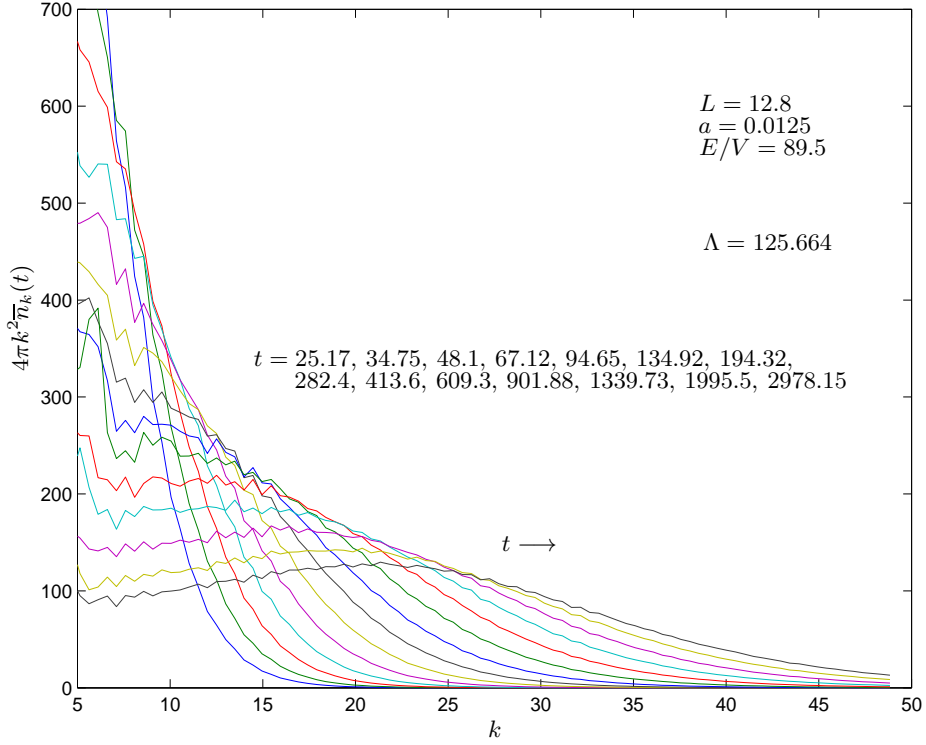}
   \caption{The number of modes over spherical shells, $ 4\pi\, k^2
     \,\overline{n}_k(t) $, vs. $k=|\bds k|$ at the different times
     indicated. The initial conditions are infrared random plane waves. The
     UV cascade is clearly seen as $ k^2 \; \overline{n}_k(t) $ is depleted
     for low $ k $ while it grows for large $ k $ as time grows.}
\label{number}
\end{figure}

We plot in fig. \ref{number} the number of modes over spherical shells,
$4\pi\, k^2\,\overline{n}_k(t) $, vs. $ k=|\bds k| $ at different times.
Fig. \ref{number} should be compared with figs.~\ref{pw2a025e89} and 
\ref{fourpw2} since $ \overline{n}_k(t) $ and $ \pivk2 $ are related
through eq.(\ref{eq:nk2}). One sees that the main dependence of 
 $ \overline{n}_k(t) $ on $ k $ comes from $ \pivk2 $.

Notice that eq.~(\ref{eq:nk2}) can be written in terms of the effective
frequency $ \Omega_k(t) $ as
\begin{equation*}
 \overline{n}_\vk(t) = \frac{\pivk2}{\Omega_k(t)} \,
\end{equation*}
which is the classical equilibrium occupation number with the temperature
replaced by $ \pivk2 $. In fact, in thermal equilibrium the power spectrum
and the temperature are related by eq.(\ref{Tflat}). During the UV cascade
we are in a situation of effective equilibration for times $ t $ later than
$ t_0 \sim 500 $ as shown by fig. \ref{ptvsmc} and the results of
section~\ref{eqcasc}. However, $ \pivk2 $ depends both on time and on the
wavenumber as depicted in figs. \ref{rpw2ta0125}, \ref{tail125} and
\ref{logtrpwa064}.  In addition, the $k$-modes with $ k^2 <
\overline{\phi^2}(t) $ only thermalize for times of the order $ 10^6 $ as
we see in fig. \ref{logtrpwa064}. This makes it awkward to try and
interpret $\pivk2$ as a $k-$dependent effective temperature since an
equilibrium or quasi--equilibrium state should depend only on few
macroscopic parameters (with one playing the role of temperature) varying
slowly in time. Notice however that $ \pivk2 $ {\bf decreases} both with
time {\bf and} with the wavenumber $ k $ for $ k^2 > \overline{\phi^2}(t) $
[see figs. \ref{rpw2ta0125}, \ref{tail125} and \ref{logtrpwa064}] and does
that in a very smooth and regular way. This suggests indeed the existence
of few slowly time--dependent parameters, not dependent on the details of
the initial conditions, which govern the evolution of the cascade for $ k^2
> \overline{\phi^2}(t) $. On the contrary $ \pivk2 $ increases as well as
decreases with $ k $ and $ t $ for the modes $ k^2 < \overline{\phi^2}(t) $
in a much more erratic way strongy dependent on the detail of the initial
conditions [see figs. \ref{rpw2ta0125}, \ref{logrpwa064} and
\ref{logtrpwa064}].  

\subsection{Effective Mass squared}

The time averaging defined by eq.(\ref{promT}) is intended to eliminate the
microscopic oscillations of the field. We illustrate such microscopic
behaviour in fig. \ref{osce100n32a01} showing the field $ \phi(\vx,t)$ 
averaged over the {\bf space}. The frequency of the time oscillation can be
therefore considered as the  {\bf effective mass} $ M_{\rm eff}(t) $ 
of the field. Moreover, by Fourier transforming the  field $ \phi(\vx,t) $ 
we obtain
frequencies $\Omega_\vk$ numerically. This procedure is more costly than 
eq.~(\ref{eq:leffom2}) but we find an excellent agreement between both.
Notice that the use
of lattice expressions like eq.~(\ref{eq:leffom2}), valid in principle to
all orders in $a$, is very efficient and convenient to compare results on
lattices with different lattice spacings.

It must be noticed that the fast oscillations displayed in fig. 
\ref{osce100n32a01} are erased by the time averaging. Thanks to such averaging
the slow dynamics becomes visible through figs. 
\ref{logopf4}-\ref{logtrpwa064}. 

\medskip

We plot in  fig. \ref{mef} $ [M_{\rm eff}^2(t) - 1] $ vs. the logarithm of
time for different values of $\rho \equiv E/V$. 
Notice that $ M_{\rm eff}^2(t) $
monotonically decreases with time while the UV cascade develops. We
see that  $ M_{\rm eff}^2(t)>1 $ and that it decreases with
$\rho$. Indeed, $ M_{\rm eff}^2(t) \to 1 $ (its linearized value) for
$\rho \to 0$. We find that  $ M_{\rm eff}^2(t) - 1 $ is approximately
proportional to  $ {\overline{ \phi^2}}(t) $. 

We depict in fig. \ref{r} the ratio 
\be \label{R}
R \equiv \frac{ M_{\rm eff}^2(t) - 1 }{ {\overline{ \phi^2}}(t)} \; ,
\ee

as a function of $ \log t $. We see that $ R $ is approximately time
independent and that it {\bf decreases} with $ E/V $. It should be noticed
that $R$ always stays below the value $ R_{Hartree} = 3 $
predicted by the Hartree approximation \cite{uno} and well above the value  $
R_{Large N} = 1 $ corresponding to the large number of components limit.

In summary, we find that $ M_{\rm eff}^2(t) $ can be written as
\begin{equation} \label{mef2}
  M_{\rm eff}^2(t) = 1 + R(\rho) \;  {\overline{ \phi^2}}(t) \; ,
\end{equation}
where $  1.5 <R(\rho)<3 $ is approximately time independent and decreases 
with $\rho$. 

We analyze the cnoidal solution in appendix C and compute its effective mass.
The parameter $R$ for the  cnoidal solution turns to be between $\frac32$ and 
$\frac{\pi}{2}$ and is depicted in fig. \ref{cno}.

As discussed in sec. \ref{frecefe} the modes with $ k^2 
\gg \overline{\phi^2}(t) $ are governed by the Hartree mass $ 1 +
3 \; \overline{\phi^2}(t) $ while the $ k=0 $ mode oscillates according
to $  M_{\rm eff}(t) $. The Hartree mass and $  M_{\rm eff}^2(t) $ are
substantially different as remarked above.

Moreover, $  M_{\rm eff}(t) $ governs the long range behaviour of the 
correlators is real space. The two-point correlator tends to zero for 
long distances as,
\begin{equation*}
C(x) \buildrel{x \gg a}\over= C_0 \; \frac{e^{-M_{\rm eff}(t) \; x}}{x} \; ,
\end{equation*}
where  $C_0$ is a constant.

In summary, modes with $ k^2 $ below and above $ \overline{\phi^2}(t) $ 
are governed by different masses. The first by $ M_{\rm eff}^2(t) \simeq 1
+ 1.5 \;\overline{\phi^2}(t) $ and the later by the Hartree mass 
$ 1 + 3 \; \overline{\phi^2}(t) $. Clearly, the long-range behaviour of the 
correlators is governed by the low momentum mass  $  M_{\rm eff}(t) $.

\subsection{The slow dynamics of the infrared modes and
effective thermalization}\label{conde}

The picture of the averaged dynamics for late times  in the lattice model
is as follows:

\begin{itemize}
\item{The modes with $ k^2 >\overline{\phi^2}(t) $ effectively 
thermalize with a time dependent temperature while they obey the Hartree 
approximation.
That is, their dynamics is weakly nonlinear.}  
\item{The modes with $ 0 \leq k^2 \lesssim \overline{\phi^2}(t) $ 
keep interacting quite strongly between them. They thermalize much later 
than the rest of the modes}
\item{Both set of modes keep interacting between each other}
\end{itemize}

There are definitely two time scales for thermalization. The shorter one
$ t_0 \sim 500 $ describes the effective thermalization of the Hartree modes
while time scale for the effective thermalization of the infrared modes is much
longer than $ t_0 $ and of the order $ \sim 10^6$.

The modes with $ 0 \leq k^2 \lesssim \overline{\phi^2}(t) $ manifest in the
IR behaviour of the correlator as a long--lived inhomogeneous
condensate. This condensate keeps interacting with the higher $k$ modes
which behave as a thermal bath in contact with it.  Notice that the mode
distribution becomes time independent on the lattice for late times. Hence,
the field correlator is a static function described by the Fourier
transform of
\begin{equation}
  \label{eqterL} 
  |{\tilde\phi}_k|^2 =\frac{T \; Z_k}{k^2 + 1 + 3
    \,\overline{\phi^2}(t)} \; .  
\end{equation}
where $T$ is the asymptotic time limit of $ \pik2 $.  

\medskip 
While $ M_{\rm eff}^2 $ decreases with a time scale $ \sim t_0 $ [see fig.
\ref{mef}], $ Z_k $ varies with an even longer time scale $ \sim
10^6 $. It can be interpreted as a wavefunction renormalization which
becomes unit for $ k^2 \gg \overline{\phi^2}(t) $.

It should be stressd that the inhomogeneous condensate is not a classical
solution of the $ \phi^4 $ equations (\ref{eqnofmot}).  It is a classical
statistical configuration that can only be seen through the correlators
like $ \overline{\phi\phi}(\vx,t) $.  Moreover, the average value of $ \phi
$ just vanishes for late times as shown in fig. \ref{fi}.

Notice that the local observables as $ \overline{\pi^2}(t), \;
\overline{(\nabla\phi)^2}(t), \; \overline{\phi^2}(t) $ and $
\overline{\phi_+^4}(t) $ are dominated by the modes with $ k^2 \gtrsim
\overline{\phi^2}(t) $. This explains that the thermal equilibrium curve
approximately agrees (to $90\%$) with time averages in fig. \ref{ptvsmc}
although the infrared modes are not yet completely thermalized.

\medskip

Extensive numerical calculations varying the initial conditions showed us that
the thermalization dynamics including the UV cascade posseses an universal
character. That is, the UV cascade exhibit the {\bf same} features
for different kinds of initial conditions (plane waves, localized packets, etc.)
with initial power in the infrared modes ($ k \lesssim 30 \, \pi/L \ll 
\pi/[2 \, a] $) and energy density not too small ($ \rho \gtrsim 1$).

\bigskip

In the continuum theory the ultraviolet cascade continues forever
and therefore, for finite energy densities, the system will reach thermal
equilibrium for infinite times at zero temperature. 
The inhomogeneous condensate
thus disappears for extremely long times in the continuum theory.

\bigskip

As the inhomogeneous condensate is an infrared phenomena, it should be present
in quantum theory for large occupation numbers and weak coupling.
In other words, the low $k$ behaviour of $ Z_k $ should not change in 
quantum theory when such conditions are fulfilled.

\section{Discussions and Conclusions}

The present work shows that a UV cascade enjoying universal properties is
the basic mechanism of thermalization both in $3+1$ and $1+1$ 
dimensions\cite{uno}.
Although we are not in a position to derive the properties of this cascade 
from the microscopic equation of motion (\ref{eqnofmot}),
it is easy to see from eq.(\ref{eqnofmot}) that
energy should flow towards higher wavenumbers. Assume we start from a
plane wave initial condition
$$
\phi(0,\vx) = A \cos(\vk \cdot \vx) \quad , \quad
 {\dot \phi}(0,\vx) =  A \sin(\vk\cdot\vx)\; ,
$$
where $A$ is a constant. The nonlinear term $ \phi^3 $ in Eq. (\ref{eqnofmot})
will immediately generate higher harmonics: $ \cos(3 \; \vk\cdot\vx) , \;
\cos(9 \;\vk\cdot\vx),  \; \cos(5 \;\vk\cdot\vx) $ etc. 
That is, energy is moving to higher wavenumbers.

Moreover, if we consider a superposition of plane waves as initial condition,
 $$
\phi(0,\vx) = \sum_{a=1}^n A_a \cos(\vk_a\cdot\vx) \quad , \quad
{\dot \phi}(0,\vx) = \sum_{a=1}^n A_a \sin(\vk_a\cdot\vx)\; ,
$$
the interaction term  $ \phi^3 $ generates sums and differences:
$ \vk_a \pm \vk_b\pm \vk_c $ etc, for $ a,b,c = 1, \ldots , n $.
This implies energy flowing {\bf both} for increasing {\bf and} for decreasing 
wavenumbers.

Notice that this mode mixing takes place very fast, at the microscopic time
scale. That is, a unit time scale in dimensionless variables.
The UV cascade we observe happens in a much longer time scale.
This means that these fast microscopic processes combine in a nontrivial way
resulting on a  UV cascade with a slow time scale.

\bigskip

It is important to estimate in which extent the present results can be
applied in quantum theory. A necessary condition for  the validity of
the classical approximation is that the occupation numbers must be
large. The relevant occupation numbers decrease during the UV cascade due to
the modes flow towards unoccupied high wavenumber slots. Hence, if the
classical approximation is valid for late times, it will be valid
earlier (as it is the case $1+1$ dimensions \cite{uno}).
A simple criterion for the validity of the classical
approximation in QFT at thermal equilibrium is that
\begin{equation}\label{condQ}
T_p \gg \omega_p(k_p)
\end{equation}
where $\omega_p(k_p)= \sqrt{k_p^2+ M_{p, eff}^2}$
are the dimension-full frequencies.  

In dimensionless variables eq.\eqref{condQ} takes the form,
\begin{equation}\label{conQd}
\frac{T}{\lambda} \gg \sqrt{k^2+ M_{\rm eff}^2} >  M_{\rm eff}
\end{equation}
Following eq.(\ref{mef2}) we obtain as an estimate for $M_{\rm eff}^2$  
from our numerical results 
\be \label{meff}
M_{\rm eff}^2 \sim 1 + A \; \sqrt{\rho}
\ee
where $ \rho = \frac{E}{V} $ and $ A \sim 0.2 - 0.5 $. 

For $ \rho \gtrsim 1 $ we thus find that the classical approximation applies 
for 
\be\label{cond1}
 \lambda \ll  (2a)^3 \;  \rho^{\frac34}
\ee
where we used that $ T = (2a)^3 \;  \rho $ for small $a$ [eq.(\ref{TV})]. 

For $ \rho \lesssim 1 $ we instead find that the classical approximation 
applies for
\be\label{lamax}
 \lambda \ll  (2a)^3 \;  \rho
\ee
Both eq.(\ref{cond1}) and (\ref{lamax})  leads to small values of the coupling 
$\lambda$ since the spacing $ a $ 
must be itself small to avoid lattice effects.  However, in inflationary 
theories very small values of $\lambda \sim 10^{-12} $ are customary (in order 
to agree with the smallness of the CMB anisotropy) which leaves room 
for the use of the classical approximation. 

\bigskip 

As stated above the validity of the classical approximation decreases with 
time. Let us estimate the time where it ceases to be valid.

The total energy density can be estimated in terms of the average occupation 
number $  \overline{ n_k } $ (where the average is here over the $\vec k$ 
modes) as follows
$$
\rho \sim  \overline{k}^3 \;  \overline{ n_k } \; 
\sqrt{\overline{k}^2 + M_{\rm eff}^2 }  \; .
$$
The classical approximation holds if the occupation numbers are large for the 
relevant modes. Therefore, a necessary condition is
$$
 \frac{\overline{ n_k }}{\lambda}  \gg 1  \; , 
$$
since the coupling $ \lambda $ has been absorbed in the field redefinition
Eq. (\ref{adim}). 

In the ultrarelativistic regime $ \overline{k} \gg  M_{\rm eff} $ we have
\be \label{ur}
\overline{k}^4 \ll \frac{\rho}{\lambda} \quad 
\mbox{and~therefore,} \quad \overline{k} \ll  
\left(\frac{\rho}{\lambda}\right)^{\frac14} \; ,
\ee
while in the non-relativistic domain  $ \overline{ k} \ll  M_{\rm eff} $ 
we have,
$$
 M_{\rm eff} \; \overline{ k}^3 \ll \frac{\rho}{\lambda}
$$
and hence using eq.(\ref{meff})
\be\label{56}
\overline{k}^3 \ll \frac{\rho^{\frac34}}{\lambda} \quad 
\mbox{and~therefore,} \quad 
\overline{k} \ll \frac{\rho^\frac14}{\lambda^\frac13} \; .
\ee
Since $ \lambda \ll 1 $,
$$
\frac1{\lambda^\frac13} 
\gg \frac1{\lambda^\frac14} \quad , \quad {\rm thus} \quad
\left(\frac{\rho}{\lambda}\right)^{\frac14} \ll 
\frac{\rho^\frac14}{\lambda^\frac13} \; ,
$$
and the ultrarelativistic condition eq.(\ref{ur}) is more stringent
than the non-relativistic bound  eq.(\ref{56}). We can therefore 
always use the condition eq.(\ref{ur}) for the validity of the 
classical approximation.

Using now eq.(\ref{expk}) for $ \overline{ k}(t) $ yields in both regimes that 
the classical approximation is valid for times $t$
\be \label{tmax}
t\ll  t_{max} 
\simeq  \frac{\rho^{\frac34}}{\lambda^{\frac34} \; k_0^3} \; .
\ee
The conditions eq.(\ref{lamax}) and (\ref{ur}) are compatible since $ 
\overline{k} < \frac{\pi}{2a} $. Therefore, eq.(\ref{ur}) implies
$$
\frac{\pi}{2a} \ll  \left(\frac{\rho}{\lambda}\right)^{\frac14} \quad 
\mbox{and ~ then} \quad 
(2a)^3 \; \rho^{\frac34} \gg \pi^3 \; \lambda^{\frac34} \; \mbox{or}  \; 
\frac{\rho}{\lambda} \gg \left(\frac{\pi}{2a}\right)^4  \; ,
$$
which is more stringent than eq.(\ref{cond1}).

In summary, the classical approximation is valid
for times earlier than  $ t_{max} $ [given by eq.(\ref{tmax})] provided 
\be \label{conval} 
\frac{\rho}{\lambda} \gg \left(\frac{\pi}{2a}\right)^4  \; .
\ee
That is, {\bf high} density (large occupation numbers) 
and/or {\bf small} coupling. 
This condition constraints the initial conditions which fix $ \rho $. 
Notice that the coupling $\lambda$ {\bf does have} an intrinsic meaning in 
the quantum theory.

A condition for the validity of the classical dynamics in QFT is
derived for high-density in ref. \cite{berR} in the context of the 
2PI-1/N approach. For further studies about the validity of the classical 
approximation in different contexts see \cite{A,preter,lisref}.

\bigskip

Let us now comment about the character of the universal stage.
The dimensionality of the space plays a crucial r\^ole in this phenomena.
In one space dimension effective thermalization takes place much faster 
in analogous conditions\cite{uno}. 
Here, in $3+1$ dimensions a second and even longer time scale appears 
characterizing
the thermalization of infrared modes with $ k < \sqrt{\overline{\phi^2}}(t) $. 
The infrared modes keep interacting for far longer times than the higher 
$k$ modes. 
In $3+1$ dimensions the infrared modes thermalize the last while in  
$1+1$ dimensions they thermalize the first\cite{uno}.

The UV cascade is clearly less  efficient in three space dimensions 
to fill the higher $k$-modes pumping energy from the low  $k$-modes. 
Clearly, in one space dimension the phase space is dramatically small
making the UV cascade very efficient. 

The same phase space effect ($k^2$) in three space dimensions makes the
classical statistical mechanics of the $ \phi^4 $ theory divergent due to
the ultraviolet catastrophe. 

\medskip

All this suggest that thermalization is reached in quantum theory too as
recent works (including memory effects) indicate \cite{ber,berR,gre}. Moreover,
for large initial occupation numbers and small coupling the classical
regime should correctly describe the evolution of the theory till a time
$ t_{max} $ [see eq.(\ref{tmax})]. 

\bigskip

A remarkable feature of the thermalization mechanism is that even starting
from a completely classical regime (large occupation numbers at low
wavenumbers), the UV cascade depletes these modes and fills the high ones.
Therefore, at some time $ \sim t_{max} $ (that could be very long) quantum
physics unavoidable shows up. This is the out of equilibrium counterpart to
the fact that classical statistical mechanics is ill defined due to the
ultraviolet catastrophe which can only be cured by the quantum treatment.

Thermalization implies to forget everything about the initial conditions
except (obviously) the conserved quantities like energy, momentum, angular
momentum. The thermalization is therefore expected to substantially change 
for integrable theories where the number of conserved quantities equals
the number of degrees of freedom.

\bigskip

In this paper we solve the exact microscopic dynamics and then we averaged
on time intervals and space in order to derive the slow dynamics we are
interested in. This is perfectly correct but a lot of information is first
obtained and then dumped in the averaging process. Alternatively a
transport approach could be followed deriving equations for the observables
in macroscopic time scales and then solve them. We mean to derive transport
equations of the Boltzmann or Fokker-Planck type.

\bigskip

{\bf Acknowledgments:} Its a pleasure to thank D. Boyanovsky for
useful discussions. The numerical calculations were performed on 
clusters of personal computers at LPTHE-Paris and at the Physics Department of 
Milano-Bicocca; the simulations with largest lattices were 
performed on the Avogadro cluster of CILEA.

\appendix
\section{Average over discrete directions}\label{app:A}

The space discretization and the finite size spoil the rotational
invariance of continuum infinite--volume Hamiltonian.  This implies the
same invariance for the classical field equation ad for the equilibrium
averages (see section~\ref{sec:Thereq}). Rotational invariance should be
recovered at short distances when $a\to0$ and at long distances when
$L\to\infty$. Thus any lattice observable on the wavenumbers cube $(2\pi/L)
\; C_N$ that corresponds to a rotational invariant observable of the
continuum infinite--volume theory, should depend only on $k^2$ when $a\to0$
and $kL\gg 1$ or when $L\to\infty$ and $ka\ll 1$ (a similar argument would
apply for the $x-space$ cube $2a\,C_N$). To verify the onset of rotational
invariance in non--perturbative or numerical lattice calculations is
usually quite costly and often not even necessary. In fact, by turning the
argument around and assuming that rotational invariance will be recovered
in the appropriate limits, one can greatly reduce fluctuations by
performing an average over all directions on discrete observables. This can
be done as follows.

Given any array $f_\vn$ over $C_N$, we consider its average over discrete
directions ${\bar f}_n$ as
\begin{equation}\label{anglavg}
  {\bar f}_n = \frac1{S_n} \sum_{\vn\in C_N} 
  \theta(n\le|\vn|<n+1) \,f_\vn \;,\quad n=0,1,2,\dots, \sqrt{3}N/2 \; .
\end{equation}
where $\theta(\ldots)=1 \; (0)$ if its arguments is true (false) and $S_n$ is
the number of points of $C_N$ at a distance $d$ from the origin
such that $n\le d <n+1$, that is
\begin{equation*}
  S_n = \sum_{\vn\in C_N} \theta(n\le|\vn|<n+1)
\end{equation*}
Up to purely geometrical fluctuations, $S_n$ grows like $4 \; \pi \;  n^2$ for
$n$ up to $N/2$, while for $N/2<n \le \sqrt{3} \; N/2 $, in the corners
of the cube, its shrinks to zero. To fix a specific value of the radius of
this spherical shells, we choose the average distance of all lattice
points within the shell
\begin{equation*}
  r_n = \frac1{S_n} \sum_{\vn\in C_N } \theta(n\le|\vn|<n+1)\,|\vn|
\end{equation*}
When plotted against $r_n$, $S_n$ still has purely geometrical
fluctuations w.r.t. the continuum expression, most noticeably at $n=0$,
where $S_0=1$ while $4\pi r_0^2=0$. Of course, averaging over several
consecutive shells would reduce these geometric fluctuations. Such an
average is in fact automatic in the limit $L\to\infty$ of continuous
wavenumbers, since in any physical realizable measure there exists a 
finite resolution $\Delta k$ independent of $L$. On the other hand the
limit $L\to\infty$ is not feasible in numerical calculations and one has to 
find the correct balance between size and control of fluctuations.   

Suppose now that the array $f_\vn$ if the finite size version of the
continuous function $f(\vk)$ over the first Brillouin zone at infinite size
(the argument would apply also if $f_\vn$ were the lattice version of a
continuous $f(\vx)$ on the cubic volume $V$).  Then ${\bar f}_n$ provides a
finite size version of the angle average of ${\bar f}(k)$ of $f(\vk)$,
where $k=|\vk|$. In other words, the discrete plot of ${\bar f}_n$ vs.
$k_n\equiv (2\pi/L) \, r_n$ provides an approximation of the continuous plot of
${\bar f}(k)$ vs. $k$, if $L\to\infty$ at fixed $a$ {\bf and} $f_\vn$ has
indeed a continuous limit. If this limit is rotational invariant all
physical information is stored in $f$ as a function of $k$ only; thus the
average over all solid angle in the continuum has no effect at all, $\bar f
= f$, while the average over discrete shells $S_n$ greatly reduce the
statistical fluctuations.

\section{Linearized discrete dynamics}\label{app:B}

It is instructive to study first the discrete field equation
(\ref{recursion}) at the linearized level. Thus we consider the case
of a uniform field $\phi(\vx,t)=\phi(t)$, namely the recursion
\begin{equation}\label{uniform}
   \phi(t+a) + \phi(t-a) = 
 \frac{2\,\phi(t)}{1 + \tfrac12 \; a^2 \; [1 + \phi^2(t)]}
\end{equation}
Given $\phi(0)$ and $\phi(a)$, this recursion determines $\phi(t)$
for the discrete times $t=na$, $n=2,3,\ldots$, yielding the discrete
version of the well-known cnoidal uniform solution of the continuum
equation (\ref{eqnofmot}).

We now linearize the generic $\phi(\vx,t)$ around $\phi(t)$ by setting
$\phi(\vx,t)=\phi(t)+\eta(\vx,t)$, where $\eta(\vx,t)$ 
has vanishing space integral; 
this yields the linear recursion
\begin{equation}\label{linearec}
  \eta(\vx,t+a) + \eta(\vx,t-a) = 
 \frac14 \; B(\phi) \; \sum_{\bds\sigma} \eta(\vx+ a\bds\sigma,t) 
\end{equation}
where   $ B(\phi) $ is given by eq.(\ref{eq:eqdisp})
In terms of Fourier modes we have
\begin{equation*}
\eta(\vx,t) = {V^{-1/2}} \; \sum_{\vk\neq\bds 0}  
{\tilde\phi}_\vk(t) \; e^{i\vk\cdot\vx}
\end{equation*}
and eq.~(\ref{linearec}) entails, for $\vk\neq\bds 0$,
\begin{equation}\label{phikrec}
  {\tilde\phi}_\vk(t+a) + {\tilde\phi}_\vk(t-a) = 
  2\,{\tilde\phi}_\vk(t)\,\cos\omega_\vk(t)
\end{equation}
where the instantaneous frequency $ \omega_\vk(t) $ is fixed by 
the dispersion rule
\begin{equation}\label{disprule}
\cos\left[\omega_\vk(t) \; a \right]= B[\phi(t)] \; \prod_{j=1}^D \cos k_ja
\end{equation}
as function of the wavenumber vector $\vk\neq\bds 0$. Eq.~(\ref{phikrec}) is
our lattice version of Lam\'e equation governing the linearization of the
continuum field equation over the cnoidal. In fact, in the continuum limit
$a\to0$, for any {\bf fixed} $\vk$ we have
\begin{equation*}
  \cos\left[\omega_\vk(t) \, a \right]= 1 -\tfrac12 a^2 [k^2 + 1 +3 \; 
 \phi^2(t)]+{\cal O}(a^4)
\end{equation*}
so that eq.~(\ref{linearec}) becomes indeed
\begin{equation}\label{lame}
\Big[\frac{d^2}{dt^2} + k^2 + 1 +3 \; \phi^2(t)\Big]\,  {\tilde\phi}_\vk(t) = 0
\end{equation}
with $\phi(t)$ solving now $\ddot\phi+\phi+\phi^3=0$, that is the cnoidal solution.

In the trivial case $\phi=0$, which corresponds to the free field case,
the general solution of eq.~(\ref{phikrec}) is straightforward and 
can be written 
\begin{equation*}
  {\tilde\phi}_k(t) = A_\vk\,e^{-i\omega_\vk t} + 
  A_{-\vk}^\ast\,e^{i\omega_\vk t} 
\end{equation*}
in terms of the constant amplitudes $A_\vk$ fixed by the initial
conditions. In this case $\omega_\vk$ solves eq.~(\ref{disprule}) with
$B(0)=(1+a^2/2)^{-1}$ and is constant in time. Then recalling the general
relation eq.~(\ref{pikphik}) we obtain
\begin{equation*} 
  {\tilde\pi}_\vk(t) \simeq  -i \frac{\sin \left(\omega_\vk a\right)}{a} \left(
  A_\vk\,e^{-i\omega_\vk t} - A_{-\vk}^\ast\,e^{i\omega_\vk t} \right)
  \mp \frac{a^2}2\,{\tilde\phi}_\vk(t) \cos\left(\omega_\vk a\right) \; .
\end{equation*}
Finally, since time averages over several mode oscillations 
kill the interference between the positive frequency and the negative
frequency terms in the power spectra, we find in this $\phi=0$ case  
the following lattice version of the well known virial theorem for
harmonic oscillations
\begin{equation}\label{fvirial}
  \overline{|{\tilde\pi}_\vk|^2} = \frac1{a^2}\Big[1 - \big(1-\tfrac14 
  a^4\big)\, \cos^2\left(\omega_\vk a\right)\Big] \;
\overline{|{\tilde\phi}_\vk|^2} =
  \frac1{a^2}\Big[1 - \frac{1-\tfrac12 \, a^2}{1+\tfrac12 a^2} \; 
  \prod_{j=1}^D \cos^2 k_ja \Big] \;\overline{|{\tilde\phi}_\vk|^2}
\end{equation}
where the dispersion relation eq.~(\ref{disprule}) was used with
$\phi=0$. In the limit $a\to0$ eq.~(\ref{fvirial}) reduces to
\begin{equation*}
  \pik2 = (k^2 + 1) \;\phik2
\end{equation*} 
as expected.

When $\phi \neq 0$ we may still write the general solution of the
recursion~(\ref{linearec}) as
\begin{equation*}
  {\tilde\phi}_k(t) = A_\vk\,z_\vk(t) + 
  A_{-\vk}^\ast\,z_\vk^\ast(t)
\end{equation*}
where the amplitudes $A_\vk$ are constant in time while
$z_\vk(t)=z_{-\vk}(t)$ solve eq.~(\ref{linearec}) with initial conditions
\begin{equation*}
  z_\vk(0) = 1 \;,\quad   z_\vk(a) = e^{-i\omega_\vk(0)a} \; .
\end{equation*}

\bigskip

Eq.(\ref{lame}) admits close form solutions when  $\phi$ is given by the cnoidal solution 
eq.(\ref{scno}). In particular, the forbidden band for $ k^2 > 0 $ corresponds to
\begin{equation}\label{banda}
\frac12 \; \phi_0^2 + 3 \leq k^2 \leq \sqrt{ \frac13 \phi_0^4 + 
2 \, \phi_0^2 + 4} + 1 \; .
\end{equation}
(see for example \cite{cosmoV}).

\section{The Hartree approximation from the late time exact 
behaviour}\label{sec:hart}

We learn  how to obtain the Hartree frequency on the lattice: we
consider the discrete dynamics linearized as in Appendix~\ref{app:B}
and replace the background $\phi_0^2(t)$ with
$\overline{\phi^2}(t)$ in the dispersion relation eq.~(\ref{disprule}),
obtaining
\begin{equation}\label{eq:hartreedisp}
  \cos(a\,\Omega^{\xi\mathrm{H}}_\vk) = B[\xi\,\overline{\phi^2}(t)]
  \prod_{j=1}^3 \cos k_ja
\end{equation}
where $  B(\phi) $ is given by Eq.(\ref{eq:eqdisp}) and where 
for future use we introduced the parameter $\xi$ to control the
coupling with the background: if $\xi=1$ we have the  Hartree
frequency, if $\xi=0$ we have the free massive frequency 
eq.(\ref{freedisprule}). 
Intermediate values $ 0 < \xi < 1 $ will actually be ruled out below by 
fitting the numerical data.

$\Omega^{\xi\mathrm{H}}_\vk$ is a function of time through the background
$\overline{\phi^2}(t)$. On the lattice $\overline{\phi^2}(t)$ will tend to
a nonzero limit as $t\to\infty$. According to the Hartree dominance of the
equilibrium $\ephik2$, from eq.~(\ref{eq:eqom}) we than see that $\Omega_\vk
\simeq \Omega^{\mathrm{H}}_\vk$, {\bf provided} the full effective
frequency $\Omega_\vk$ tend to $\Omega^\mathrm{eq}_\vk$ as $t\to\infty$. It
is therefore very important to compare $\Omega^{\xi \, \mathrm{H}}_\vk$ with
the full effective frequency $\Omega_\vk$. Recall that the Hartree
approximation is a good approximation if the modes are weakly interacting,
since it neglects direct wave scatterings and allows energy transfer only
through the uniform self-consistent background $\overline{\phi^2}(t)$.  

\begin{figure}[htbp]
  \centering
  \psfrag{L12.8}{$L=12.8$ } 
  \psfrag{a0.064}{$a=0.064$ }
  \psfrag{EoV569.5}{$E/V=569.5$}
  \psfrag{tvals}{$t=$}
  \psfrag{xi0}{$\xi=0$}
  \psfrag{xi0.5}{$\xi=0.5$}
  \psfrag{xi1}{$\xi=1$}
  \psfrag{xi2}{$\xi=2$}
  \psfrag{Lambda}{$\Lambda=24.543$} 
  \psfrag{logt}{$\log t$}
  \psfrag{kvar}{$k$}
  \psfrag{theratio}{$1/Z_k^{\xi}$}
  \includegraphics[width=12.25cm,height=10cm]{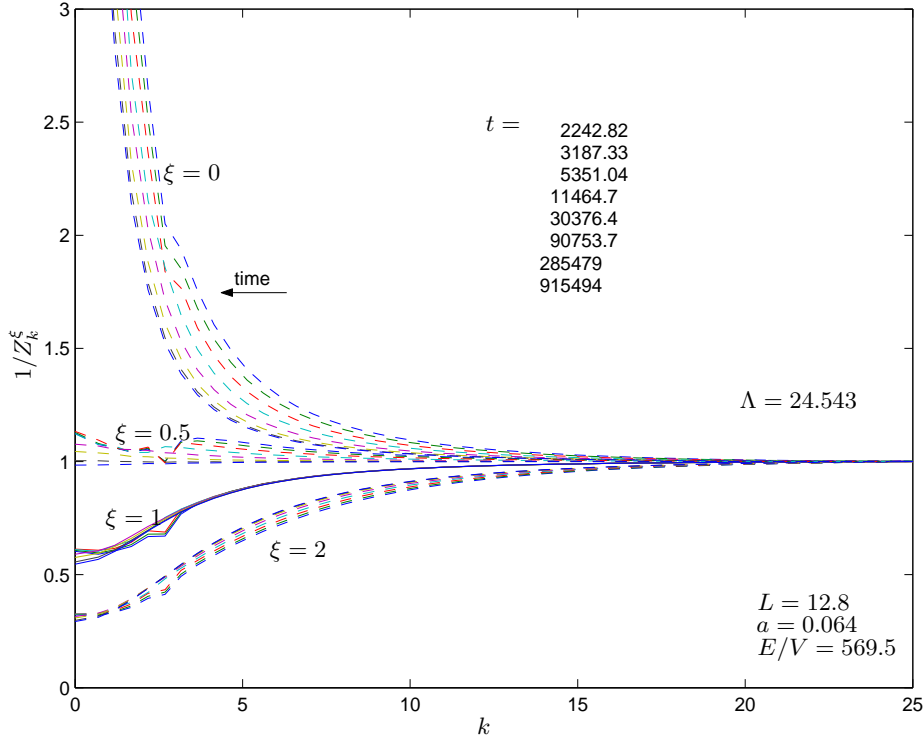}
\caption{The inverse of the ratio $ Z_k^\xi $ [defined by eq.(\ref{eq:R})] 
vs. the radial wavenumber for the values of time and of $\xi$ indicated. 
All parameters are as in fig.~\ref{opa064}.} \label{ratio}
\end{figure}

To perform the comparison we numerically average over discrete directions
and study the ratio
\begin{equation} \label{eq:R}
  Z_k^{(\xi)} \equiv 
\frac{\overline{|{\tilde\phi}_k|^2}}{a^2\,\overline{|{\tilde\pi}_k|^2}}
\Big[1 -\big(1-\tfrac14 a^4\big)\cos^2(a\,\Omega^{\xi\mathrm{H}}_k)\Big]
\end{equation}
In fig. \ref{ratio} we plot $Z_k^{(0)}$, $Z_k^{(0.5)}$, $Z_k^{(1)}$ and
$Z_k^{(1.5)}$ for several late times in the long evolution when $\pi$
thermalized (see previous section). The Hartree value $\xi=1$ clearly
stands out as the most accurate in the time dependence: the curves at all
different times almost perfectly collapse in a single curve, except that
for very small wavenumbers.  All curves collapse for large $k$, regardless
of the value of $\xi$, simply because when $k^2\gg \overline{\phi^2}(t)$
any effect on the dispersion relation due to the background is negligible.

\section{The cnoidal solution}\label{sec:cnoidal}

We discuss here the cnoidal solution of the $ \phi^4 $ equation 
(\ref{eqnofmot}), that is the space-independent solution of 
${\ddot \phi} + \phi + \phi^3 = 0$ explicitly given by 
\begin{equation}\label{scno}
  \phi(t) = \phi_0 \; \mbox{cn}\Big(t\,\sqrt{1+\phi_0^2},\, k\Big) 
\quad \mbox{with}\quad
k \equiv \frac{\phi_0}{\sqrt{2(1+\phi_0^2)}} \; ,
\end{equation}
where cn$(x,k)$ stands for the Jacobi cosinus of {\em module} $k$. This is a
(doubly) periodic function satisfying $\phi(t)= - \phi(t+ P)$, where $
P=2 \; K(k)/\sqrt{1+\phi_0^2} $ is the half-period and $ K(k) $
stands for the complete elliptic integral.

\begin{figure}[htbp]
  \psfrag{R vs. rho for the cnoidal solution}[r][r]
  {$R$ vs. $\rho$ for the cnoidal solution}
  \includegraphics[width=8cm,height=12.72cm,angle=270]{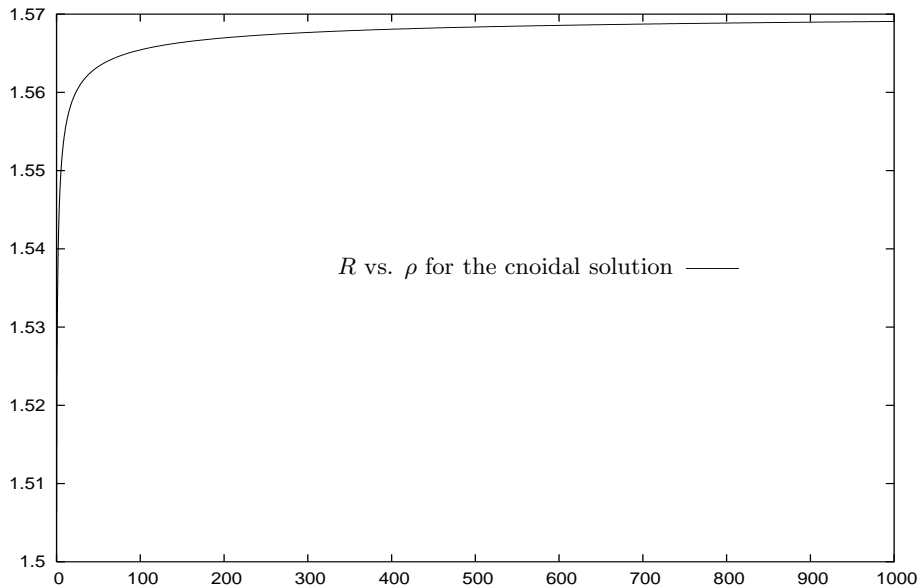}
  \caption{ $ R(\phi_0) =[ M_{\rm eff}^2 - 1]/ < \phi^2 > $ vs. $\rho$
    for the cnoidal solution Eq. (\ref{scno}). Here $<...>$ stands for the
    time average over one period}
\label{cno}
\end{figure}

The effective mass associated to this solution is therefore the basic frequency
\begin{equation}\label{mefcno}
  M_\mathrm{eff} = \frac{2 \; \pi}{2 \; P} = \frac{\pi \; 
    \sqrt{1+\phi_0^2}}{2 \; K(k)} \; .
\end{equation}
The average value of the squared field over a period can be computed with the 
result (see eq.~(3.40) in ref.\cite{Ngran}),
\begin{equation}\label{fimcno}
  \avg{\phi^2} = \frac1{2 \; P}\int_0^{2 \; P} dt \; \phi^2(t)=
  (1+\phi_0^2)\left[ \frac{2 \; E(k)}{K(k)} - 1 \right] - 1 \; .
\end{equation}
We plot in fig. \ref{cno} the ratio $ R(\phi_0) = (M_\mathrm{eff}^2 -
1)/\avg{\phi^2} $ using Eqs. (\ref{mefcno}) and (\ref{fimcno}) [compare
with eq.(\ref{R})]. We plot it as a function of the energy density $ \rho =
\frac12 \; \phi_0^2 \left( 1 + \frac12 \; \phi_0^2 \right) $.

We can compute analytically the function $ R(\phi_0) $ for small and large 
arguments with the result
$$
 R(\phi_0) \buildrel{\phi_0 \to 0 }\over= \frac32 + {\cal O}(\phi_0^2)
\quad , \quad R(\phi_0) \buildrel{\phi_0 \to \infty }\over= \frac{\pi}2 \; .
$$
In summary,  $ R(\phi_0) $ grows slowly and monotonically from the value 
$\frac32$ to the value $\frac{\pi}2$ when  $\phi_0$ varies from zero to 
infinity.

\bigskip

The average value of $ \dot\phi^2 $ can be computed analogously with the
 result (see eq.~(3.60) in ref.\cite{Ngran}),
\begin{equation}\label{fipmcno}
  \avg{\dot\phi^2} = \frac13
  (1+\phi_0^2)\left[\phi_0^2+2 -\frac{2 \; E(k)}{K(k)}\right] \; .
\end{equation}
We find from Eqs. (\ref{fimcno}) and  (\ref{fipmcno}) using energy conservation,
\begin{equation}\label{fi4mcno}
 \avg{\phi^4} = \frac13\left(\phi_0^4 + 6 \, \phi_0^2 + 8\right)
-\frac{8 \; E(k)}{3\; K(k)}\; (1+\phi_0^2)\; .
\end{equation}
Inserting Eqs. (\ref{fimcno}), (\ref{fipmcno}) and (\ref{fi4mcno})
for this homogeneous solution in the virial theorem Eq.(\ref{virial2}) shows that
this theorem is identically satisfied.

\end{document}